\documentclass[aps,prd,groupedaddress]{revtex4}
\usepackage{graphicx}
\usepackage{epsfig}
\usepackage{amsfonts}
\usepackage{amssymb}
\usepackage{epsf}
\usepackage{color}
\newcommand{\insertplot}[5]{\begin{figure}
 \hfill\hbox to 0.05in{\vbox to #5in{\vfill
 \inputplot{#1}{#4}{#5}}\hfill}
 \hfill\vspace{-.1in}
 \caption{#2}\label{#3}
 \end{figure}}
\newcommand{\inputplot}[3]{
 \special{ps: plotfile #1}
}

\newcounter{fig}

\begin{document}

\title{Properties of ultra-compact 
particle-like solutions in Einstein-scalar-Gauss-Bonnet theories}

\author{Burkhard Kleihaus}
\email[]{b.kleihaus@uni-oldenburg.de}
\author{Jutta Kunz}
\email[]{jutta.kunz@uni-oldenburg.de}
\affiliation{Institut f\"ur Physik, Universit\"at Oldenburg,D-26111 Oldenburg, Germany}
\author{Panagiota Kanti}
\email[]{pkanti@cc.uoi.gr}
\affiliation{Division of Theoretical Physics, Department of Physics,
             University of Ioannina, GR-45110, Greece}

\date{\today}
\pacs{04.70.-s, 04.70.Bw, 04.50.-h}
\begin{abstract}
Besides scalarized black holes and wormholes,
Einstein-scalar-Gauss-Bonnet theories allow also for particle-like solutions. 
The scalar field of these particle-like solutions diverges at the origin, 
akin to the divergence of the Coulomb potential
at the location of a charged particle.
However, these particle-like solutions possess a globally regular metric, 
and their effective stress energy tensor is free from pathologies, as well.
We determine the domain of existence for particle-like solutions in a number of 
Einstein-scalar-Gauss-Bonnet theories,
considering dilatonic and power-law coupling functions, and we analyze the 
physical properties of the solutions. Interestingly, the solutions may possess pairs of lightrings,
and thus represent ultra-compact objects. We determine the location of these lightrings,
and study the effective potential for the occurrence of echoes in the gravitational-wave spectrum.
We also address the relation of these particle-like solutions to the respective wormhole 
and black-hole solutions, 
and clarify the limiting procedure to recover the Fisher solution 
(also known as Janis-Newman-Winicourt-Wyman solution).
\end{abstract}
\maketitle

\section{Introduction}

While our current experiments and observations are all in agreement with
predictions of general relativity (GR), there are profound reasons
to investigate alternative theories of gravity,
that might represent low-energy effective theories, 
obtainable from some yet unknown fundamental theory of gravity
(see e.g., \cite{Will:2005va,Capozziello:2010zz,Berti:2015itd,Sotiriou:2015lxa}).
One such class of attractive alternative theories of gravity
is constituted by the so-called 
Einstein-scalar-Gauss-Bonnet (EsGB) theories.
EsGB theories amend GR by including quadratic curvature terms, 
in the form of a topological invariant, the Gauss-Bonnet (GB) term,
that is non-minimally coupled to a scalar field $\phi$.
These theories lead to second-order equations of motion,
and do not suffer from the Ostrogradski instability or ghosts
\cite{Horndeski:1974wa,Charmousis:2011bf,Kobayashi:2011nu}.

The EsGB action arises naturally in the framework of low-energy effective string theories,
in which case the scalar field represents the dilaton
\cite{Zwiebach:1985uq,Gross:1986mw,Metsaev:1987zx}.
The coupling function $F(\phi)$ is then of exponential type,
$F(\phi)=\alpha \exp{(-\gamma \phi)}$, with coupling constants
$\alpha$ and $\gamma$ (and string theory value $\gamma=1$).
The black-hole solutions of the string-theory motivated EsGB action
have been studied since a long time
\cite{Kanti:1995vq,Torii:1996yi,Guo:2008hf,Pani:2009wy,Pani:2011gy,Kleihaus:2011tg,Ayzenberg:2013wua,Ayzenberg:2014aka,Maselli:2015tta,Kleihaus:2014lba,Kleihaus:2015aje,Blazquez-Salcedo:2016enn,Cunha:2016wzk,Zhang:2017unx,Blazquez-Salcedo:2017txk,Konoplya:2019hml,Zinhailo:2019rwd}.
The EsGB equations of motion with this dilatonic coupling function do not
allow for the Schwarzschild or Kerr solutions of GR. 
Instead, all these black holes carry dilatonic hair,
evading the no hair theorem of GR 
(see e.g. \cite{Chrusciel:2012jk,Herdeiro:2015waa})
due to the presence of the GB term.
Interestingly, the GB term also allows for dilatonic wormholes, since the GB term
gives rise to an effective stress-energy tensor permitting
the violation of the null energy condition (NEC)
\cite{Kanti:2011jz,Kanti:2011yv,Cuyubamba:2018jdl}.
Unlike the wormholes in GR
\cite{Ellis:1973yv,Ellis:1979bh,Bronnikov:1973fh,Kodama:1978dw,Morris:1988cz,Visser:1995cc,Kleihaus:2014dla,Chew:2016epf},
these dilatonic wormholes do not need any type of exotic matter.

In recent years, EsGB theories with different coupling functions $F(\phi)$ moved into the focus of interest,
when it was realized that an appropriately chosen coupling function would allow for
curvature-induced spontaneous scalarization of black holes
\cite{Antoniou:2017acq,Doneva:2017bvd,Silva:2017uqg}.
To allow for spontaneous scalarization, the GR black holes should remain solutions
of the EsGB theories with a vanishing scalar field, and become unstable to the emergence of
a non-trivial scalar field at some critical value(s) of the GB coupling strength
\cite{Antoniou:2017acq,Doneva:2017bvd,Silva:2017uqg,Antoniou:2017hxj,Blazquez-Salcedo:2018jnn,Doneva:2018rou,Minamitsuji:2018xde,Silva:2018qhn,Brihaye:2018grv,
Myung:2018jvi,Bakopoulos:2018nui,Doneva:2019vuh,Myung:2019wvb,Cunha:2019dwb,
Macedo:2019sem,Hod:2019pmb,Bakopoulos:2019tvc, Collodel:2019kkx,Bakopoulos:2020dfg,Blazquez-Salcedo:2020rhf},
a mechanism reminiscent of matter-induced spontaneous scalarization in neutron stars
\cite{Damour:1993hw}.
The coupling function then needs a vanishing first derivative $dF/d\phi=0$ for vanishing scalar field $\phi=0$, 
and a positive second derivative. Other coupling functions will also allow for hairy black holes,
albeit similar to the dilatonic case \cite{Sotiriou:2013qea,Sotiriou:2014pfa,Antoniou:2017acq,Delgado:2020rev}.
As one might have expected, all these EsGB theories also allow for
wormhole solutions without the presence of any exotic matter
\cite{Antoniou:2019awm}.

Black holes and wormholes represent highly compact objects which allow to scrutinize the effects of
strong gravity predicted by GR and alternative theories of gravity, that give rise to distinct characteristic
gravitational radiation in merger events observable by current and future gravitational-wave
observatories (see e.g. \cite{Berti:2015itd,Cardoso:2016ryw,Berti:2018cxi,Berti:2018vdi}).
In particular, the presence of echoes in the ringdown may signal the absence of 
an event horizon in the final compact object
\cite{Cardoso:2016rao,Cardoso:2016oxy,Cardoso:2017cqb}.
Likewise, the shadows of highly compact objects may lead to 
distinctive observable effects depending significantly on the gravitational theory and on the type of
\textit{ultra-compact object} (UCO). Though current observations of the shadow of the supermassive
black hole at the center of M87 are in agreement with the shadow being produced
by a Kerr black hole \cite{Akiyama:2019eap}, they still provide new observational bounds.

Recently, we have realized that EsGB theories possess another class of interesting compact
objects, which represent particle-like solutions \cite{Kleihaus:2019rbg}.
These static and spherically symmetric  solutions 
possess a globally regular, asymptotically flat spacetime, but 
their scalar field diverges at the origin as $r^{-1}$. 
This is akin to the divergence of the Coulomb potential of a charged particle
located at the origin.
The effective stress-energy tensor of these particle-like solutions is
everywhere regular, and yields simple expressions at the origin.
Many of these particle-like solutions possess lightrings, and thus qualify as 
%\textit{ultra-compact objects} (UCOs) \cite{Cardoso:2017cqb}.
UCOs \cite{Cardoso:2017cqb}.
In fact, being horizonless, they always feature a pair of lightrings
in agreement with general arguments \cite{Cunha:2017qtt}.
The absence of a horizon also implies that these particle-like solutions will feature 
a sequence of echoes in a gravitational-wave signal
\cite{Cardoso:2016rao,Cardoso:2016oxy,Cardoso:2017cqb}.

Here, we provide a detailed discussion of these new particle-like solutions,
considering EsGB theories for a set of coupling functions
$F(\phi)$ of dilatonic and power-law type. 
We also allow for different boundary conditions of the scalar field
at spatial infinity, yielding two distinct classes of solutions: 
those which possess an asymptotically vanishing scalar field,
$\phi_\infty=0$, and those that feature a finite asymptotic value 
of the scalar field, $\phi_\infty \ne 0$.
A finite asymptotic value can be interpreted as a cosmological
value, since it will subsist in solutions describing the evolution
of the Universe. 
On the other hand, 
a vanishing asymptotic value in a theory with spontaneous scalarization will 
pass current constraints from binary mergers \cite{Sakstein:2017xjx},
since the scalar field may be set to zero in the cosmological context,
yielding an evolution of the Universe in agreement with the standard cosmological
$\Lambda$CDM model.

The structure of this paper is as follows:
In section II we present the theoretical setting, including the action,
the line-element and scalar-field ansatzes, and the equations of motion.
We discuss the expansions at the origin and at infinity, including regularity of
the effective stress-energy tensor and the curvature tensor,
for the particle-like solutions, in section III.
Here, we also discuss the limiting procedure that allows to recover the Fisher solution, 
also known as Janis-Newman-Winicourt-Wyman (JNWW) solution
\cite{Fisher:1948yn,Janis:1968zz,Wyman:1981bd,Agnese:1985xj,Roberts:1989sk}.
The numerical approach and the solutions themselves are presented in section IV.
In section V, we present the domain of existence of the solutions,
and address their relation to the black holes and wormholes of the respective theories.
We discuss possible observational effects including lightrings and echoes in section VI,
and we conclude in section VII.

\section{Theoretical setting}

We here consider the following effective action describing
a class of EsGB theories,
%motivated by the low-energy heterotic string theory
\begin{eqnarray}  
S=\frac{1}{16 \pi}\int d^4x \sqrt{-g} \left[R - \frac{1}{2}
 \partial_\mu \phi \,\partial^\mu \phi
 + F(\phi) R^2_{\rm GB}   \right],
\label{act}
\end{eqnarray} 
where $R$ is the curvature scalar,
$\phi$ is the scalar field with the coupling function $F(\phi)$,
and
\begin{eqnarray} 
R^2_{\rm GB} = R_{\mu\nu\rho\sigma} R^{\mu\nu\rho\sigma}
- 4 R_{\mu\nu} R^{\mu\nu} + R^2 
\end{eqnarray} 
is the quadratic Gauss-Bonnet correction term. 

Variation of the action with respect to the metric and the scalar field
leads to the Einstein equations and the scalar field equation,
\begin{eqnarray}
G^\mu_\nu & = & T^\mu_\nu \ , 
\label{Einsteq}\\
\nabla^\mu \nabla_\mu \phi & + & \dot{F}(\phi) R^2_{\rm GB}=0 \ ,
\label{scleq}
\end{eqnarray}
respectively, with the effective stress-energy tensor given by the expression
\begin{equation}
T_{\mu\nu} =
-\frac{1}{4}g_{\mu\nu}\partial_\rho \phi \partial^\rho \phi 
+\frac{1}{2} \partial_\mu \phi \partial_\nu \phi
-\frac{1}{2}\left(g_{\rho\mu}g_{\lambda\nu}+g_{\lambda\mu}g_{\rho\nu}\right)
\eta^{\kappa\lambda\alpha\beta}\tilde{R}^{\rho\gamma}_{\phantom{\rho\gamma}\alpha\beta}\nabla_\gamma \partial_\kappa F(\phi) \ .
\label{tmunu}
\end{equation}
Above, we have used the definitions
$\tilde{R}^{\rho\gamma}_{\phantom{\rho\gamma}\alpha\beta}=\eta^{\rho\gamma\sigma\tau}
R_{\sigma\tau\alpha\beta}$ and $\eta^{\rho\gamma\sigma\tau}= 
\epsilon^{\rho\gamma\sigma\tau}/\sqrt{-g}$,
and the dot denotes the derivative with respect to the scalar field $\phi$.

To obtain static, spherically-symmetric solutions we assume 
the line-element in the form 
\begin{equation}
ds^2 = -e^{f_0} dt^2 +e^{f_1}\left[dr^2 
+r^2\left( d\theta^2+\sin^2\theta d\varphi^2\right) \right]\ ,
\label{met}
\end{equation}
where the metric functions $f_0$ and $f_1$ are functions of the isotropic radial coordinate $r$ only. 
We also assume that the scalar field $\phi$ depends only on $r$.

Substitution of the ansatz for the metric and the scalar field into
the Einstein and scalar-field equations yields four coupled, nonlinear,
ordinary differential equations,
where three of them are of second order and one is of first order,  %constraint,
\begin{eqnarray}
0  & = & \dot{F} \left[
 \phi'(f_1'^3 r^2 - 4 f_1' f_1'' r^2 - 8 f_1'  - 8 f_1'' r)
-2 \phi''  ( f_1'^2 r^2 +4 f_1'  r)\right]/(2 e^{2f_1} r^2) 
\nonumber\\
& &
- \ddot{F} f_1' \phi'^2 \left[f_1' r + 4\right]/(e^{2f_1} r) 
+ (f_1'^2 r + 8 f_1' + 4 f_1'' r + \phi'^2 r)/(4 e^{f_1} r) \ , 
\label{odeq1}\\
0  & = &  \dot{F} f_0' \phi' \left[ - 3 f_1'^2 r^2 - 12 f_1' r - 8\right]/(2 e^{2f_1} r^2) 
+ (2 f_0' f_1' r + 4 f_0' + f_1'^2 r + 4 f_1' - \phi'^2 r)/(4 e^{f_1} r) \ ,
\label{odeq2}\\
0  & = &  -\dot{F} \left[\phi' (
f_0'^2 f_1' r + 2 f_0'^2 - 2 f_0' f_1'^2 r- 
2 f_0' f_1'  + 2 f_0' f_1'' +2 f_0'' f_1' + 4 f_0'' )
+ \phi''( 2 f_0' f_1'  r + 4 f_0' ) 
 \right]/(2 e^{2f_1} r) 
\nonumber\\
& &
- \ddot{F} f_0' \phi'^2 
\left[ f_1' r + 2\right]/(e^{2f_1} r) + (f_0'^2 r + 2 f_0' + 2 f_0'' r + 2 f_1' + 2 f_1'' r + 
\phi'^2 r)/(4 e^{f_1} r) \ ,
\label{odeq3}\\
0  & = & \dot{F} \left[f_0'^2 f_1'^2 r^2 + 4 f_0'^2 f_1' r - f_0' f_1'^3 r^2 + 
4 f_0' f_1' f_1'' r^2 + 8 f_0' f_1' + 8 f_0' f_1'' r + 2 f_0'' f_1'^2 r^2 + 
8 f_0'' f_1' r\right]/(2 e^{2f_1} r^2) 
\nonumber\\
& &
+ (f_0' \phi' r + f_1' \phi' r + 4 \phi' + 
2 \phi'' r)/(2 e^{f_1} r) \ .
\label{odeq4}
\end{eqnarray}
Equation (\ref{odeq3}) is now used to eliminate the second derivative $f_0''$ in Eq.~(\ref{odeq4}).
Also,  Eq.~(\ref{odeq2}) is used to eliminate $f_0'$ in % Eqs.~(\ref{odeq1}) and (\ref{odeq4}). 
Eq.~(\ref{odeq4}).
Thus, we end up with a first-order equation for the function $f_0$,
%that serves as a constraint, 
and two second-order equations for the functions $f_1$ 
and $\phi$ of the form
\begin{equation}
Q_{11} f_1'' + Q_{12} \phi'' - P_1 = 0 \ , \ \ \ 
Q_{21} f_1'' + Q_{22} \phi'' - P_2 = 0 \ , 
\label{eodes}
\end{equation}
where $Q_{kl}$ and $P_k$ depend on the functions $f_1$ and $\phi$, and their first-order derivatives.
Diagonalization of Eqs.~(\ref{eodes}) implies dividing by the determinant $\det Q$. 
If $\det Q$ possesses a node at some coordinate value $r_\star$, the respective 
solution of the equations of motion will no longer be regular but possess a cusp singularity at $r_\star$.
 
Furthermore, we note that the solutions are invariant under the scaling transformation
\begin{equation}
r \to \lambda r \ , \ \ \ F \to \lambda^2 F \ , \ \ \  \lambda > 0 \ .
\label{scalinvar}
\end{equation}

\section{Expansions and regularity}

We now discuss the expansions at infinity and at the origin for the particle-like solutions
of EsGB theories with various coupling functions, 
and we show regularity of their effective stress-energy tensor
and of their curvature tensor.
We then determine the redshift factor for these particle-like solutions,
and address the possibility of a Smarr-like mass relation.
We finally discuss the GR limit, where the Fisher (JNWW) solution, 
%or Janis-Newman-Winicourt-Wyman solution, 
is recovered
\cite{Fisher:1948yn,Janis:1968zz,Wyman:1981bd,Agnese:1985xj,Roberts:1989sk}.

\subsection{Expansion in the asymptotic region}
Since we are interested in localised objects, we look for asymptotically-flat solutions
described by power series expansions in $(1/r)$ in the asymptotic region
$r \to \infty$. Our equations of motion then yield, up to order ${\cal O}(r^{-5})$,  
%Performing the expansion in the asymptotic region $r \to \infty$  yields
%
\begin{eqnarray}
f_0 & = & - \frac{2M}{r}
          - \frac{(D^2 + 4M^2)M}{24 r^3}
          - 4\frac{\dot{F}_\infty DM}{r^4} 
	  +{\cal O}\left(r^{-5}\right) \ ,
\label{expf0} \\
f_1 & = & \frac{2M}{r}
          - \frac{D^2 + 4M^2}{8 r^2}
          + \frac{(D^2 + 4M^2)M}{24 r^3}
          - \frac{(D^2 + 4M^2)^2 - 512\dot{F}_\infty D M}{256r^4}
	  +{\cal O}\left(r^{-5}\right) \ ,
\label{expf1} \\
\phi & = & \phi_\infty
           -\frac{D}{r}
          - \frac{(D^2 + 4M^2)D}{48 r^3}
          - 4\frac{\dot{F}_\infty M^2}{r^4}
	  +{\cal O}\left(r^{-5}\right) \ ,
\label{expphi} 	  
\end{eqnarray}
where $\dot{F}_\infty = \dot{F}(\phi_\infty)$.
The constants
$M$ and $D$ denote the mass and the scalar charge of the solutions, respectively.
We note that the coefficients of all terms of higher-than-first order in $r^{-1}$
are completely determined by the mass $M$, the scalar charge $D$ and 
the asymptotic value of the scalar field $\phi_\infty$.
%There is no need to specify the coupling function $F(\phi)$ in this expansion.

%
\subsection{Expansion at the origin}
The expansion of the spacetime at the origin $r=0$ is more involved. 
Here, we need to specify the coupling functions and exploit their explicit $\phi$-dependence.
Let us, therefore,
restrict to polynomial coupling functions $F(\phi) = \alpha \phi^n$, with $n\geq 2$,
and to dilatonic coupling functions $F(\phi) = \alpha e^{-\gamma \phi}$.

We first consider polynomial coupling functions, $F(\phi) = \alpha \phi^n$, with $n\geq 2$.
We here assume a power series expansion for the metric functions,
whereas for the scalar field we assume a behavior of the type 
\begin{equation}
\phi =\phi_c -c_0/r +P_\phi(r)\ ,
\end{equation}
as $r\to 0$, where $P_\phi(r)$ is a polynomial in $r$. 
After substituting these expansions into the Einstein and scalar field equations, we can
successively determine the expansion coefficients.
We then find, that the lowest (non-zero) order 
in the expansion of the metric functions is exactly $n$,
whereas it is  $n-1$ in the expansion of the scalar-field polynomial $P_\phi(r)$.
For general $n\geq 2$, we find
\begin{equation}
f_0 = f_{0c} +\frac{e^{f_{1c}}}{16\alpha n^2} \left(-c_0\right)^{(2-n)} r^n +{\cal O}( r^{n+1}) \ , 
\ \ \ \ 
f_1 = f_{1c} +\frac{e^{f_{1c}}}{16\alpha n^2} \left(-c_0\right)^{(2-n)} r^n +{\cal O}( r^{n+1}) \ , 
\nonumber\\
\end{equation}
\begin{equation}
\phi = -\frac{c_0}{r} + \phi_c +
\frac{5-2n}{n^2(n-1)}\frac{e^{f_{1c}}}{64\alpha}\left(-c_0\right)^{(3-n)} r^{(n-1)} +{\cal O}( r^{n})\ .
\label{expgenn}  
\end{equation}
Explicitly we find, for instance, for $n=2$, the expansions
\begin{eqnarray}
f_0 & = & 
     f_{0c} 
     +\frac{e^{f_{1c}}}{64\alpha} r^2
     +\frac{e^{f_{1c}}\phi_c}{96\alpha c_0} r^3
     +{\cal O}(r^4) \ ,
\label{expf0c}     \\
f_1 & = & 
     f_{1c} 
     +\frac{e^{f_{1c}}}{64\alpha} r^2
     +\frac{\nu_3}{6} r^3
     +{\cal O}(r^4) \ ,
\label{expf1c}     \\
\phi & = & 
      -\frac{c_0}{r}
      +\phi_c
      -\frac{e^{f_{1c}} c_0}{256\alpha} r
      +\frac{32 \alpha c_0 \nu_3 - e^{f_{1c}}\phi_c}{768\alpha} r^2     
     +{\cal O}(r^3) \ ,
\label{expphic}  
\end{eqnarray}
where $f_{0c}$, $f_{1c}$, $\nu_3$, $\phi_c$, and $c_0$ are constants.
The coefficients of all higher-order powers can be expressed in terms of these constants.

In the case of a dilatonic coupling function, $F(\phi) = \alpha e^{-\gamma \phi}$, a power-series
expansion fails to lead to a local solution. 
Assuming $\gamma>0$, we instead consider a non-analytic behaviour of the type
\begin{equation} 
f_0 = f_{0c} + p_0(r) e^{-\gamma \frac{c_0}{r}} + ... \ , \ \ \ 
f_1 = f_{1c} + p_1(r) e^{-\gamma \frac{c_0}{r}} + ... \ , \ \ \  
\phi =  -\frac{c_0}{r}+\phi_c + p_\phi(r) e^{-\gamma \frac{c_0}{r}} +...  \ , \ \ \  
\label{exorgdil}
\end{equation}
where now $c_0>0$, $p_0(r)$, $p_1(r)$, and $p_\phi(r)$ are polynomials in $r$,
and the dots indicate terms of order less than $e^{-\gamma \frac{c_0}{r}}$.
Substitution into the Einstein and scalar-field equations then yields
to lowest powers in $e^{-\gamma \frac{c_0}{r}}$ and $r$
\begin{eqnarray}
f_0  &=&  f_{0c}+
\left(\frac{e^{f_{1c}}}{16\gamma^2 \alpha e^{-\gamma \phi_c}}\right)
 r^2 \left(1-2\frac{r}{\gamma c_0}\right) e^{-\gamma \frac{c_0}{r}}
 \ ,  
\label{expf0org} \\
f_1 &=& f_{1c} +
\left(\frac{e^{f_{1c}}}{16\gamma^2 \alpha e^{-\gamma \phi_c}}+\nu_3 r\right) r^2\,
e^{-\gamma \frac{c_0}{r}} \ , 
\label{expf1org} \\
\phi &=&  -\frac{c_0}{r}+\phi_c + c_0 \left[\frac{e^{f_{1c}}}{32 \gamma^2 \alpha e^{-\gamma \phi_c}} \left(1-\frac{3}{2}\,\frac{1}{\gamma c_0}\,r\right) +\frac{\nu_3}{2}\,r\right]
r e^{-\gamma \frac{c_0}{r}}.
\label{expfporg}
\end{eqnarray}

\subsection{Stress-energy tensor and curvature tensor}

We will now demonstrate the regularity of both the effective
stress-energy tensor and curvature tensor at the origin $r=0$.
In order to find the effective stress-energy tensor at the origin,
we substitute the respective expansions into the Einstein tensor and make use of the Einstein equations.
For the polynomial coupling function with $n=2$, we find, for the non-vanishing components
of the effective stress-energy tensor, the results
\begin{equation}
T^t_t(0) = \frac{3}{32\alpha} \ , \ \ \ 
T^r_r(0) = T^\theta_\theta(0) = T^\varphi_\varphi(0) = \frac{2}{32\alpha} \ .
\label{Torg}
\end{equation}
Introducing the energy density $\epsilon_0= -T^t_t(0)$ and the
pressure $p_0 =T^r_r(0) =T^\theta_\theta(0) =T^\varphi_\varphi(0)$, 
the above expressions lead to a
homogeneous equation of state $p_0(\epsilon_0) = -\frac{2}{3}\epsilon_0$
at the origin. It is interesting to note that the stress-energy tensor
at the origin does not depend on the mass or the scalar charge. 
Also, the value of the energy density at the origin is not sign-definite and, in fact,
is negative for positive coupling constant $\alpha$.
If instead we consider polynomial coupling functions with powers
$n>2$ or dilatonic coupling functions, then 
the stress-energy tensor vanishes identically at the origin, thus remaining again regular.

Let us now turn to the curvature invariants of the theory.
If we substitute the expansions  Eqs.~(\ref{expf0c})--(\ref{expphic})
into the curvature invariants, for the polynomial coupling function with $n=2$,
we obtain the expressions
\begin{equation}
R(0) = -\frac{9}{32\alpha } \ , \ \ \ 
R_{\mu\nu} R^{\mu\nu}(0) = \frac{21}{(32\alpha )^2} \ , \ \ \ 
R_{\mu\nu\kappa\lambda} R^{\mu\nu\kappa\lambda}(0) = \frac{15}{(32\alpha )^2} \ , \ \ \ 
R_{\rm GB}^2 =  \frac{12}{(32\alpha )^2} \ ,
%R(0) = -\frac{9}{32\alpha r_0^2} \ , \ \ \ 
%R_{\mu\nu} R^{\mu\nu}(0) = \frac{21}{(32\alpha r_0^2)^2} \ , \ \ \ 
%R_{\mu\nu\kappa\lambda} R^{\mu\nu\kappa\lambda}(0) = \frac{15}{(32\alpha r_0^2)^2} \ , \ \ \ 
%R_{\rm GB}^2 =  \frac{12}{(32\alpha r_0^2)^2} \ ,
\label{curvinvorg}
\end{equation}
which again depend only on the coupling constant $\alpha$. 
As was the case for the stress-energy tensor, the curvature invariants vanish at the origin
for either polynomial coupling functions with powers $n>2$ 
or for dilatonic coupling functions. 

It is also straightforward to show that all components of the stress-energy tensor
as well as all curvature invariants vanish at asymptotic infinity in accordance with the condition
of asymptotic flatness. As we will demonstrate in the following sections, where complete numerical
solutions are presented, our particle-like solutions are characterized by regularity not only
at these two asymptotic regions but over the entire radial regime.

\subsection{Redshift and Smarr-type mass relation}

The redshift constitutes an interesting observable for compact objects.
To obtain the redshift, we consider
the ratio of the wavelength of a photon as measured by an observer in the 
asymptotic region, $\lambda_{\rm asym}$, 
to the wavelength of the photon as emitted at the origin, $\lambda_{\rm emit}$,
\begin{equation}
\frac{\lambda_{\rm asym}}{\lambda_{\rm emit}}= 
\frac{\sqrt{-g_{tt}(\infty)}}{\sqrt{-g_{tt}(0)}} = e^{-f_0(0)/2} \ .
\end{equation}
The corresponding redshift factor $z$ is then given by 
\begin{equation}
z =\frac{\lambda_{\rm asym}}{\lambda_{\rm emit}} -1 =  e^{-f_0(0)/2}-1 \ .
\label{redshift}
\end{equation}
For black holes and wormholes in dilatonic EsGB theories, Smarr-type mass relations
are known to exist 
\cite{Kleihaus:2011tg,Kanti:2011jz}.
In order to obtain analogously a Smarr-type mass relation for the particle-like
solutions, we start from the Komar expression for the mass
\begin{equation}
M=\frac{1}{4\pi}\,\int_\Sigma R_{\mu\nu}\, \xi^\mu n^\nu\,dV= 
-\frac{1}{4\pi}\,\int_\Sigma R^t_{t}\,\sqrt{-g}\,d^3x\,.
\label{Komar}
\end{equation}
In the above expression, $\xi^\mu=(1,0,0,0)$ is the timelike Killing vector,
$n^\nu=(e^{-f_0/2}, 0,0,0)$ is the unit vector normal to the spacelike hypersurface
$\Sigma$, and $dV$ the natural volume element on $\Sigma$. 
Making use of the equations of motion, we may write
\begin{equation}
R^t_t=T^t_t-\frac{1}{2}\,T^\mu_\mu=\frac{1}{2}\,(T^t_t-T^r_r-2T^\theta_\theta) + 
\lambda \,\Bigl[\nabla^\mu \nabla_\mu \phi+\dot F(\phi) R^2_{GB}\bigr]\,,
\end{equation}
where, as a last term, we have added a zero in the form of the scalar-field equation
multiplied by a real constant $\lambda$. If we also multiply the above by $\sqrt{-g}$, we write
\begin{equation}
\sqrt{-g}\,R^t_t=\sqrt{-g}\,\left[\frac{1}{2}\,(T^t_t-T^r_r-2T^\theta_\theta) + 
\lambda \dot F(\phi) R^2_{GB}\right]+
\lambda \,\partial_\mu (\sqrt{-g}\,\partial^\mu \phi)\,,
\end{equation}
or, if we use the explicit expressions for the $T^\mu_\nu$ components and the
GB term,
\begin{eqnarray}
\sqrt{-g}\,R^t_t &=&-2 \sin\theta \,\partial_r \left\{e^{(f_0-f_1)/2} r \,\left[
f_1' \Bigl(1+\frac{f_1' r}{4}\Bigr) (F f_0' -F') + F' f_0'\,\Bigl(1+\frac{f_1' r}{2}\Bigr) 
-\frac{\lambda}{2}\,e^{f_1} \phi' r \right]\right\} \nonumber \\[1mm]
&+& (F + 2 \lambda \dot F)\,\partial_r\left[e^{(f_0-f_1)/2} r f_0' f_1' 
\Bigl(1+\frac{f_1' r}{4}\Bigr)\right]\,. \label{finalRtt}
\end{eqnarray}
For a nontrivial coupling function $F(\phi)$, the last term in the above expression
prevents us from writing $\sqrt{-g}\,R^t_t$ in the form of a total derivative with respect to the
radial coordinate. If, however, it holds that  $F + 2 \lambda \dot F=0$, i.e. if 
$F(\phi)=\alpha e^{-\gamma \phi}$ with $\gamma=1/2 \lambda$, the last term in
Eq. (\ref{finalRtt}) vanishes. Then, employing the expansions for the metric functions and
the scalar field at infinity and at the origin, the integral over the volume in Eq. (\ref{Komar})
leads to the relation
\begin{equation}
M+ \frac{D}{2\gamma} = 
e^{\frac{f_{0c}+f_{1c}}{2}}
\frac{c_0}{4\gamma}
\left(1-8 \alpha c_0 \nu_3 \gamma^3 e^{-f_{1c}} e^{-\gamma \phi_c}\right) \ .
\label{smarr}
\end{equation}
The above Smarr-type formula combines the global charges of the particle-like
solutions obtained at infinity, the mass $M$ and the scalar charge $D$, with an expression
of the fields evaluated at the origin. The latter here replaces the contributions from the horizon
in the case of black holes and the throat in the case of wormholes.
As in the case of the aforementioned studies, a non-integral, closed-form for
the Smarr-type relation can be derived only in the case of the dilatonic coupling function.

\subsection{GR limit}
%
%For vanishing GB coupling constant, 
For a constant GB coupling function,  the EsGB theory reduces to 
GR with a self-gravitating scalar field.
In this limit, a singular solution is known in closed form, the Fisher (JNWW) solution 
%or Janis-Newman-Winicourt-Wyman (JNWW) solution
\cite{Fisher:1948yn,Janis:1968zz,Wyman:1981bd,Agnese:1985xj,Roberts:1989sk}.
In terms of the metric (\ref{met}) with isotropic coordinates, this solution reads
\begin{equation}
e^{f_0} = \left(\frac{1-r_{s}/r}{1+r_{s}/r}\right)^{2s} \ , \ \ \
e^{f_1} = \left(1-r_{s}^2/r^2\right)^2\, e^{-f_0} 
            \ , \ \ \
\phi=\phi_\infty \pm \frac{d}{2} \, f_0 \ ,
\label{Fishersol}
\end{equation}
where $d=D/M$ is the scaled scalar charge, and $s=1/\sqrt{1+d^2/4}$.
The curvature singularity is located at $r_{s} =M/2s$.
In the limit $d\to 0$, corresponding to $s\to 1$, the Schwarzschild solution is obtained.
Note that in this limit $\phi=\phi_\infty$.

Let us now address the relation between the particle-like EsGB solutions and the
Fisher solution.
We note that in the limit $d \to \infty$ the scale symmetry, Eq.~(\ref{scalinvar}),
of the field equations
implies that the mass, the scalar charge,
and the coupling strength scale as
\begin{equation}
M \to \lambda M \ , \ \ \ D \to \lambda D \ , \ \ \ \alpha \to \lambda^2 \alpha \ .
\end{equation}
By scaling with a factor $\lambda = 1/D$, we can achieve that
the coupling functions vanish in this limit.
We are then left with a solution for a self-gravitating scalar field in GR.
Solutions with $M=0$ exist in closed form
\begin{equation}
e^{f_0} = 1\ , \ \ \ 
e^{f_1} = \left(\frac{1-\frac{D}{4r}}{1+\frac{D}{4r}}\right)^2 \ , \ \ \
%\phi   = \phi_\infty  + 2\log\left(\frac{1-\frac{D}{4r}}{1+\frac{D}{4r}}\right)  \ ,
\phi   = \phi_\infty  +f_1 \ ,
\label{dlimsol}
\end{equation}
which possess a curvature singularity at $r=D/4$.

This solution can be obtained from the Fisher solution (\ref{Fishersol})
by taking the limit $d\to \infty$ while keeping $r_{s}$ fixed. 
This implies taking the limit $s\to 0$ while keeping fixed $s\, d$.
Similarly, by  fixing $d$ and taking the limit $\hat{\alpha} \to 0$, 
where $\hat{\alpha}=8\alpha/M^2$,
we see that the particle-like solutions tend to the Fisher solution,
when the radial coordinate $r$ is larger than the value of the
radial coordinate of the Fisher singularity,  i.e.  $r> r_{\rm s}$.
In the interval $0\leq {r} \leq r_{\rm s}$ the metric components $g_{tt}$ and $g_{rr}$ 
tend to zero, leading again to a singular region.

\section{Numerical approach and solutions}

We turn now to our numerical analysis.
We first discuss the numerical approach used in constructing the particle-like solutions.
Then, we present a set of solutions, construct their embedding diagrams,
illustrate the cusp singularity and indicate the GR limit.

\subsection{Numerical approach}
In order to solve numerically the set of the second-order ordinary differential
equations given in Eqs. (\ref{eodes}), we introduce the new coordinate
\begin{equation}
\xi = \frac{r_0}{r} \ ,
\label{xi}
\end{equation}
where the constant $r_0$ is a scaling parameter. 
We also rescale the coupling constant $\alpha \to \alpha r_0^2$. 
Then, all the differential equations are independent of the parameter $r_0$.

In terms of the new coordinate $\xi$, the asymptotic expansions (\ref{expf0})-(\ref{expphi})
read 
\begin{equation}
f_0  \to -2M \xi + {\cal{O}}(\xi^2) \ , \ \ \ 
f_1  \to  2M \xi + {\cal{O}}(\xi^2) \ , \ \ \ 
\phi \to \phi_\infty-D  \xi + {\cal{O}}(\xi^2) \  \ ,
\label{asymp}
\end{equation}
where $M$ and $D$ have also been rescaled, i.~e.~$M\to r_0 M$, $D\to r_0 D$.
We recall that the coefficients of all higher order terms in $\xi$ can be expressed in 
terms of the three constants $M$,  $D$ and $\phi_\infty$. 
Thus, for a given coupling function, the solutions are then completely determined
by these three parameters.

We treat the set of ordinary differential equations as an initial value problem, for which
we employ the fourth order Runge Kutta method.
The initial values for the functions follow immediately from the expansion (\ref{asymp})
\begin{equation}
\left. f_0\right._{\rm ini} =  0 \ , \ \ \ 
\left. f_1\right._{\rm ini} =  0 \ , \ \ \ 
\frac{d f_1}{d\xi}_{\rm ini} =  2M\ , \ \ \ 
\phi_{\rm ini} =  \phi_\infty\ , \ \ \ 
\frac{d\phi}{d\xi}_{\rm ini} = -D\ . 
\label{inival}
\end{equation}
We perform the numerical calculations in the interval
$\xi_{\rm min} \leq \xi \leq \xi_{\rm max}$, choosing always 
the lower bound $\xi_{\rm min}=10^{-8}$. 
On the other hand, for the upper bound we choose $\xi_{\rm max}=100$ in the case
of the quadratic coupling function,
and $\xi_{\rm max}=10$ in the case of the dilatonic coupling function.

\subsection{Numerical solutions}

%Profile functions (examples)
\begin{figure}
\begin{center}
(a)\includegraphics[width=.45\textwidth, angle =0]{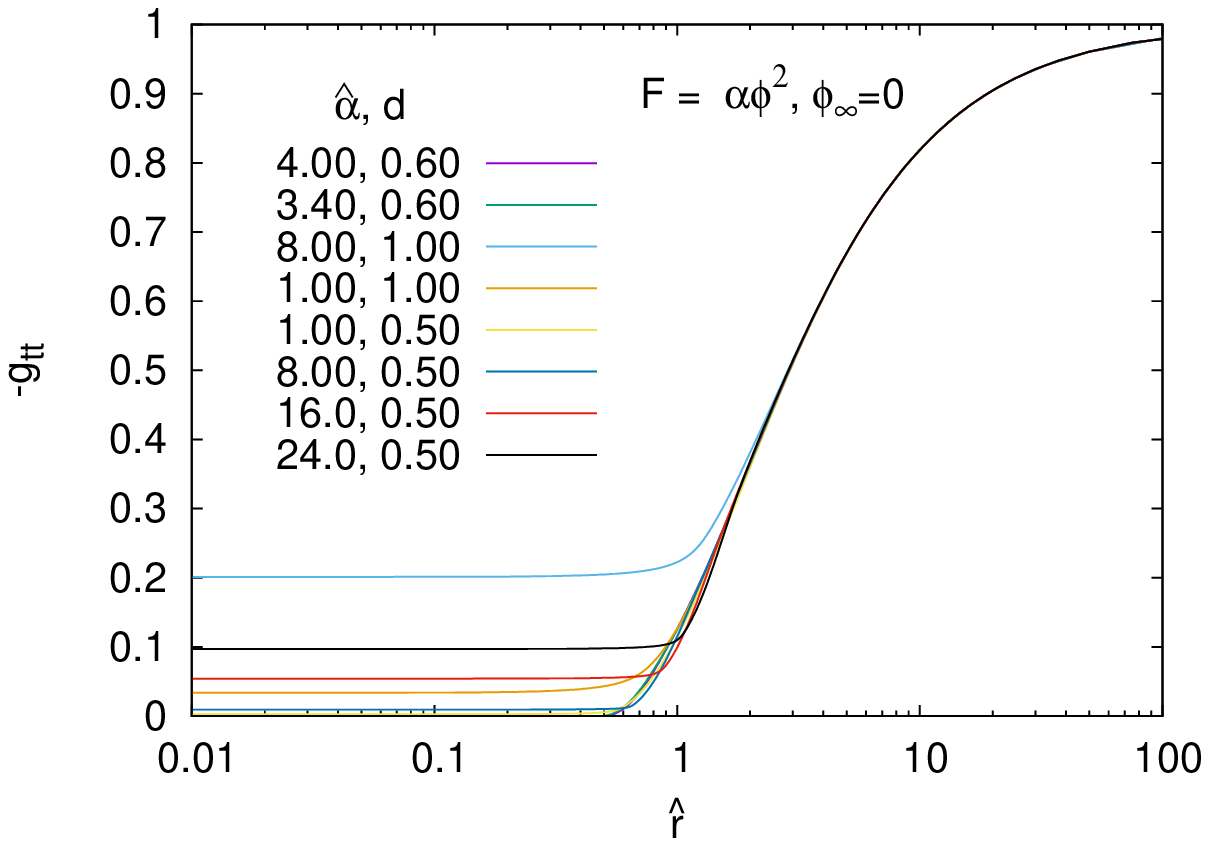}
(b)\includegraphics[width=.45\textwidth, angle =0]{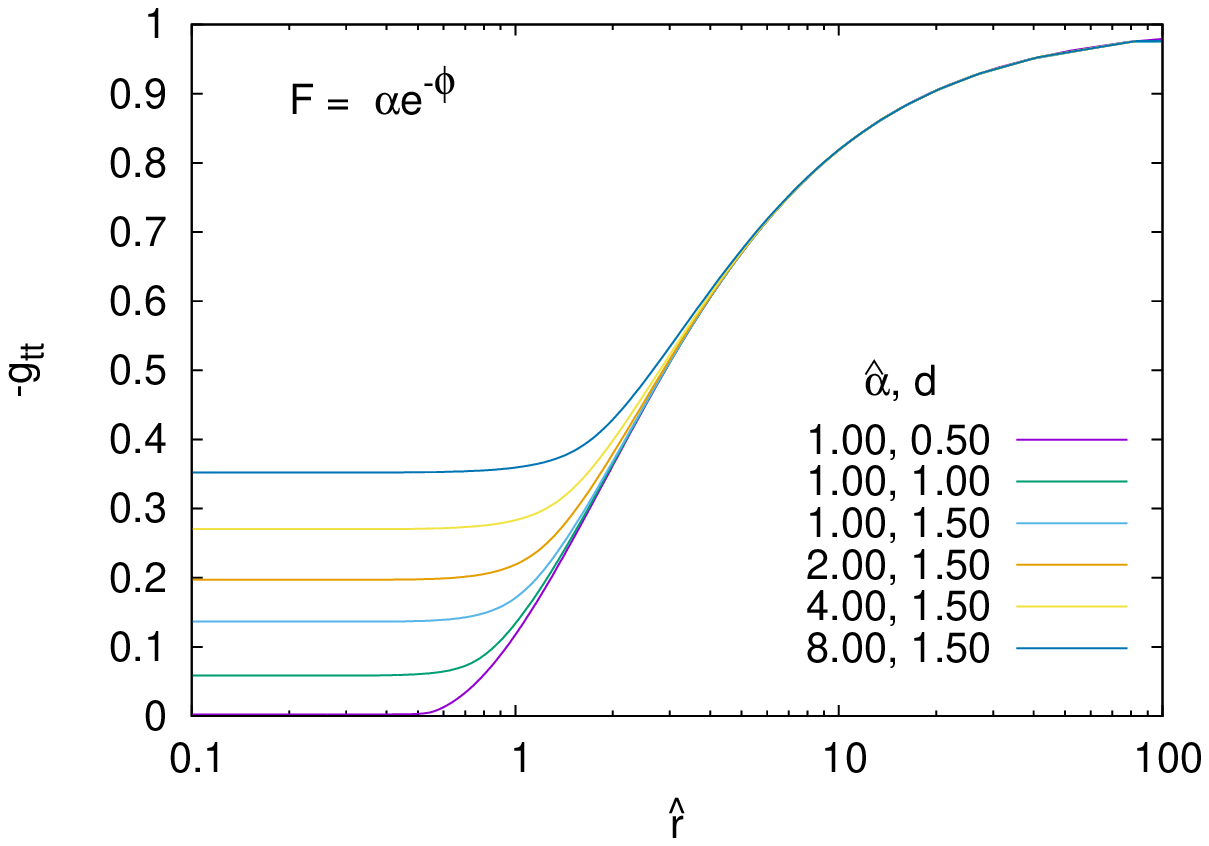}
\\
(c)\includegraphics[width=.45\textwidth, angle =0]{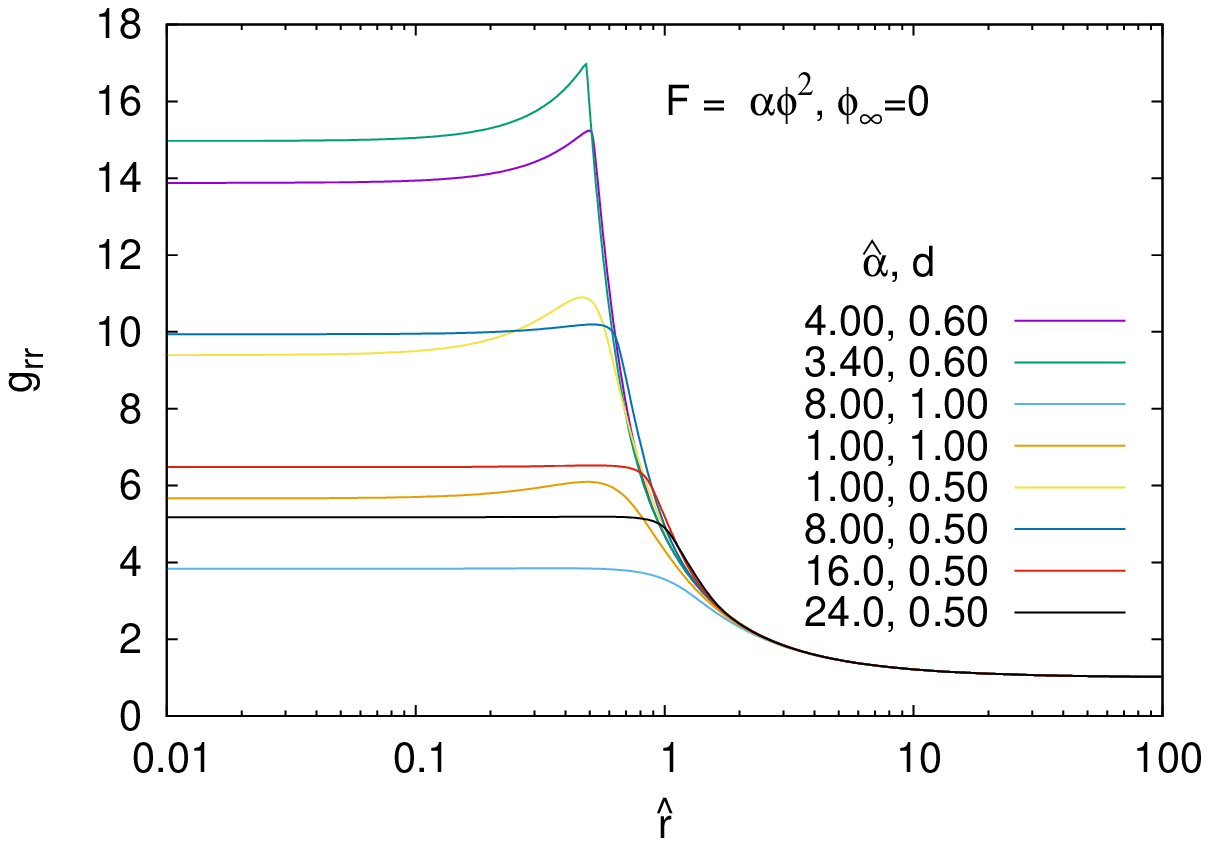}
(d)\includegraphics[width=.45\textwidth, angle =0]{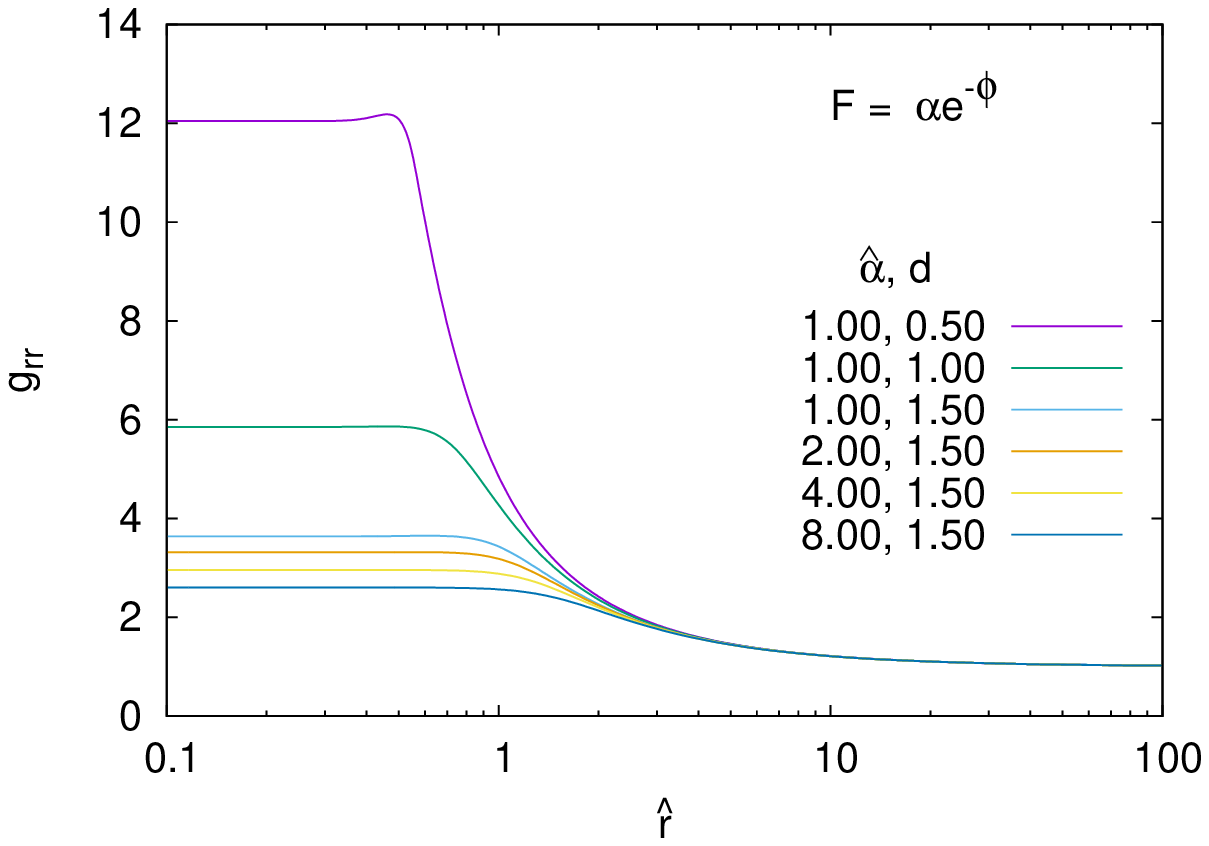}
\\
(e)\includegraphics[width=.45\textwidth, angle =0]{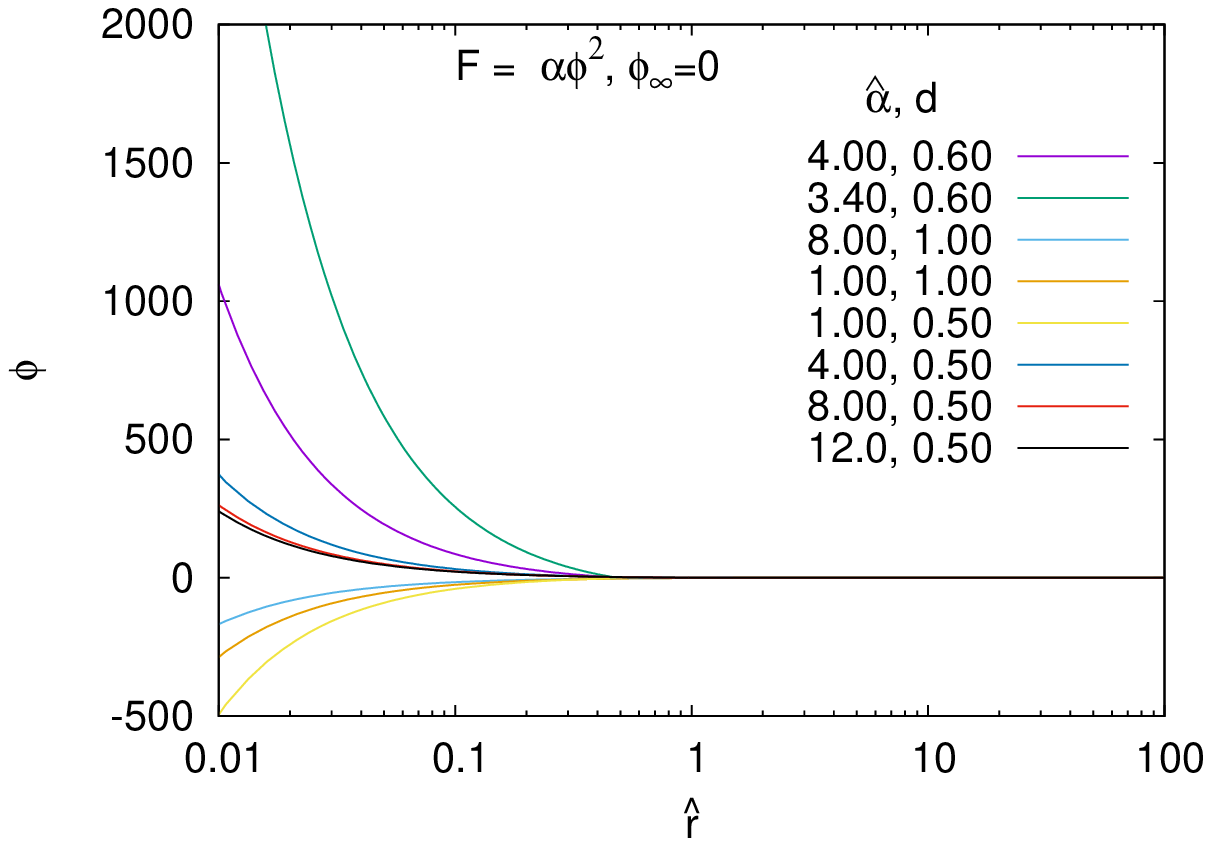}
(f)\includegraphics[width=.45\textwidth, angle =0]{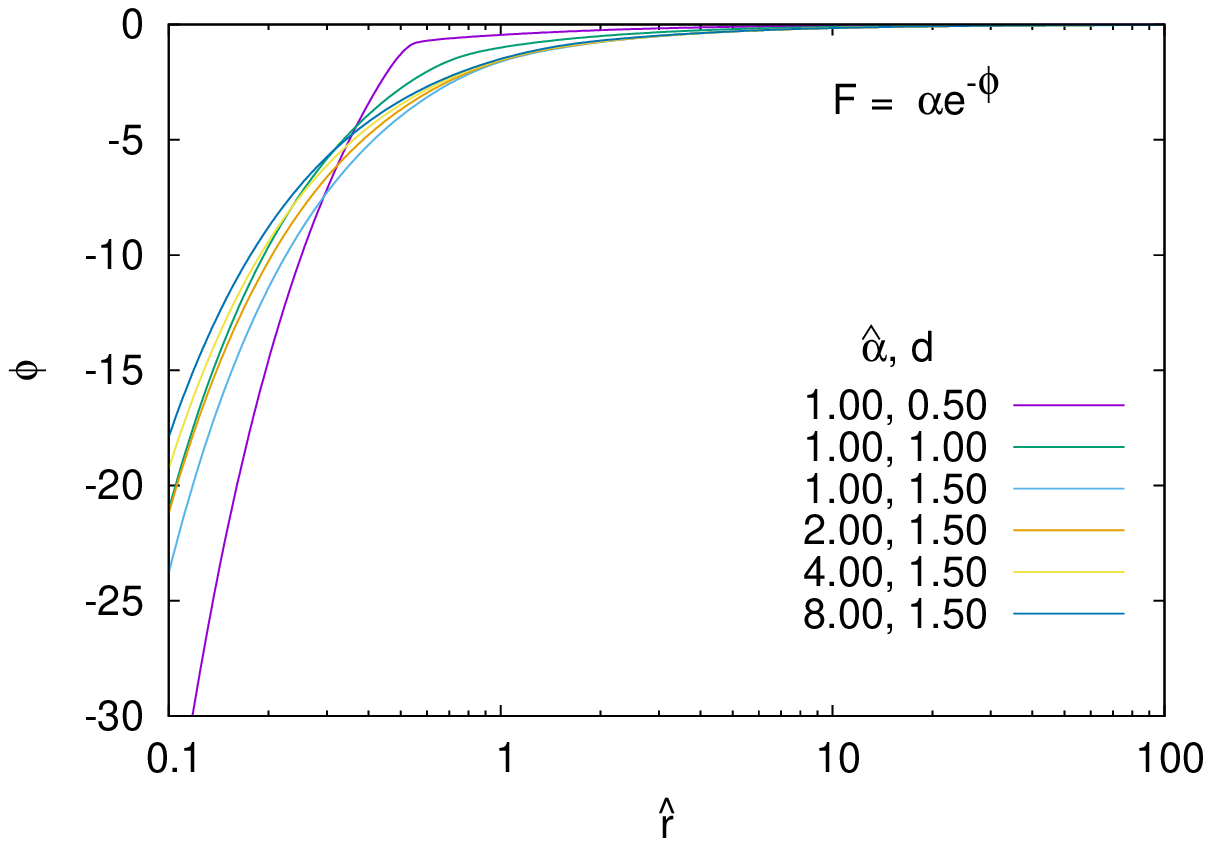}
\end{center}
\caption{
(a)-(f) The metric functions $-g_{tt}$ and $g_{rr}$ and the scalar field $\phi$ vs
the scaled radial coordinate $\hat{r}$ for the coupling functions 
$F=\alpha\phi^2$ with $\phi_\infty=0$ [left column: (a), (c), (e)] and 
$F=\alpha e^{-\phi}$ [right column: (b), (d), (f)] for several values of 
$\hat{\alpha}$ and  $d$. }
\label{fig_profile}
\end{figure}

\begin{figure}
\begin{center}
(a)\includegraphics[width=.45\textwidth, angle =0]{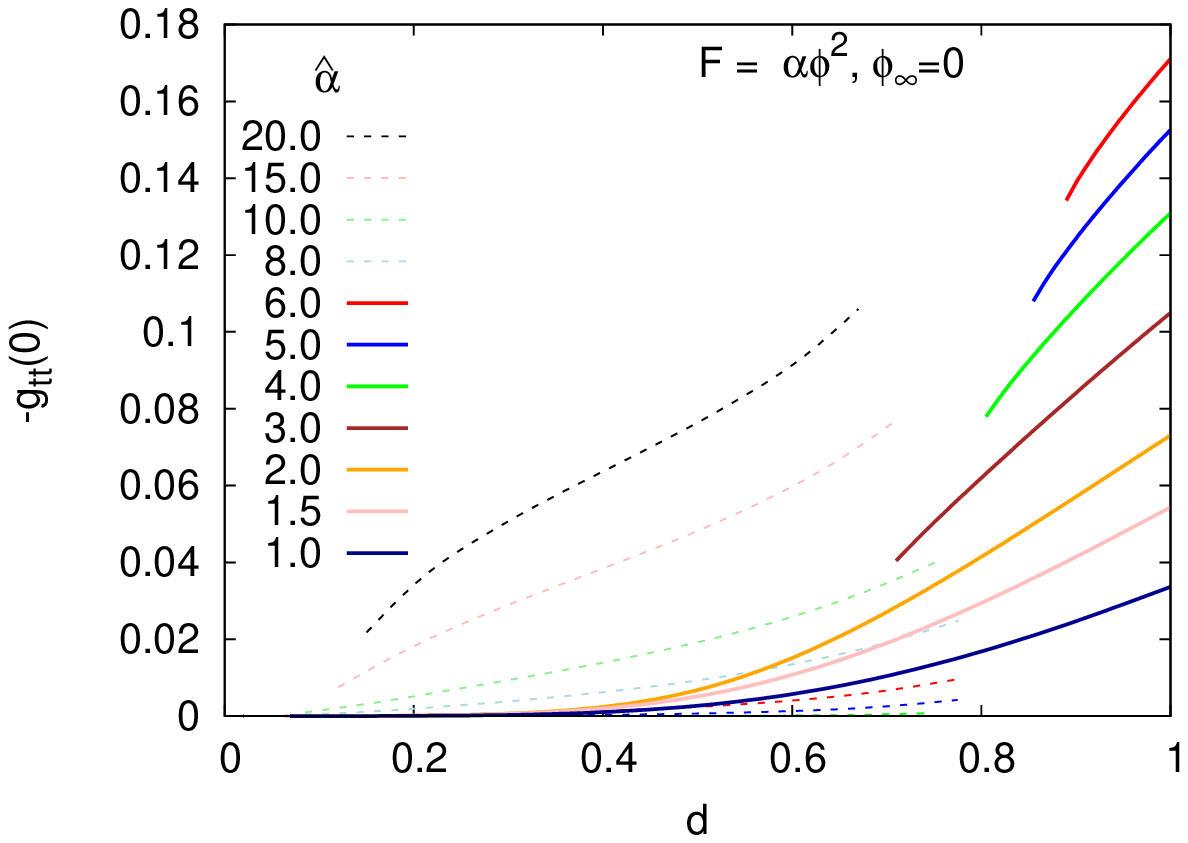}
(b)\includegraphics[width=.45\textwidth, angle =0]{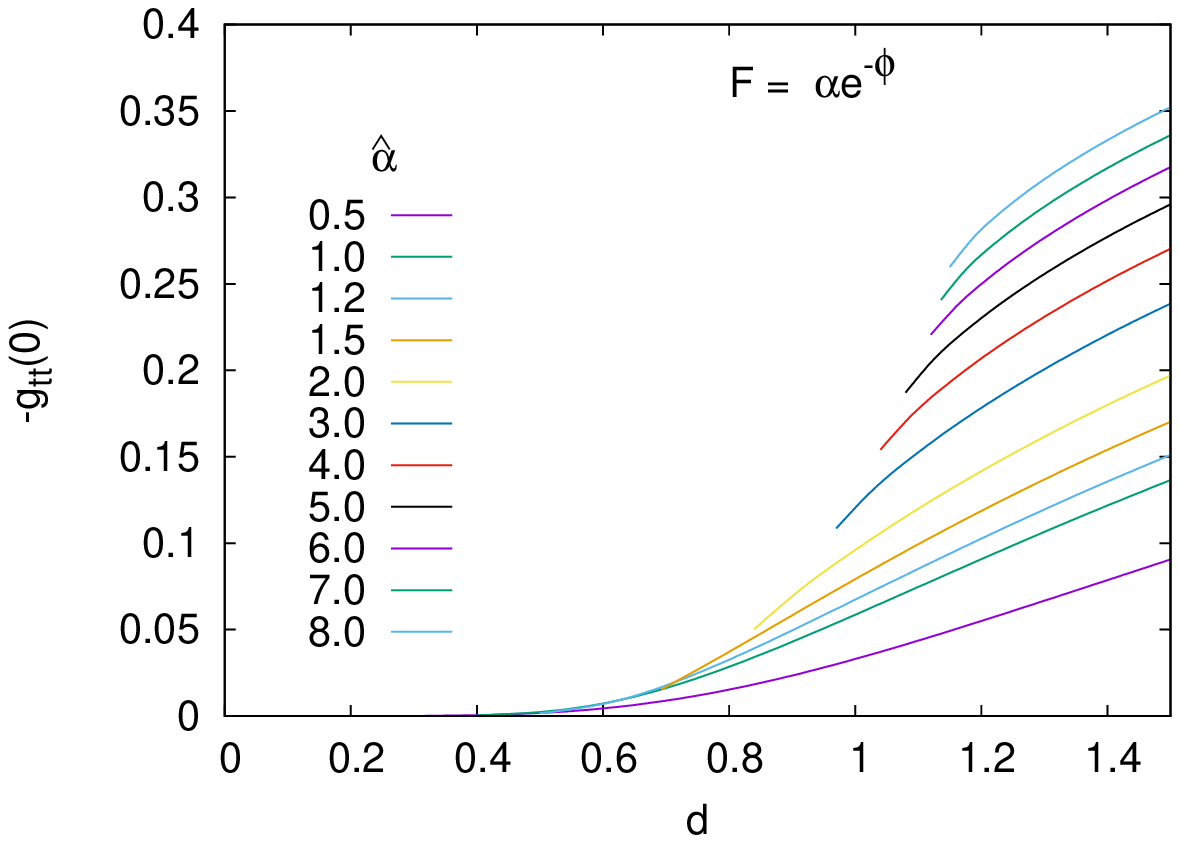}
\\
(c)\includegraphics[width=.45\textwidth, angle =0]{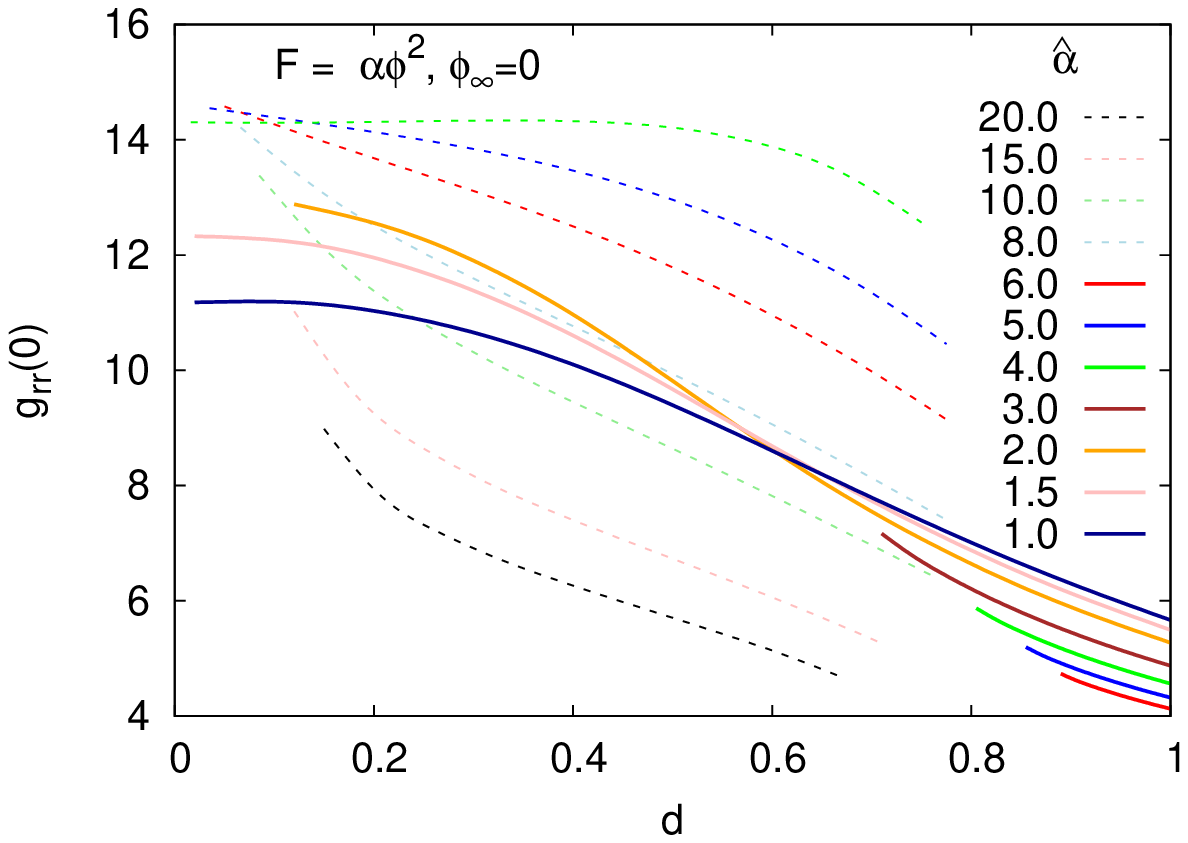}
(d)\includegraphics[width=.45\textwidth, angle =0]{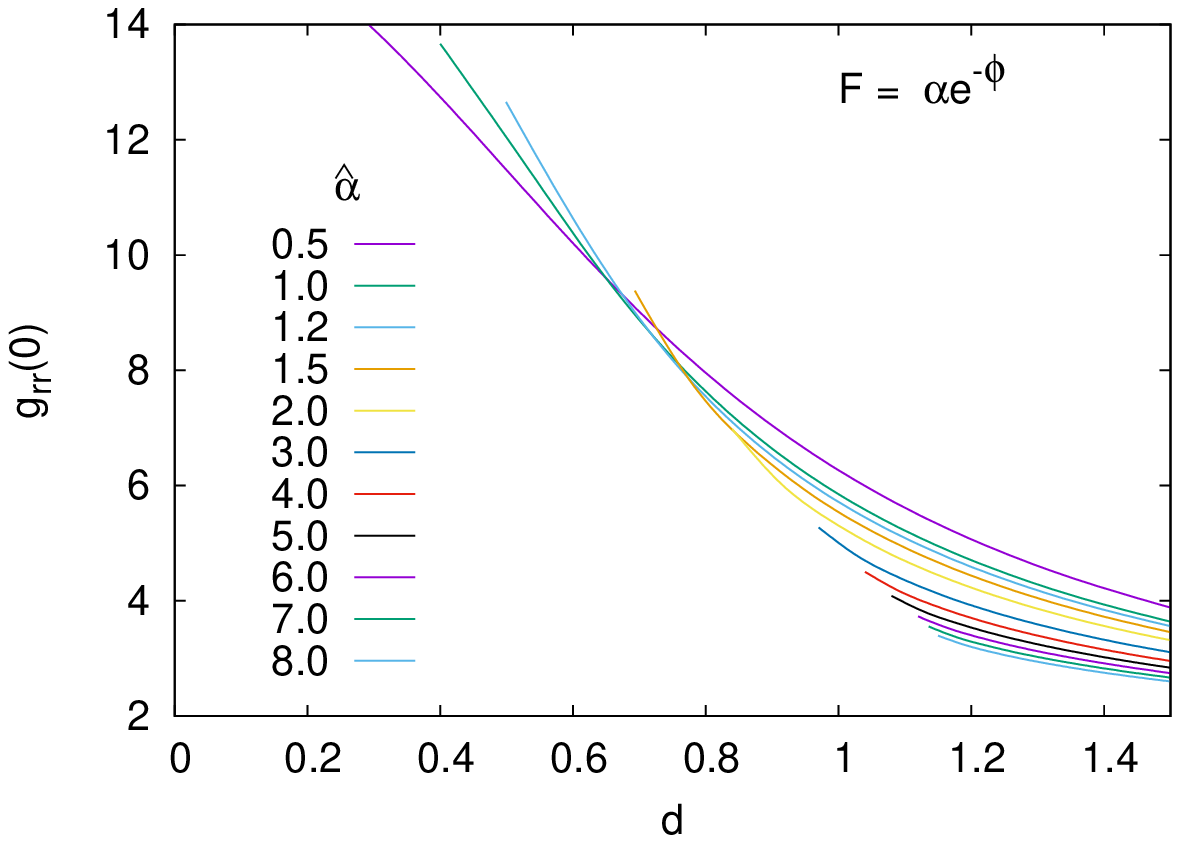}
\\
(e)\includegraphics[width=.45\textwidth, angle =0]{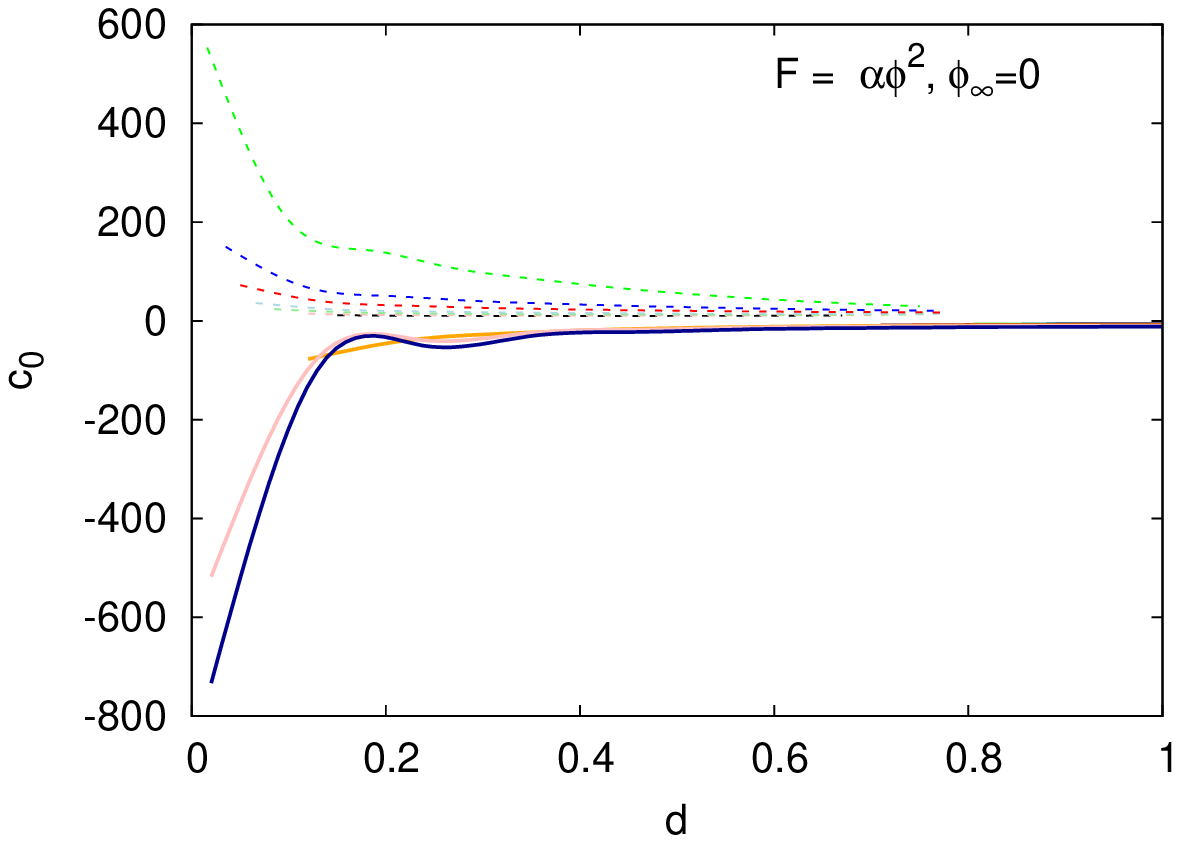}
(f)\includegraphics[width=.45\textwidth, angle =0]{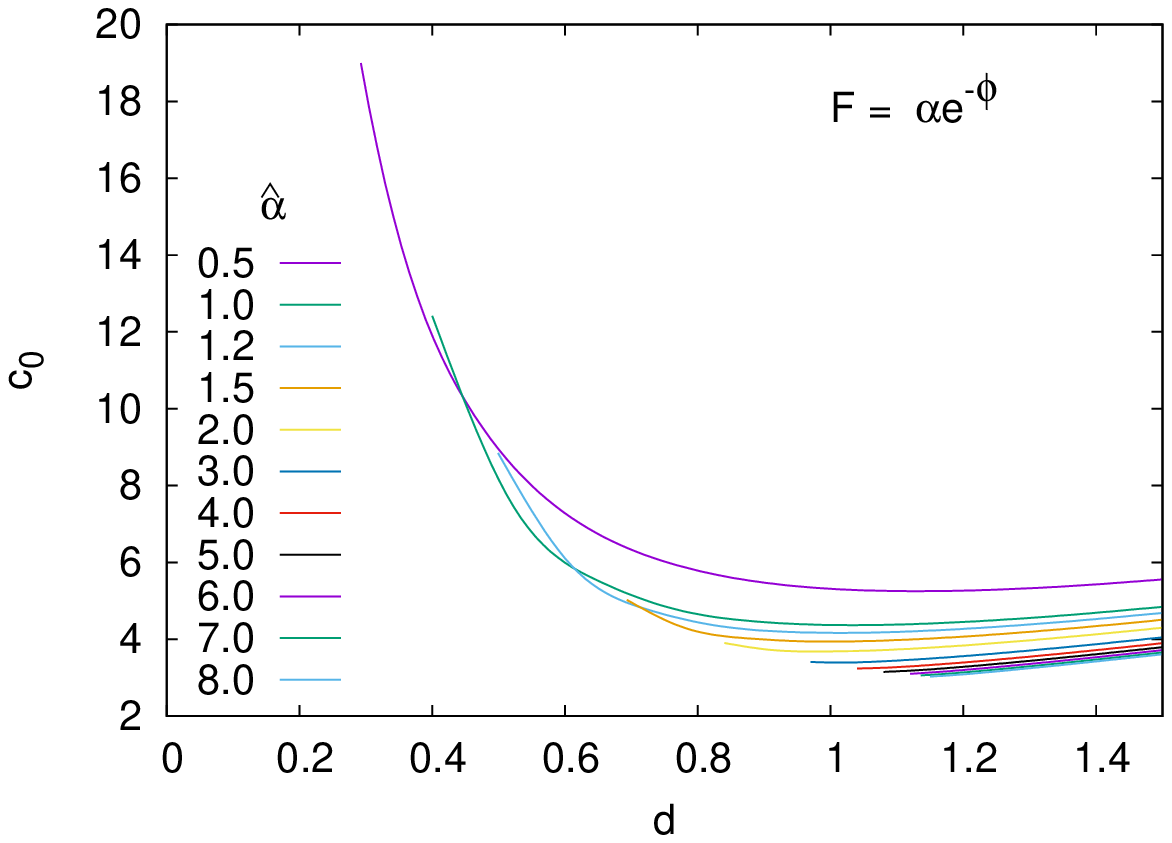}
\end{center}
\caption{
(a)-(f) The value of the metric components $-g_{tt}(0)$ and  $g_{rr}(0)$ at the origin, 
and the coefficient $c_0$ of the diverging term
in the expansion of the scalar field $\phi$ at the origin
vs the scaled scalar charge $d$ for the coupling functions 
$F=\alpha\phi^2$ with $\phi_\infty=0$ [left column: (a), (c), (e)] and 
$F=\alpha e^{-\phi}$ [right column: (b), (d), (f)] for several values of $\hat{\alpha}$.
(The values of $\hat{\alpha}$ in (e) are the same as in (a) and (c))}
\label{fig_profile0}
\end{figure}

We now present typical sets of particle-like solutions.
Figure \ref{fig_profile} shows the metric functions $-g_{tt}$ (first row)
and $g_{rr}$ (second row) as well as the scalar-field function $\phi$ (third row) versus
the scaled radial coordinate $\hat{r}=r/M$ for the coupling functions 
$F=\alpha\phi^2$ with $\phi_\infty=0$ (left column) and 
$F=\alpha e^{-\phi}$ (right column; for the needs of our numerical analyis,
we will henceforth set $\gamma=1$). Each graph presents several
solutions for different values of  the scaled GB coupling constant 
$\hat{\alpha}=8\alpha/M^2$ and the scaled scalar charge $d=D/M$. 
%-- these values are given in Figs. \ref{fig_profile}(e,f) for the scalar field
%and hold also for the corresponding metric-function plots (a),(c) and (b),(d) for the two
%coupling functions.

Figure \ref{fig_profile} clearly demonstrates the regularity of the metric
at the origin $r=0$, where the metric functions assume finite values, in full agreement
with the expansions. In the same regime, the scalar field is shown to
diverge in accordance again with its asymptotic expansions found in the previous Section.
At large distances, both metric functions obey the asymptotic flatness condition whereas
the scalar field assumes a constant, vanishing value.

The behaviour of the metric tensor and scalar field near the origin
$r=0$ is more clearly depicted in Fig. ~\ref{fig_profile0}. 
The finite asymptotic values of the two metric components
$-g_{tt}$ and $g_{rr}$ at the origin are shown in the first and second row, respectively,
for numerous sets of solutions in order
to illustrate their dependence on the GB coupling constant.
Also shown is the coefficient $c_0$ of the diverging term
in the expansion of the scalar field at the origin, for the same set of solutions.
The two columns correspond again to the two choices for the
coupling function, namely the quadratic and the exponential function.
%obtained by taking the limit $\hat r \to 0$ of $\hat r \phi$

%%%%%%%%%%%%%%%%
\begin{figure}
\begin{center}
%
%(a)\includegraphics[width=.46\textwidth, angle =0]{curvi_2_0.pdf}
%(b)\includegraphics[width=.46\textwidth, angle =0]{curvi_d.pdf}
(a)\includegraphics[width=.46\textwidth, angle =0]{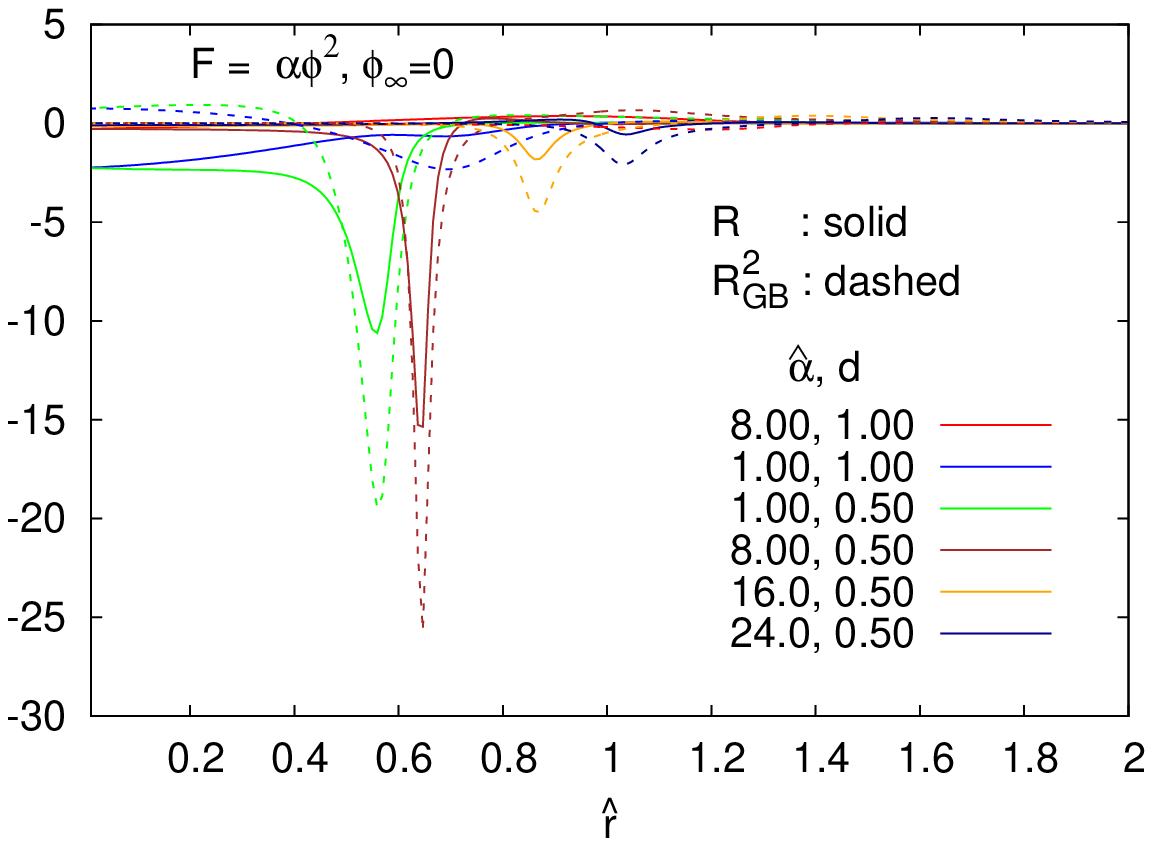}
(b)\includegraphics[width=.46\textwidth, angle =0]{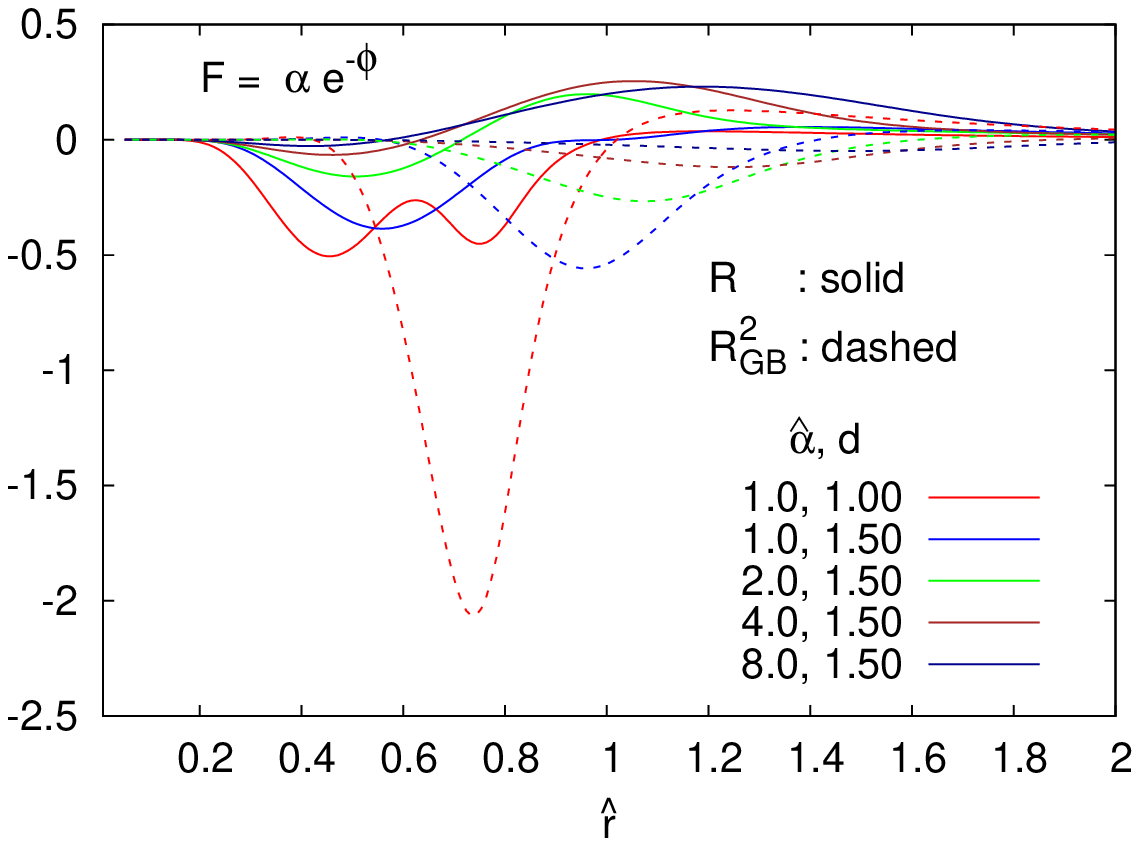}
\end{center}
\caption{
The scalar curvature $R$ (solid lines) and Gauss-Bonnet term $R^2_{GB}$ 
(dashed lines) vs
the scaled radial coordinate $\hat{r}$ for the coupling functions 
(a) $F=\alpha\phi^2$ with $\phi_\infty=0$, and (b) 
$F=\alpha e^{-\phi}$, and for several values of 
$\hat{\alpha}$ and  $d$. }
\label{fig_invar}
\end{figure}
%%%%%%%%%%%%%%%%%

In Fig. \ref{fig_invar}, we depict two scalar invariant quantities, the 
scalar curvature $R$ (indicated by solid lines) and the Gauss-Bonnet term (indicated
by dashed lines) in terms of the scaled radial coordinate $\hat r$. Their profiles are
presented for a number of solutions corresponding to different values of the parameters 
$\hat{\alpha}$ and  $d$, and clearly demonstrate the regularity of spacetime.
Both curvature invariants assume finite asymptotic values near the origin $r=0$,
reach their maximum values at some small value of $\hat r$ (where the stress-energy
tensor $T^\mu_\nu$ also reaches its maximum value, as we will later see)
and vanish at asymptotic infinity.

As we observe in Fig.~\ref{fig_profile}, the metric function $g_{tt}$ is always monotonic. 
%and the scalar function $\phi$ are always monotonic. 
However, this is not always the case for the metric function $g_{rr}$.
Depending on the parameters, the metric function $g_{rr}$
may exhibit a maximum away from the origin
and then fall sharply with increasing radial coordinate.
This behavior entails that the relation between the radial coordinate $r$
and the circumferential coordinate $R_c$, defined as
\begin{equation}
R_c = \frac{1}{2\pi}\,\int_0^{2 \pi} \sqrt{g_{\varphi \varphi}}|_{\theta=\pi/2}\, d\varphi
= r e^{f_1/2} \ ,
\label{circum-radius}
\end{equation}
may not be monotonic any more, and this may have interesting
consequences for the geometry of the spacetime:
%The monotonic behavior is lost, when 
The circumferential radius $R_c(r)$ in that case
%develops a saddle point, which upon further change of the parameters splits into 
features a local maximum and a local minimum, where
the local minimum corresponds to a throat while the local maximum represents an equator.

\subsection{Embedding diagrams}

\begin{figure}[t]
\begin{center}
(a)\includegraphics[width=.45\textwidth, angle =0]{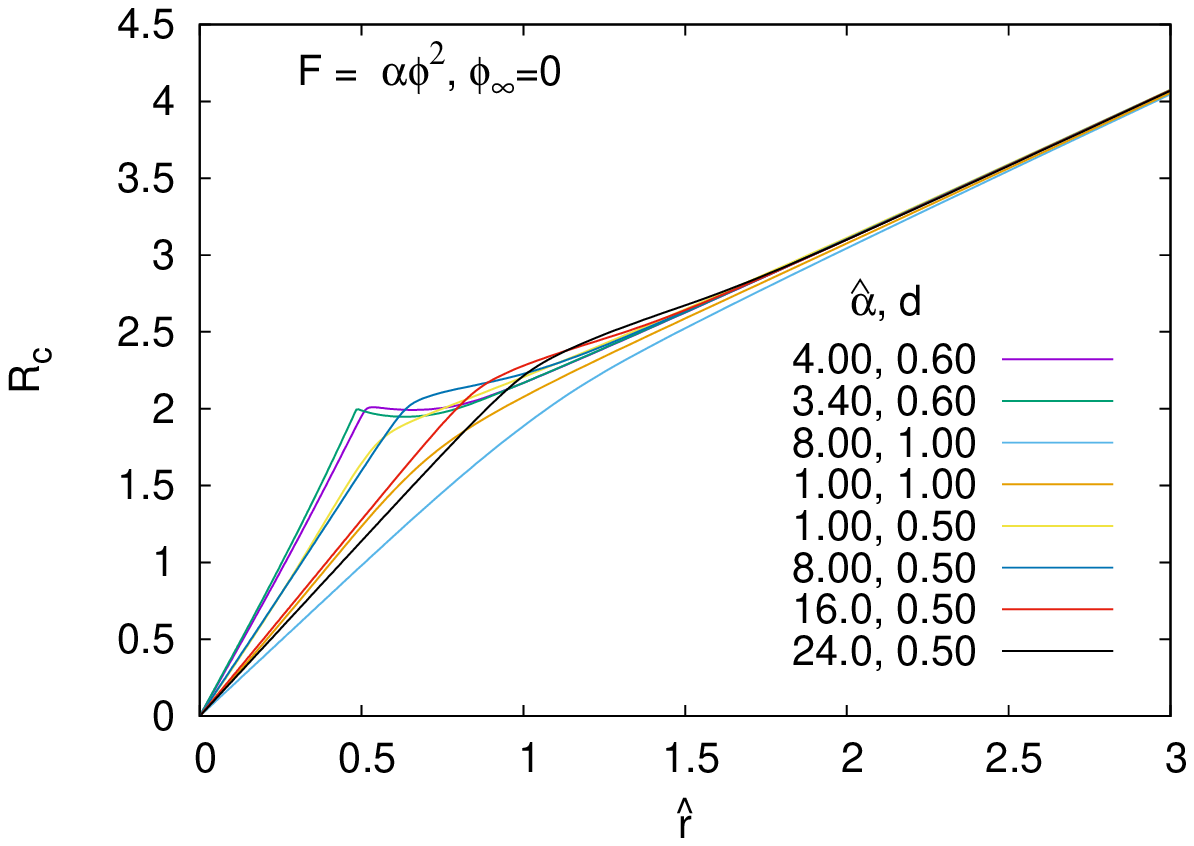}
(b)\includegraphics[width=.45\textwidth, angle =0]{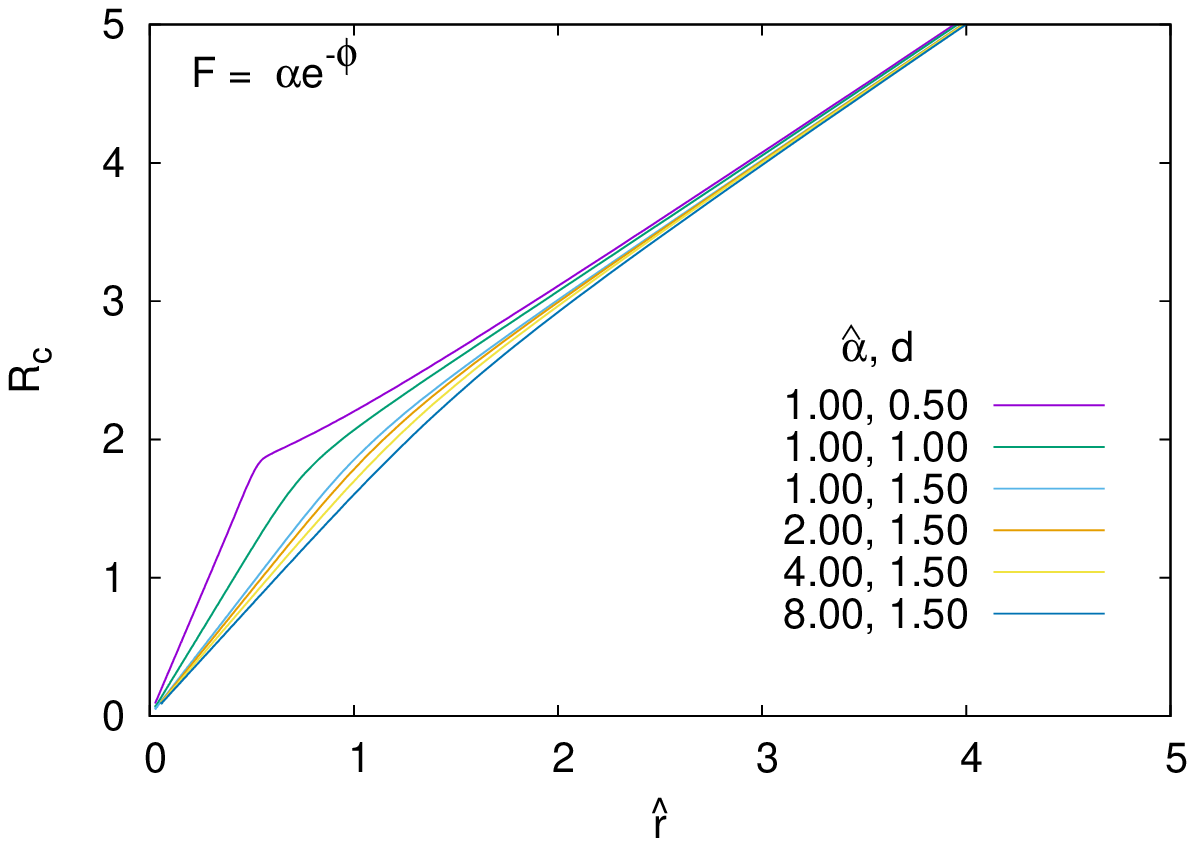}
\\[-0.5mm]
(c)\includegraphics[width=.45\textwidth, angle =0]{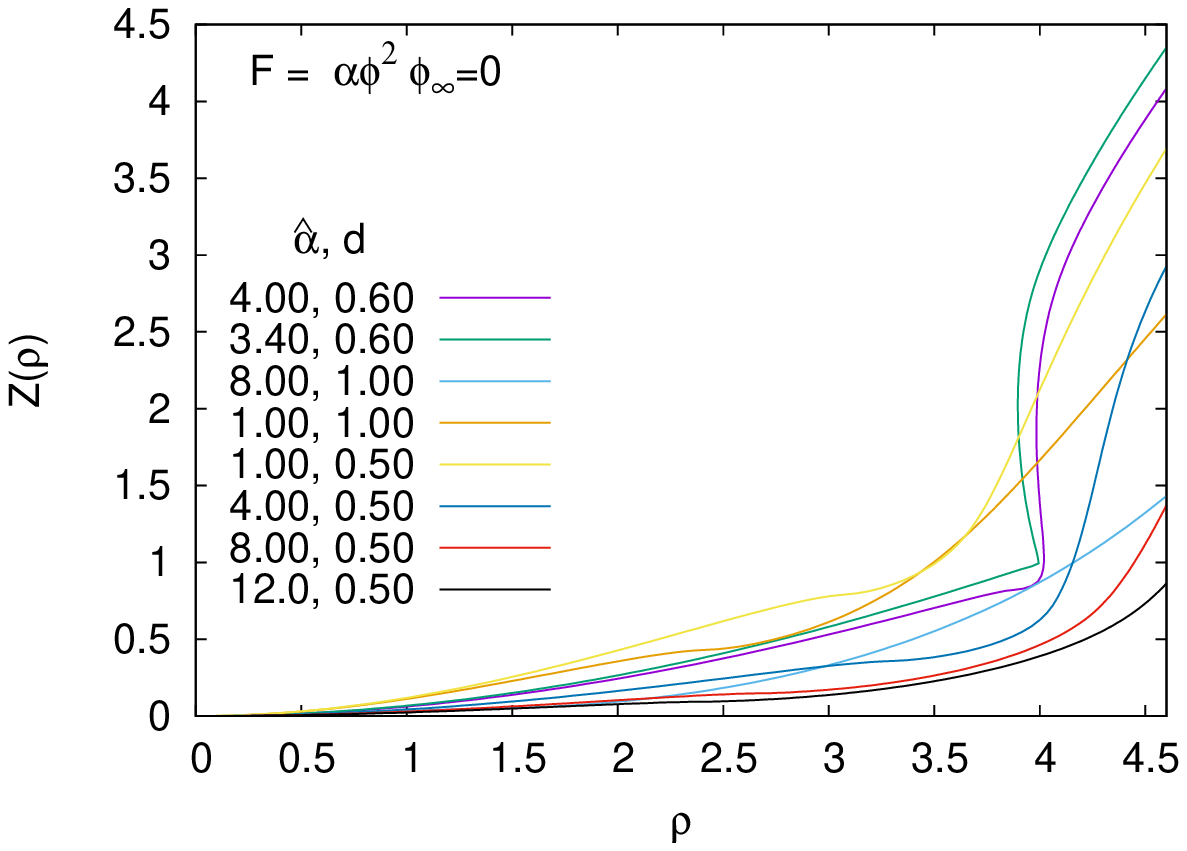}
(d)\includegraphics[width=.45\textwidth, angle =0]{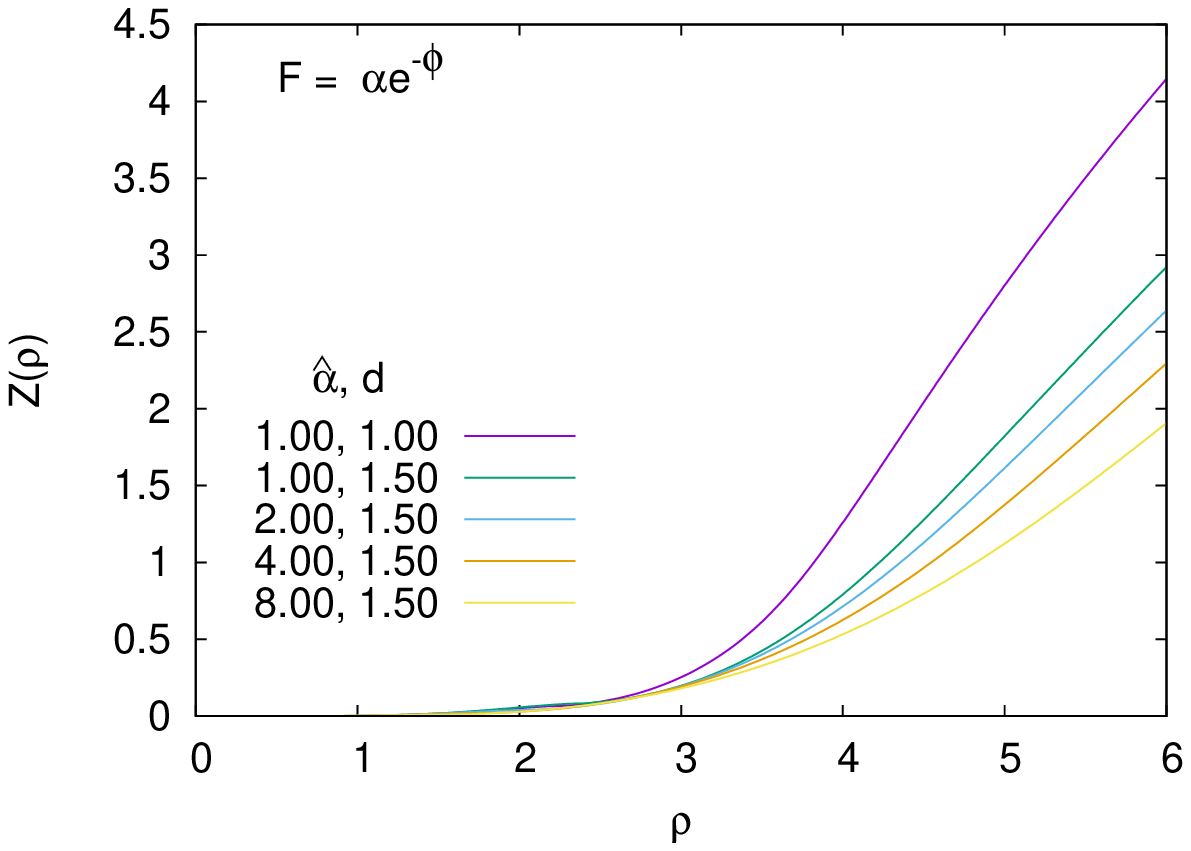}
\end{center}
\caption{
(a)-(d) The circumferential coordinate $R_c(r)$ 
and the embedding relation $ Z(\rho)$
for the coupling functions 
$F=\alpha\phi^2$ with $\phi_\infty=0$ [left column: (a), (c)] and 
$F=\alpha e^{-\phi}$ [right column: (b), (d)] 
 for several values of $\hat{\alpha}$ and  $d$.}
\label{fig_emb1}
\end{figure}

\begin{figure}
\begin{center}
\vspace*{-0.5cm}
(a)\includegraphics[width=.4\textwidth, angle =0]{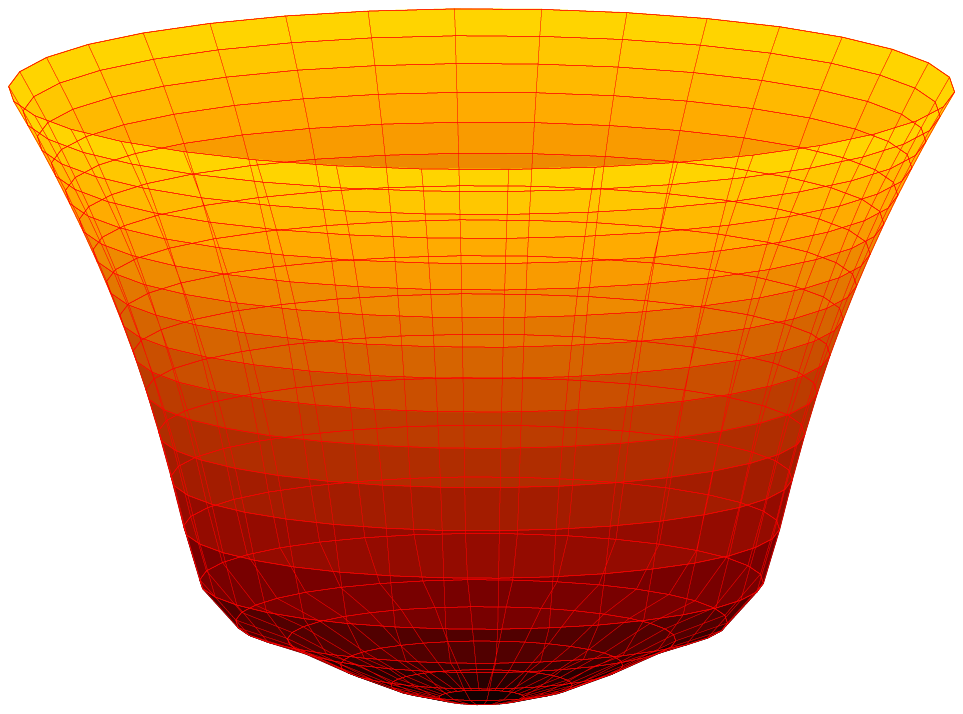}
(b)\includegraphics[width=.4\textwidth, angle =0]{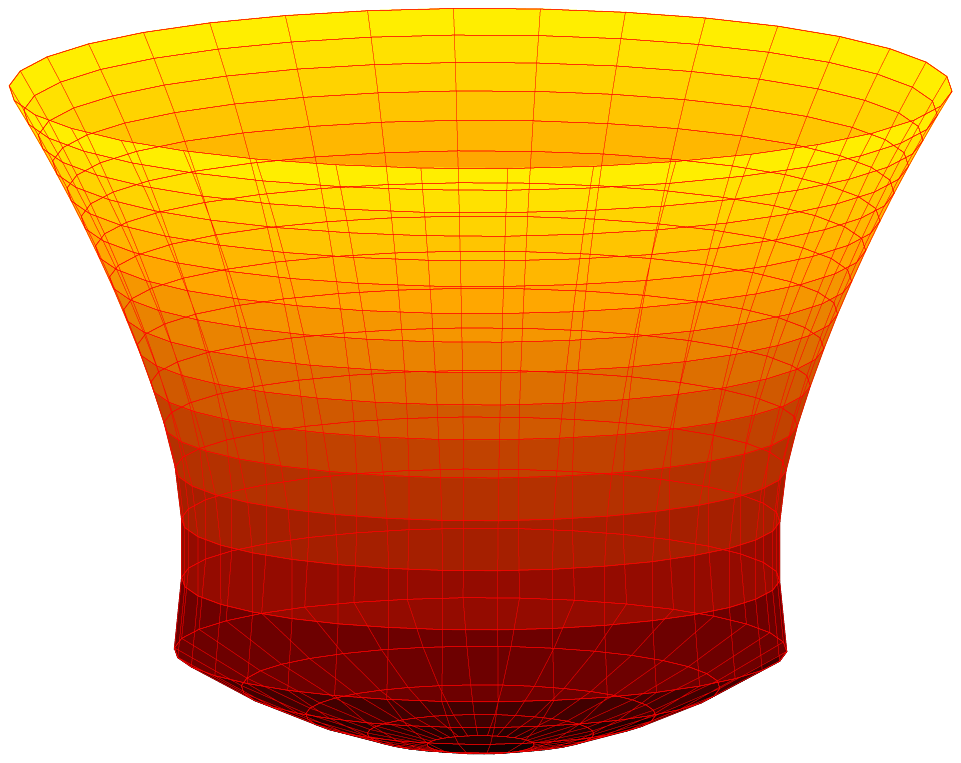}
\\[-5mm]
(c)\includegraphics[width=.4\textwidth, angle =0]{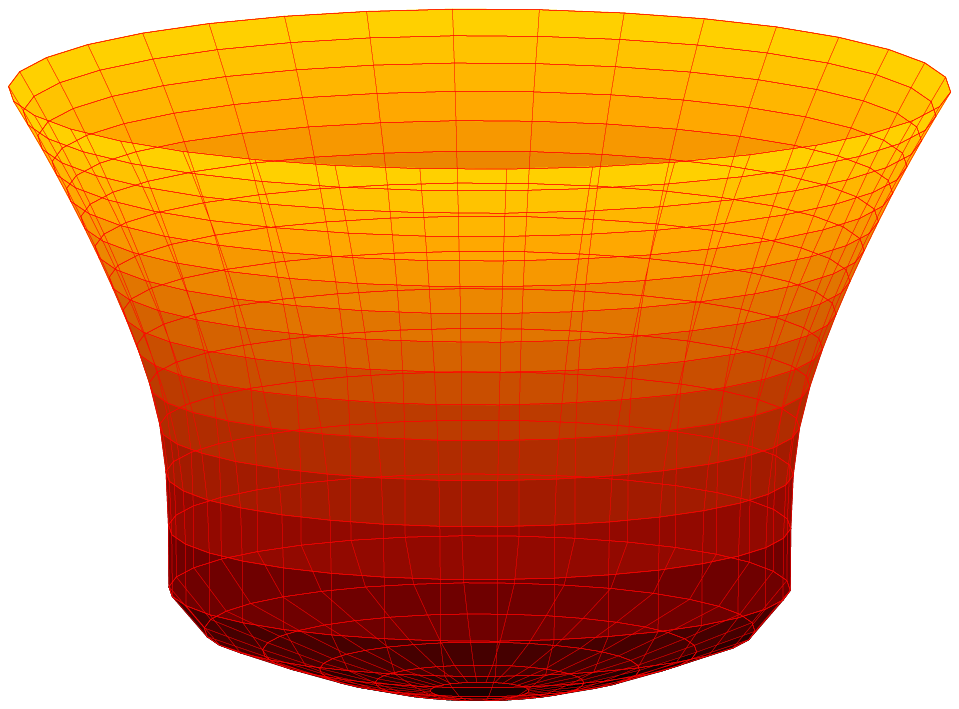}
(d)\includegraphics[width=.4\textwidth, angle =0]{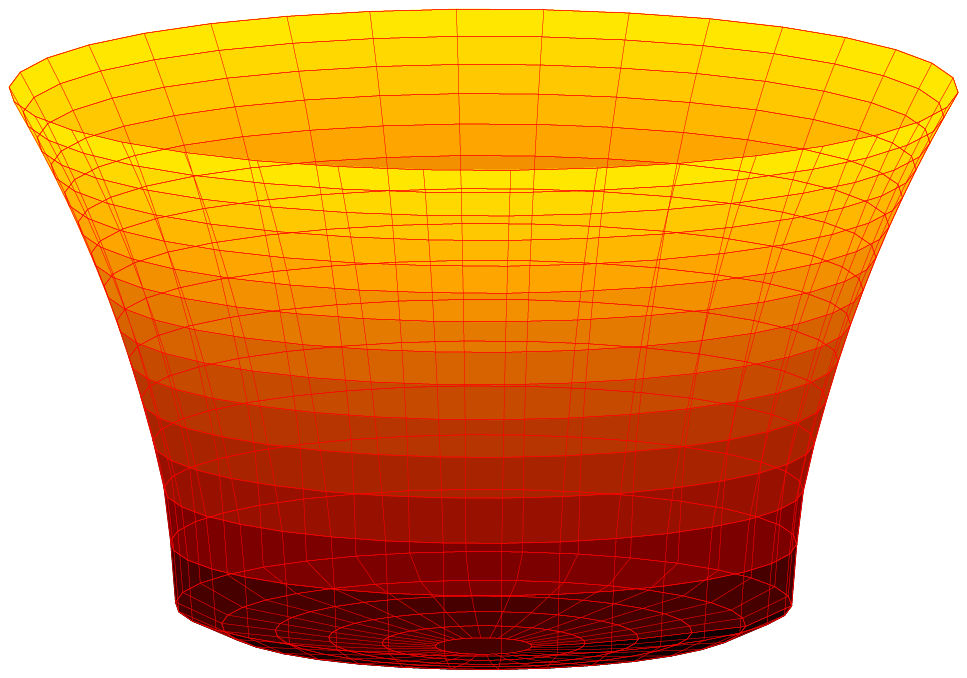}
%(a)\includegraphics[width=.4\textwidth, angle =0]{emb_2_alp=0_5_q=-0_5.eps}
%(b)\includegraphics[width=.4\textwidth, angle =0]{emb_2_alp=1_7_q=-0_6.eps}
%\\[-5mm]
%(c)\includegraphics[width=.4\textwidth, angle =0]{emb_2_0_alp=0k585_q=0k6.eps}
%(d)\includegraphics[width=.4\textwidth, angle =0]{emb_2_0_alp=3k5_q=0k15.eps}
%(c)\includegraphics[width=.4\textwidth, angle =0]{emb_2_alp=2_0_q=-0_6.eps}
%(d)\includegraphics[width=.4\textwidth, angle =0]{emb_2_alp=12_0_q=-0_5.eps}
%\\
%(e)\includegraphics[width=.45\textwidth, angle =0]{emb_2_alp=1_7_q=-0_6.eps}
%(f)\includegraphics[width=.45\textwidth, angle =0]{emb_2_alp=2_0_q=-0_6.eps}
%\\
%(g)\includegraphics[width=.45\textwidth, angle =0]{emb_2_alp=8_0_q=-0_5.eps}
%(h)\includegraphics[width=.45\textwidth, angle =0]{emb_2_alp=12_0_q=-0_5.eps}
\end{center}

\vspace*{-0.3cm}
\caption{
The isometric embedding of the equatorial plane for particle-like solutions 
for the coupling function $F=a\phi^2$ with $\phi_\infty=0$
for: 
(a) $\hat{\alpha} = 1, d=0.5$;
(b) $\hat{\alpha} =3.4, d=0.6$;
(c) $\hat{\alpha} =4.68, d=0.6$;
(d) $\hat{\alpha} =28, d=0.15$.
%(c) $\hat{\alpha} =4.0, d=0.6$;
%(d) $\hat{\alpha} =24, d=0.5$ (?)}.
}
\label{fig_emb2}
\end{figure}

Let us now visualize the spatial geometry of these particle-like solutions,
by illustrating the dependence of their circumferential coordinate $R_c(r)$ 
on the radial coordinate and by embedding these solutions into
Euclidean space. In particular, we take the line-element of our solutions Eq.~(\ref{met})
at $t=const.$ and $\theta=\pi/2$, and set it equal to the line-element
of the three-dimensional Euclidean space 
\begin{equation}
e^{f_1} \left( dr^2 + r^2 d \varphi^2 \right) =
 d \rho^2 +\rho^2 d\varphi^2 +dZ^2 \ , \label{emb}
\end{equation}
%%%%%%%%%%%%%%%%%
where $\rho$, $\varphi$ and $Z$ are the Euclidean coordinates.
We then consider $r$ and $Z$ as functions of $\rho$.
Identifying $\rho=r e^{f_1/2}$ from the $g_{\varphi\varphi}$ terms, we obtain $Z$ via the integral
%
%\begin{equation}
%   Z(\rho)=\int_0^\rho \sqrt{e^{f_1(\tilde \rho)} -1}\,d\tilde \rho \ . \label{zeq}
%\end{equation}
%
%

\begin{equation}
   Z(\rho)=\pm \int_0^\rho \sqrt{\left(\frac{dr}{d\tilde \rho}\right)^2 e^{f_1(\tilde \rho)} -1}\,d\tilde \rho \ . 
   \label{zeq}
\end{equation}

This then gives us a parametric representation of the embedded 
($\theta = \pi/2$)-plane for a fixed value of the $\varphi$ coordinate.
The geometry of the particle-like solutions is then visualized
by considering the respective surface of revolution.

Figure \ref{fig_emb1} shows the circumferential coordinate $R_c(r)$, Eq.~(\ref{circum-radius}),
for the same sets of solutions as those in Fig.~\ref{fig_profile},
and the embedding relation $ Z(\rho)$, Eq.~(\ref{zeq}).
In particular, in Fig.~\ref{fig_emb1}(c) we notice, that the geometry
can develop a throat and an equator. 
However, in contrast to wormhole solutions
\cite{Antoniou:2019awm},
the second asymptotic infinity is missing in these particle-like spacetimes,
since the origin is a regular point.
We note that regular solutions with a throat and an equator have also
been obtained in different contexts
\cite{Frolov:1988vj,Guendelman:2010pr,Chernov:2007cm,Dokuchaev:2012vc}.

We demonstrate the wide variety of different geometries in Fig.~\ref{fig_emb2},
for a set of particle-like solutions with a quadratic coupling function
and an asymptotically vanishing scalar field.
In Figs.~\ref{fig_emb2}(a)-(d), we see a typical particle-like
solution, a solution with a throat and an equator, a solution with a degenerate
throat and an excited solution (emerging for large values of $\hat \alpha$, as
discussed in Sec. V), respectively. 
For a dilatonic coupling function, on the other hand, we do not
find solutions featuring a throat and an equator. The presence or absence of
such solutions will be reflected in the domain of existence of the particle-like solutions,
which is discussed in the next section.

\newpage

\subsection{Cusp singularity}

\begin{figure}
\begin{center}
(a)\includegraphics[width=.45\textwidth, angle =0]{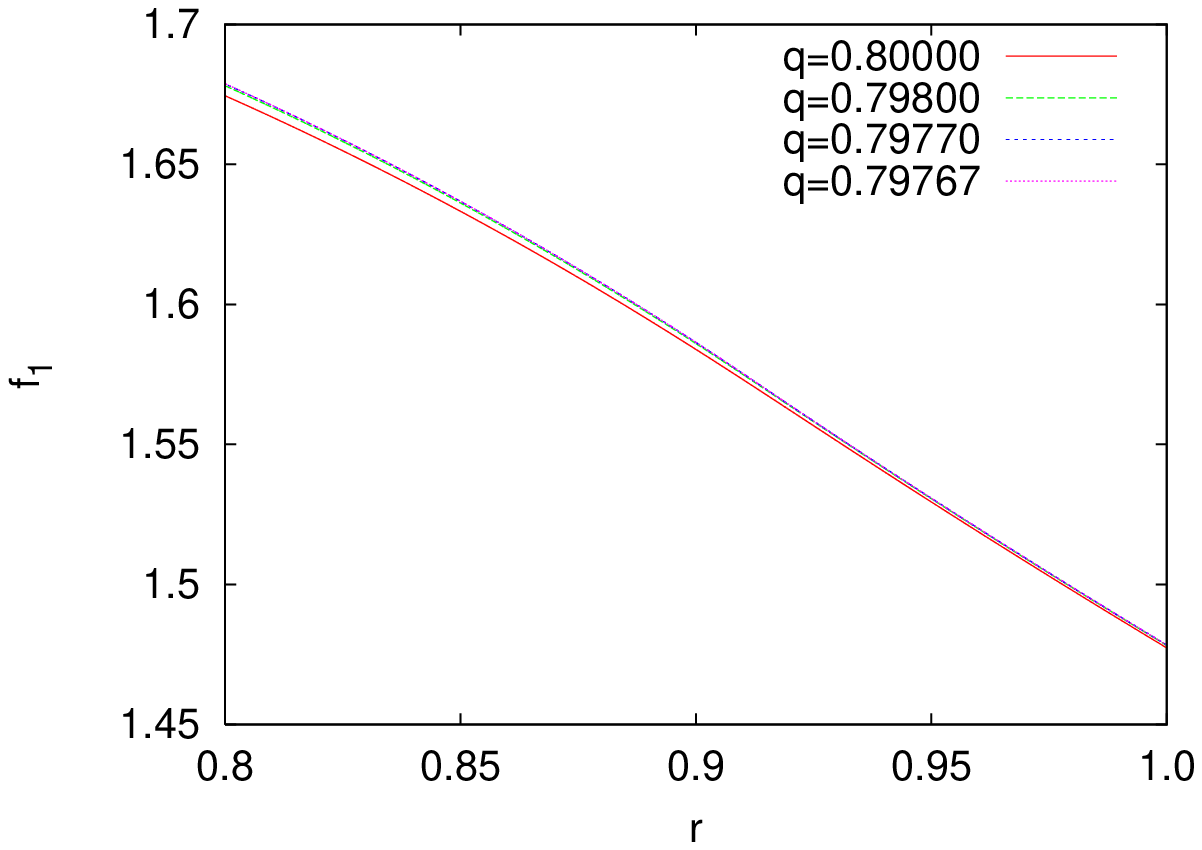}
(b)\includegraphics[width=.45\textwidth, angle =0]{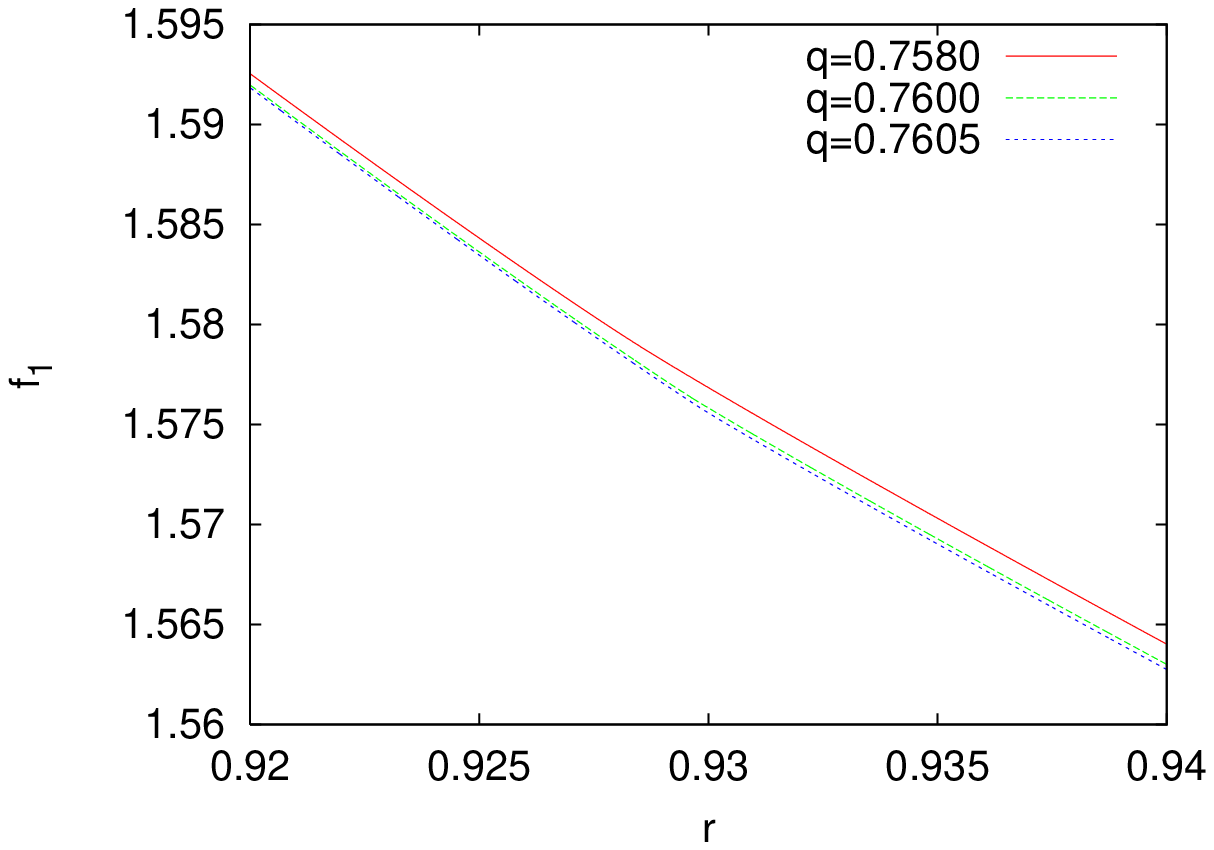}
\\
(c)\includegraphics[width=.45\textwidth, angle =0]{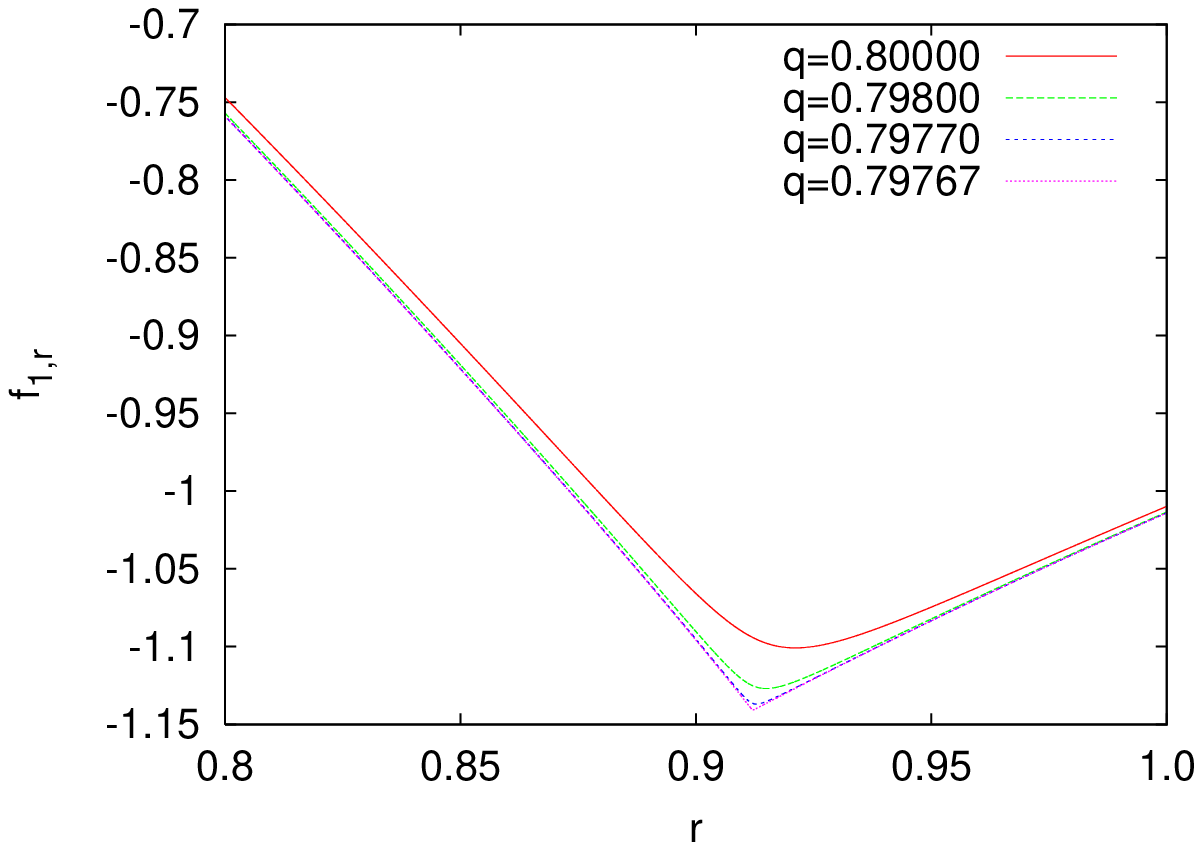}
(d)\includegraphics[width=.45\textwidth, angle =0]{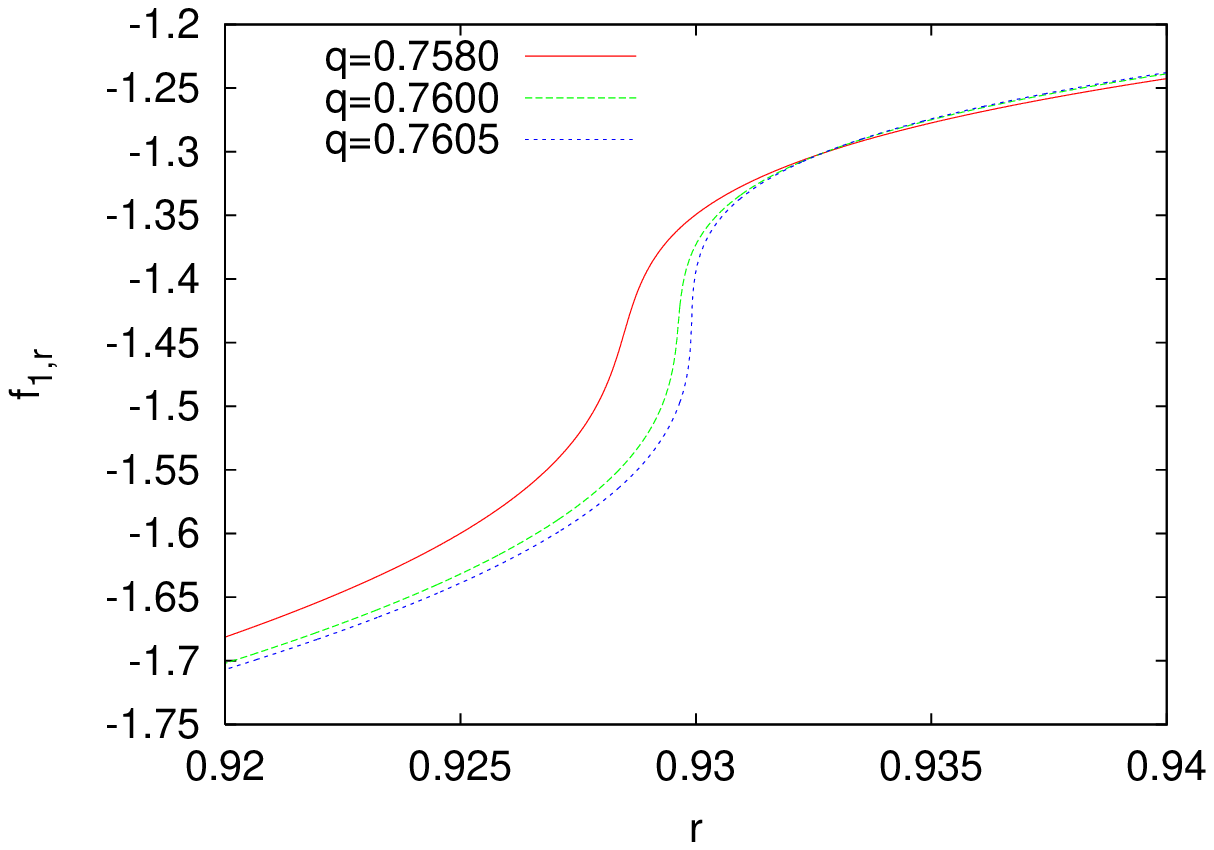}
\\
(e)\includegraphics[width=.45\textwidth, angle =0]{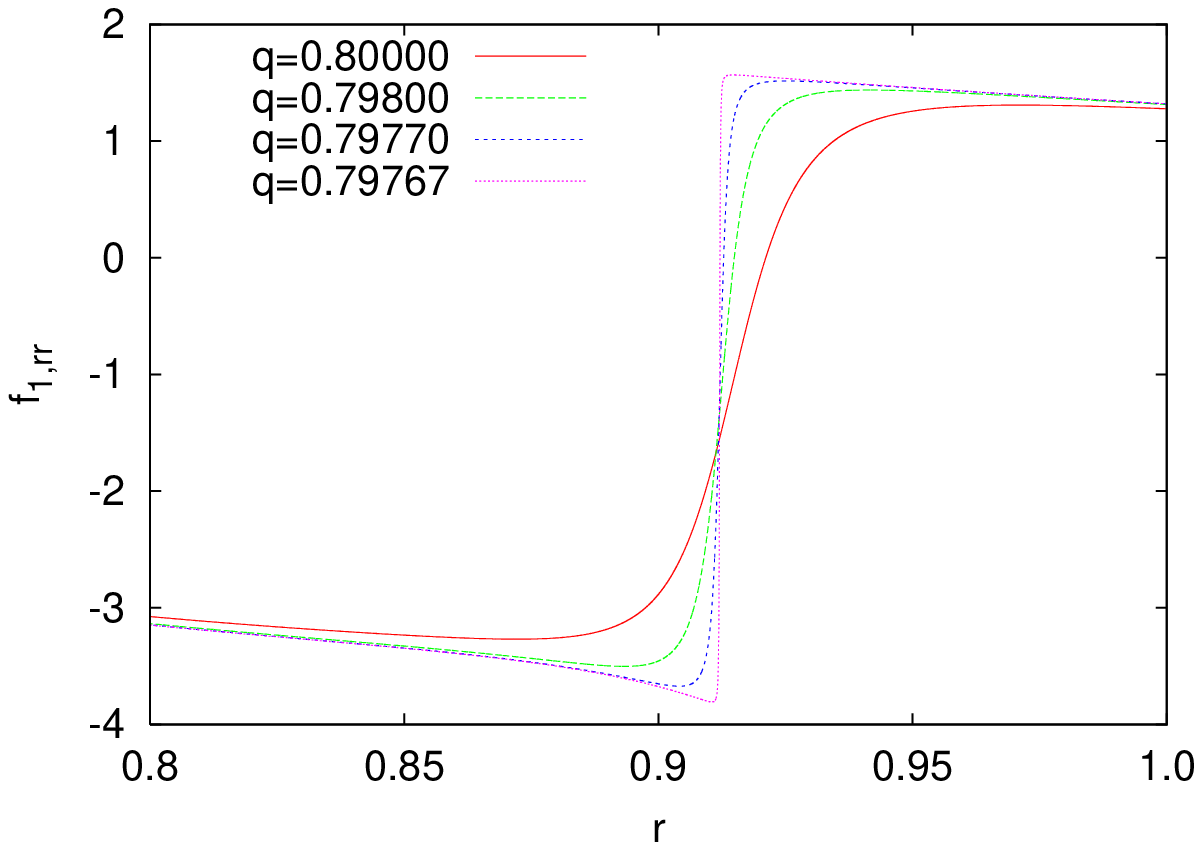}
(f)\includegraphics[width=.45\textwidth, angle =0]{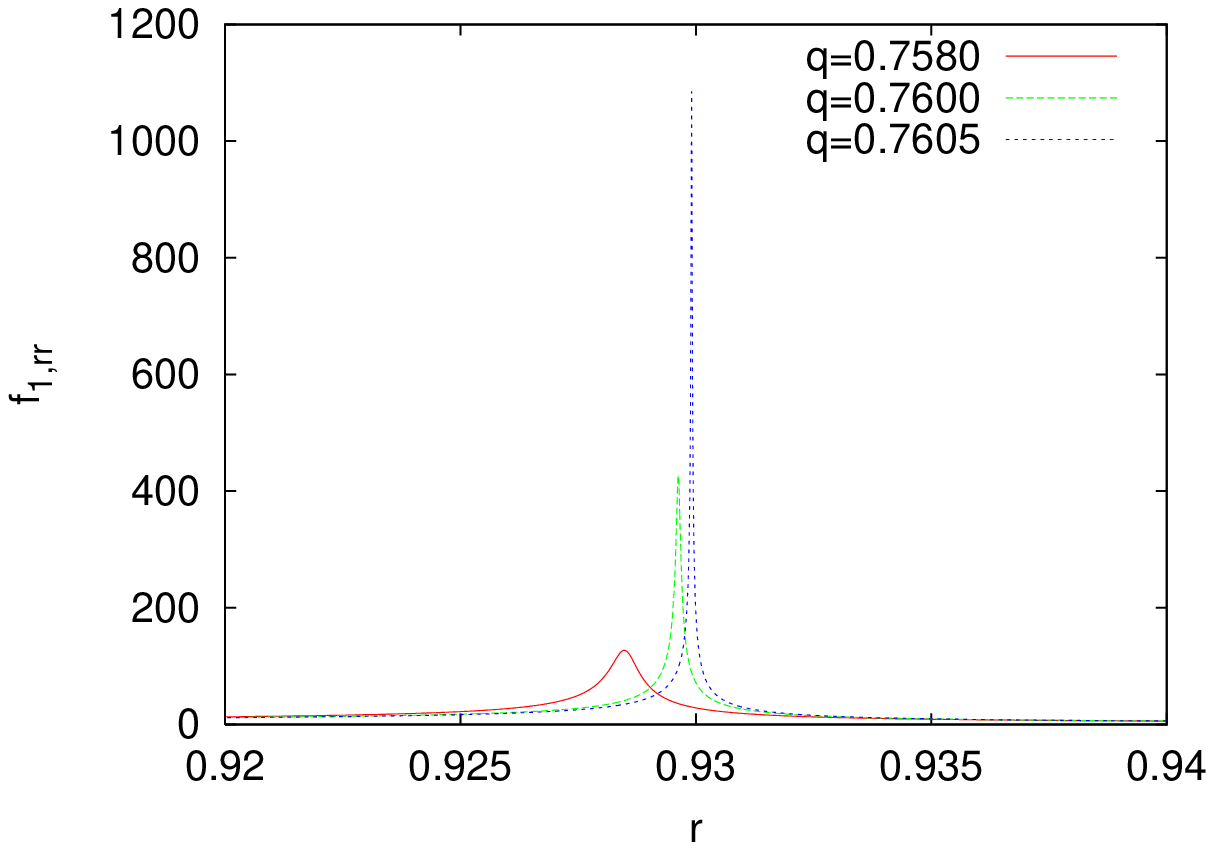}
\end{center}
\caption{
(a)-(f) The emergence of a cusp singularity for the function $f_1$ in a family of solutions
for the coupling function $F=a\phi^2$ with $\phi_\infty=0$
for $\hat{\alpha}=0.5$ and varying scalar charge $d$.
In type (i), the second derivative $f_{1,rr}$ develops a jump
at a critical $r_\star$ [left column:  (a), (c), (e)], whereas in type (ii),
$f_{1,rr}$ diverges at $r_\star$ [right column: (b), (d), (f)].
}
\label{fig_cusp}
\end{figure}

As noted in Section II, for the numerical procedure we need to diagonalize
the second-order differential equations with respect to the
second derivatives of the metric function $f_1$
and the scalar-field function $\phi$. 
This then introduces the determinant of the matrix of coefficients $\det Q$
in the denominator of the diagonalized equations.
When this determinant develops a node at some coordinate value $r_\star$, 
this leads to a cusp singularity in the respective solution.

In Fig.~\ref{fig_cusp}, we demonstrate how such a cusp singularity forms.
For that purpose, we exhibit the metric function $f_1$, its first derivative
$f_{1,r}$ and its second derivative $f_{1,rr}$  for a family of solutions
for the coupling function $F=a\phi^2$ with $\phi_\infty=0$
for $\hat{\alpha}=0.5$ and varying scalar charge $d$,
approaching the critical charge $d_c$ of the solution with the cusp singularity.
We note that we find two types of cusp singularities.
In type (i), the second derivative $f_{1,rr}$ (and likewise $\phi_{rr}$) develops a jump
at a critical $r_\star$ [left column:  (a), (c), (e)].
Here, the denominator behaves as $(r-r_\star)$.
In type (ii),
$f_{1,rr}$ diverges at $r_\star$ [right column: (b), (d), (f)].
Here, the denominator behaves as $(r-r_\star)^g$ with $g<1$.
In our numerical procedure, we monitor the determinant and stop the computation
if this quantity changes sign.

\subsection{GR limit}

%Fig. 6: JNWW limit ($\hat{\alpha} \to 0$)
\begin{figure}
\begin{center}
(a)\includegraphics[width=.45\textwidth, angle =0]{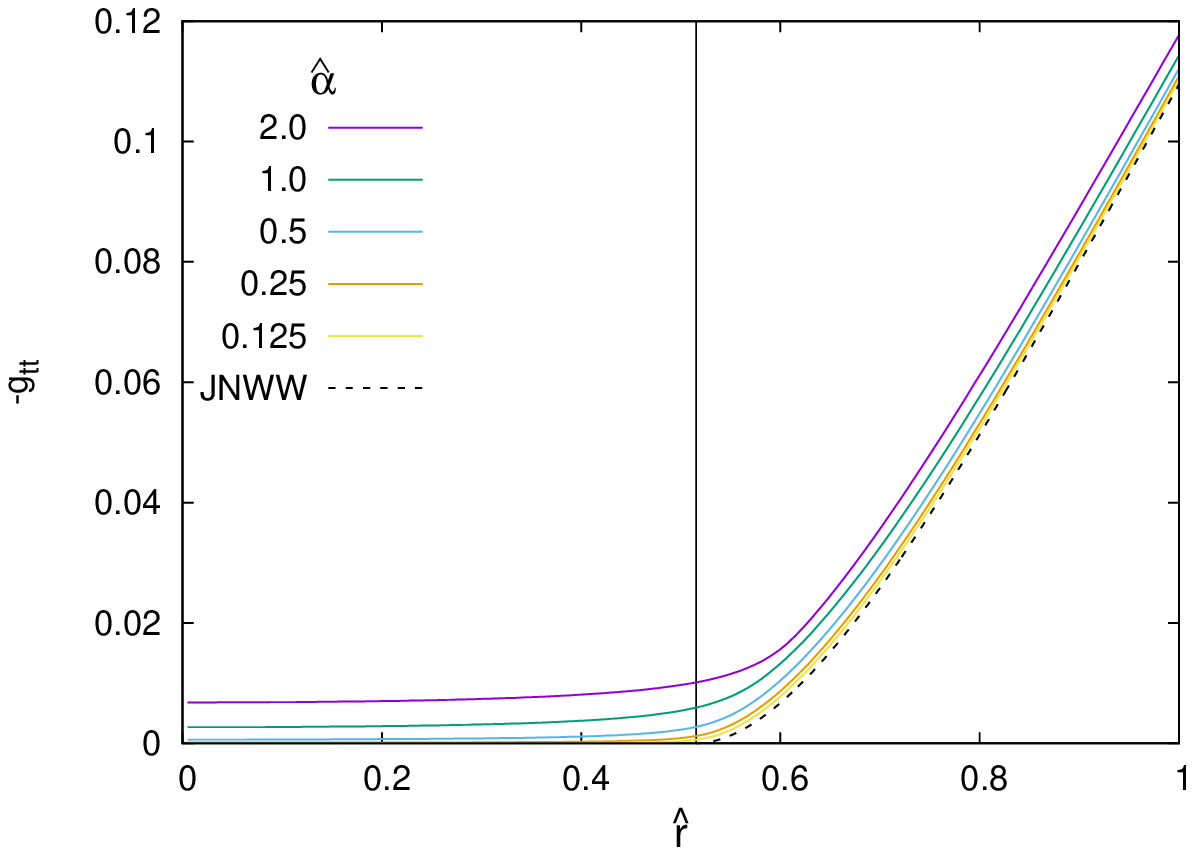}
(b)\includegraphics[width=.45\textwidth, angle =0]{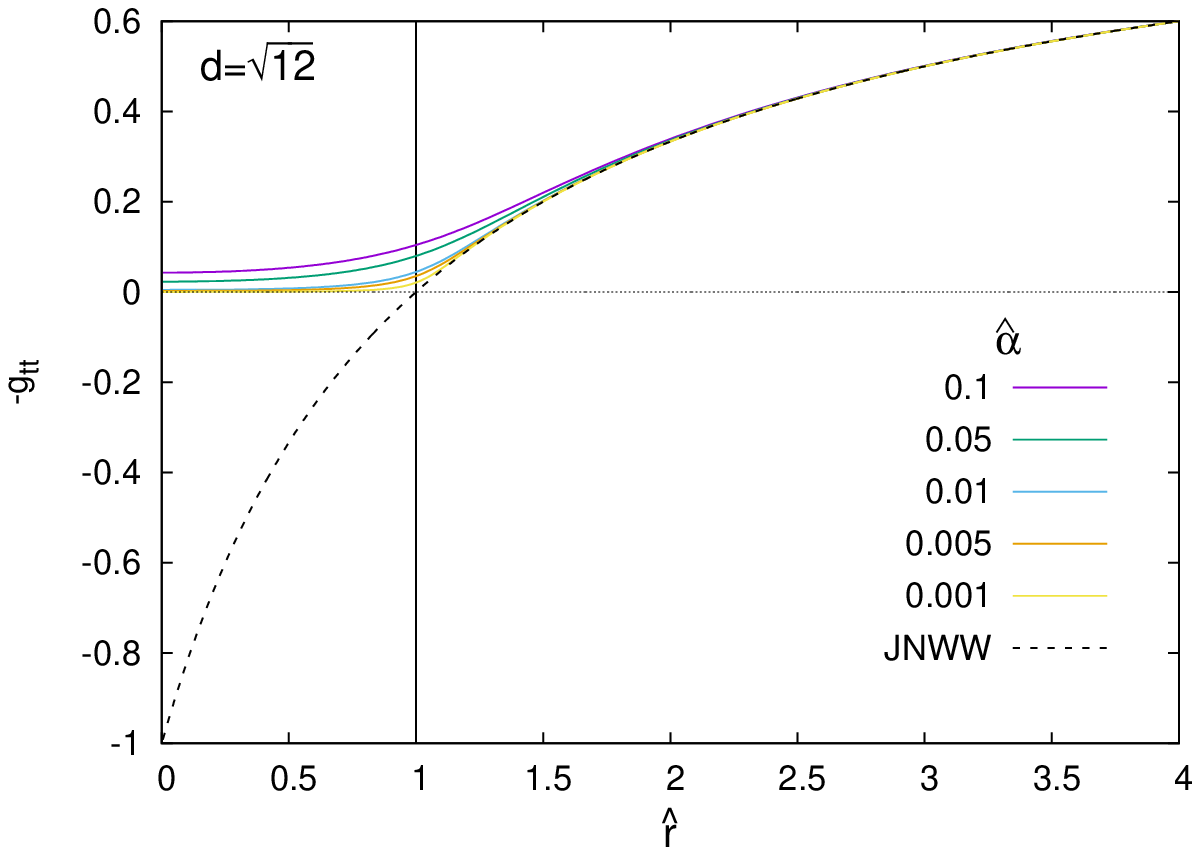}
\\.
(c)\includegraphics[width=.45\textwidth, angle =0]{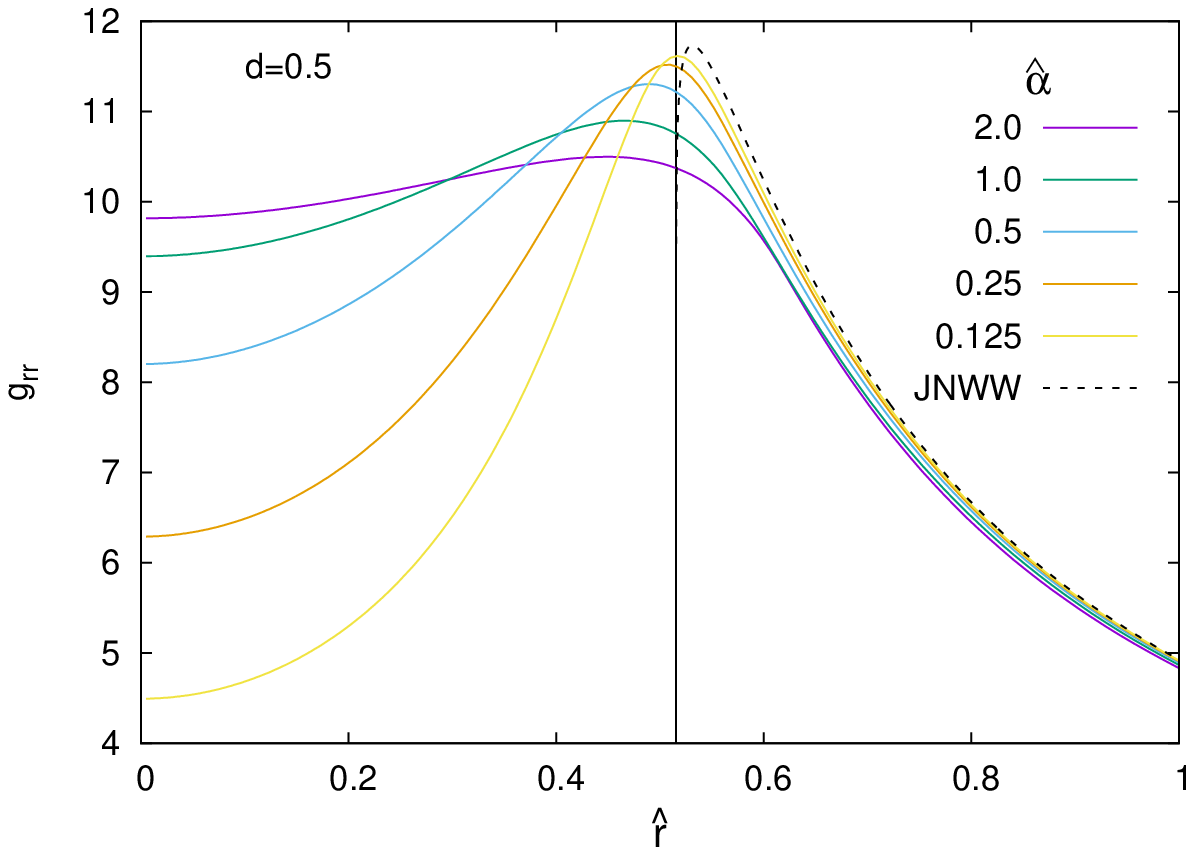}
(d)\includegraphics[width=.45\textwidth, angle =0]{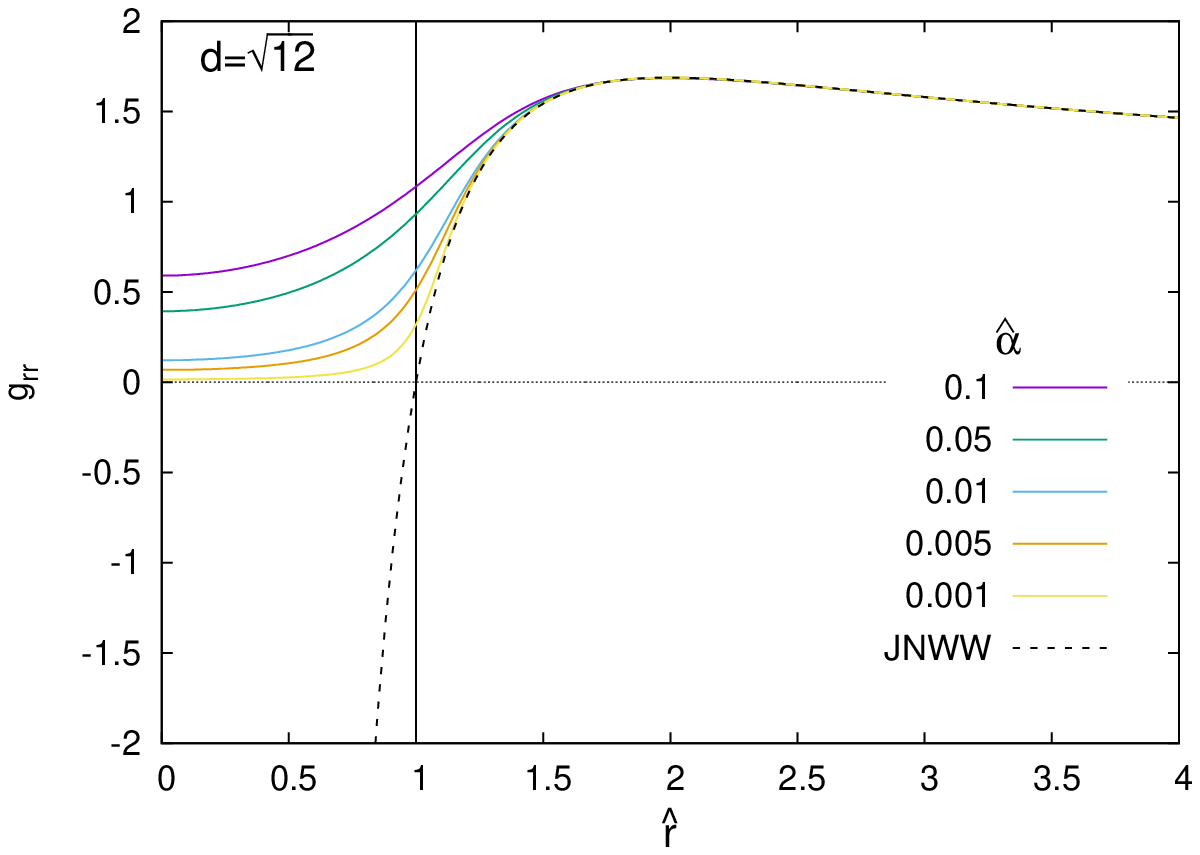}
\\
(e)\includegraphics[width=.45\textwidth, angle =0]{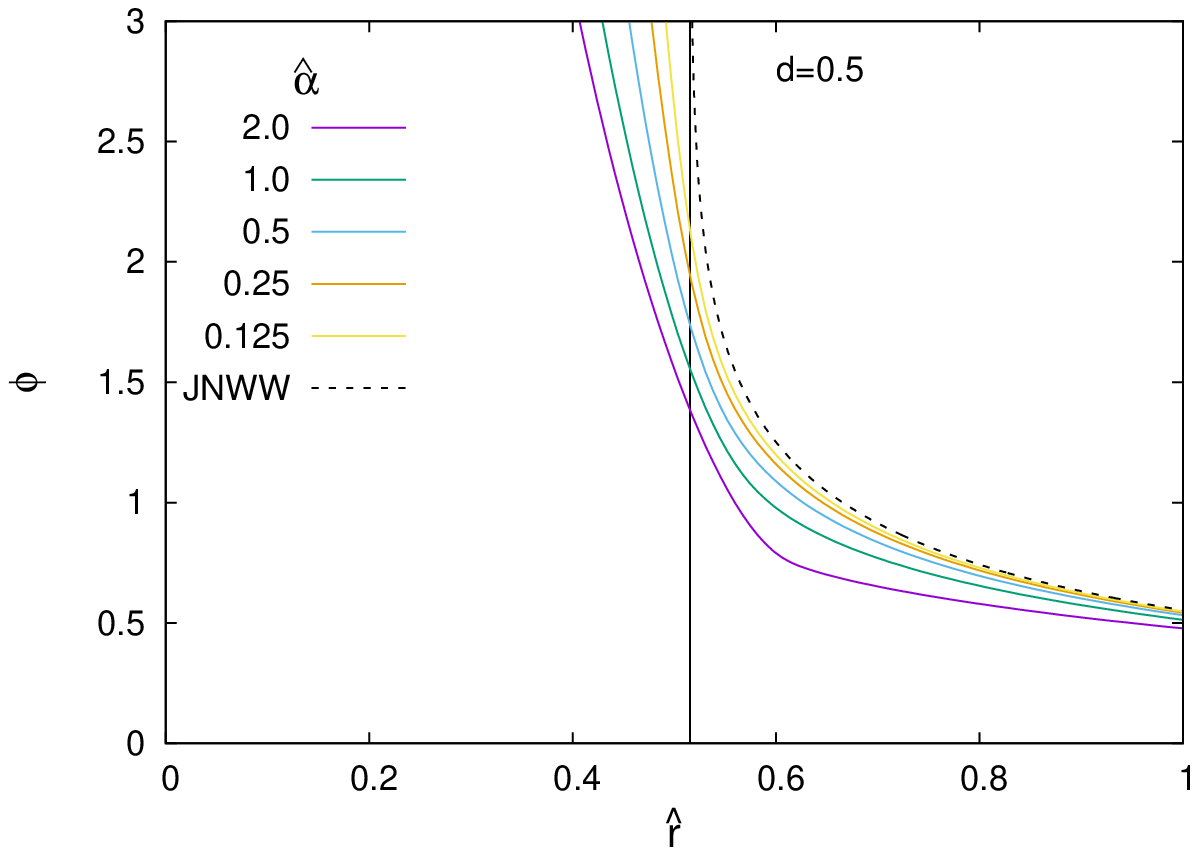}
(f)\includegraphics[width=.45\textwidth, angle =0]{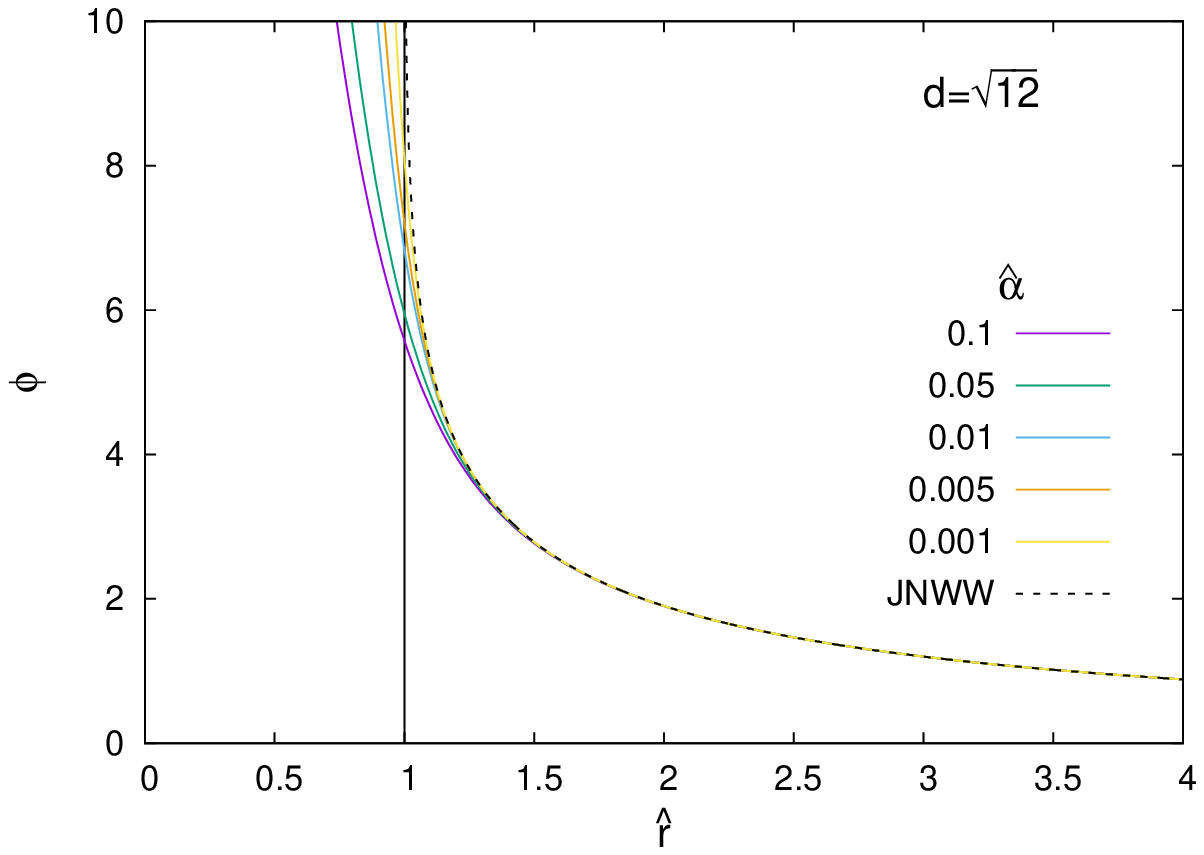}
\end{center}
\caption{
(a)-(f) The metric functions $-g_{tt}$ and $g_{rr}$ and the scalar field $\phi$ vs
the scaled radial coordinate $\hat{r}$ for the coupling function 
$F=\alpha\phi^2$ with $\phi_\infty=0$
in the GR limit $\hat{\alpha} \to 0$ for scaled dilaton charge $d=1/2$
[left column: (a), (c), (e)] and  $d=\sqrt{12}$
[right column: (b), (d), (f)].
The limiting Fisher (JNWW) solution is also indicated (dotted)
together with the location of its curvature singularity (vertical line). }
\label{fig_JNNW}
\end{figure}

We demonstrate in Fig.~\ref{fig_JNNW} how the particle-like solutions 
change as the limiting GR solution, i.e., the Fisher (or JNWW) solution,
is approached. 
Here, the coupling function is $F=\alpha \phi^2$ with $\phi_\infty=0$.
The functions of the GR solution are indicated by dotted lines,
while the location of its curvature singularity is marked by the vertical line.
As predicted in Sec. II, in the exterior region $r>r_s$, all functions
of the particle-like solutions indeed tend towards the corresponding
functions of the GR solutions,
while, in the interior region $r<r_s$, the metric functions tend to zero,
leading to a singular interior region.

\section{Domain of existence}

We now present the domain of existence of the particle-like EsGB solutions
considering,  in particular, quadratic, cubic and dilatonic coupling functions.
We also address their relation to the black holes and wormholes of the respective theories.

\begin{figure}
\begin{center}
(a)\includegraphics[width=.45\textwidth, angle =0]{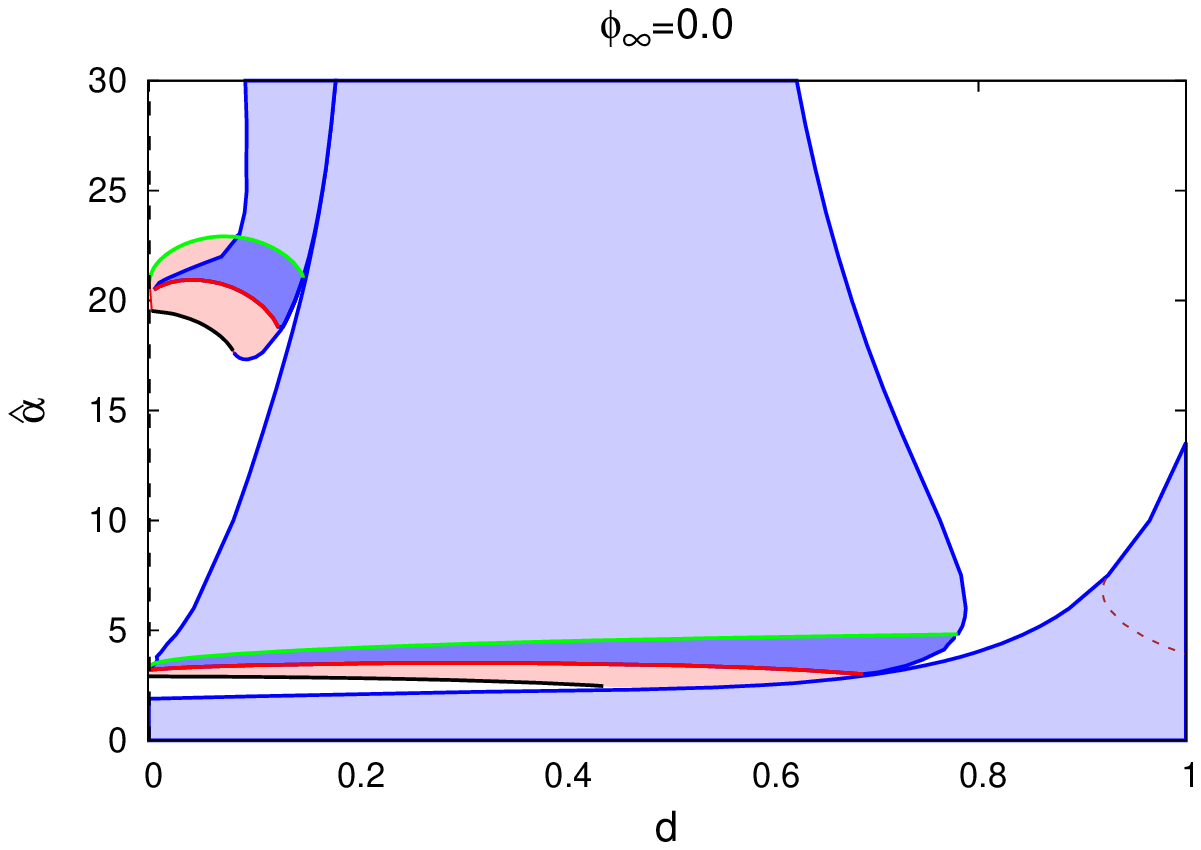}
%(b)\includegraphics[width=.45\textwidth, angle =0]{doer_dil_aM2vDMph.pdf}\\
%(a)\includegraphics[height=.40\textheight, angle =0]{doer_2_0_aM2vDM.pdf}\\
(b)\includegraphics[width=.45\textwidth, angle =0]{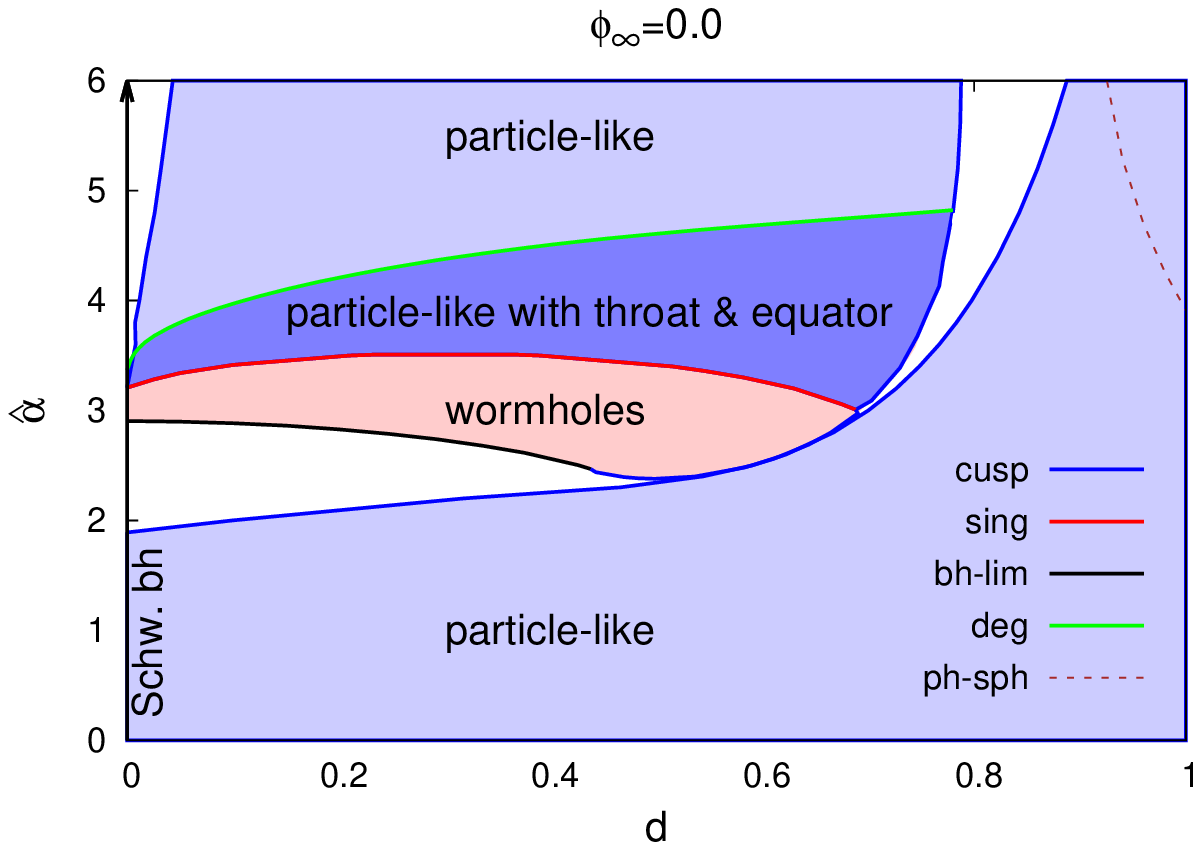}\\
(c)\includegraphics[width=.45\textwidth, angle =0]{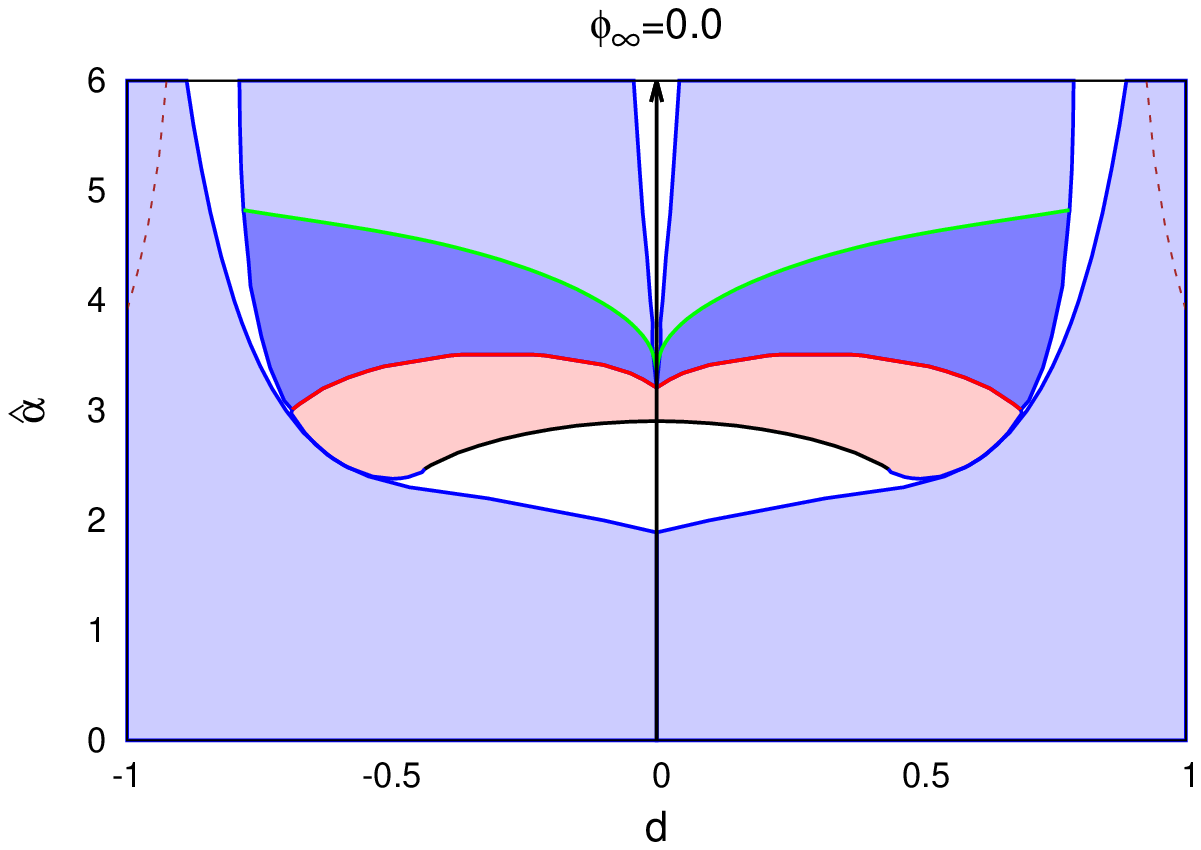} 
(d)\includegraphics[width=.45\textwidth, angle =0]{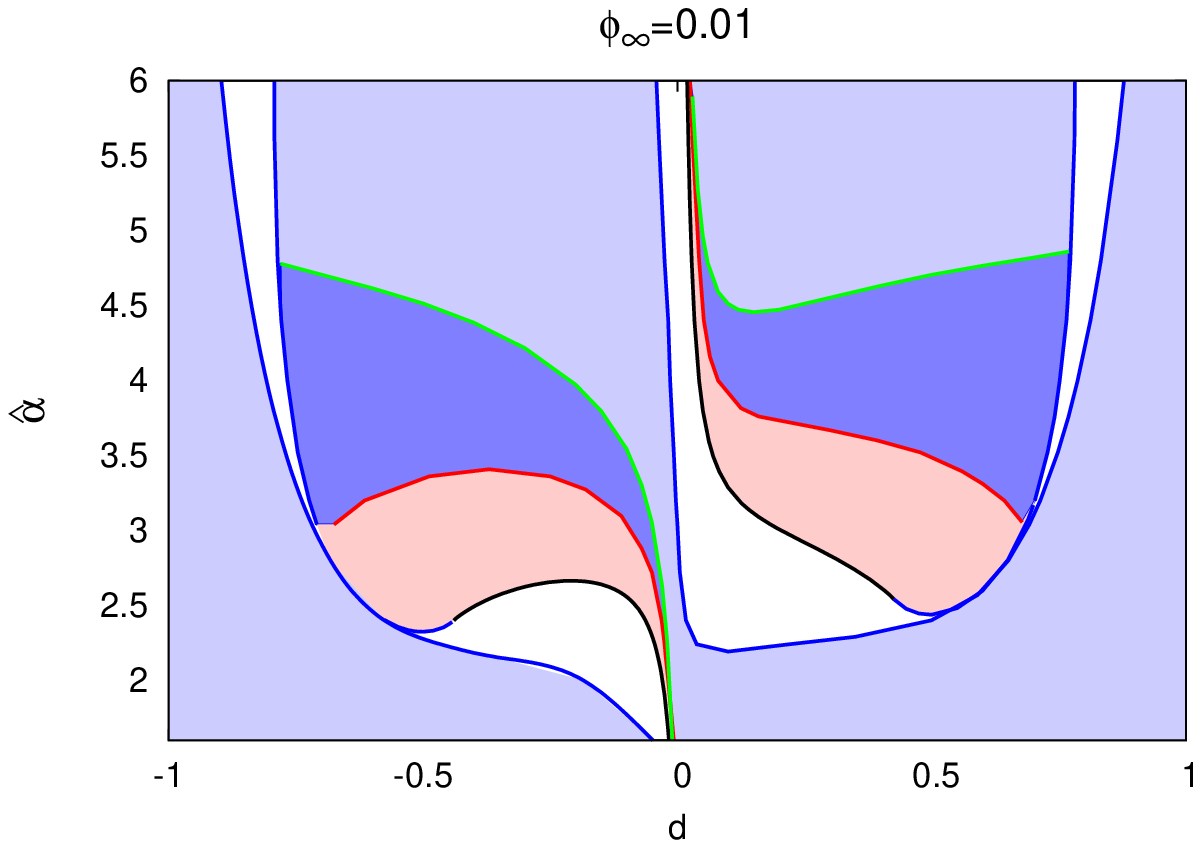}\\ 
(e)\includegraphics[width=.45\textwidth, angle =0]{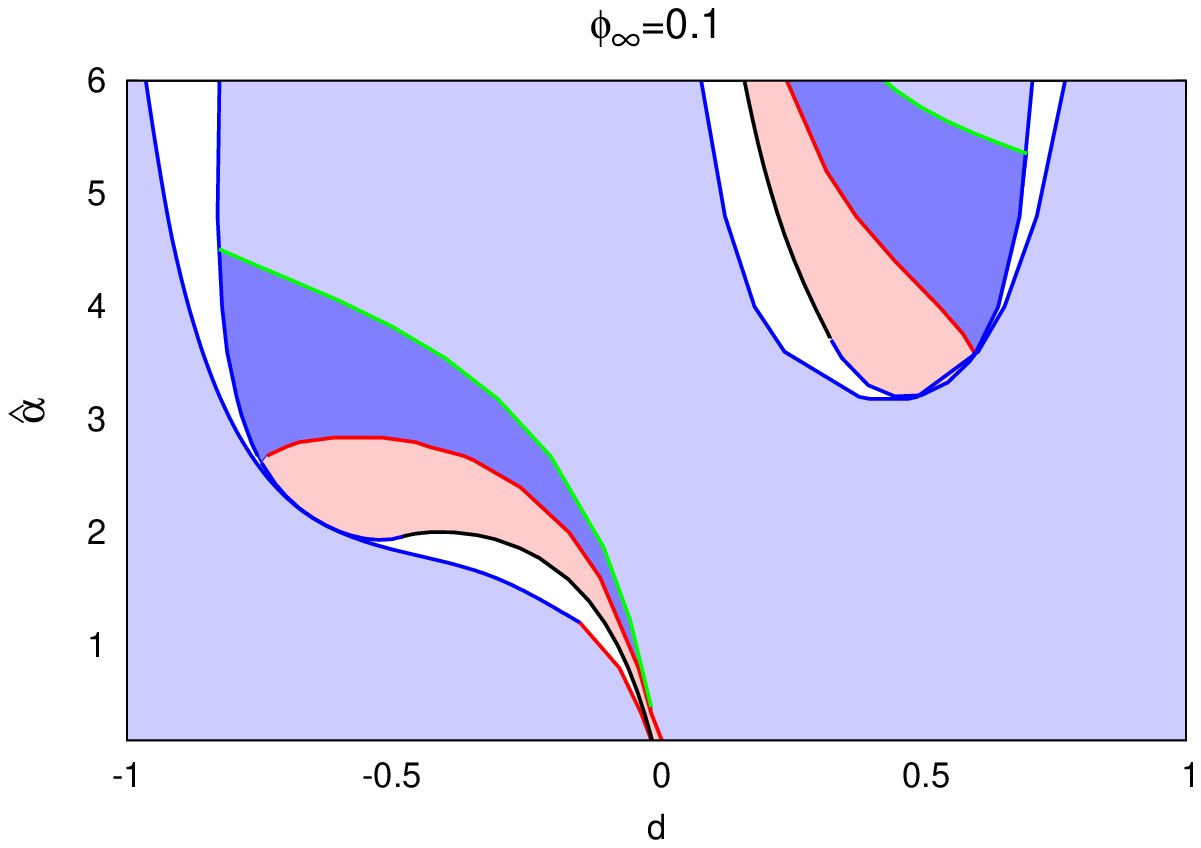} 
(f)\includegraphics[width=.45\textwidth, angle =0]{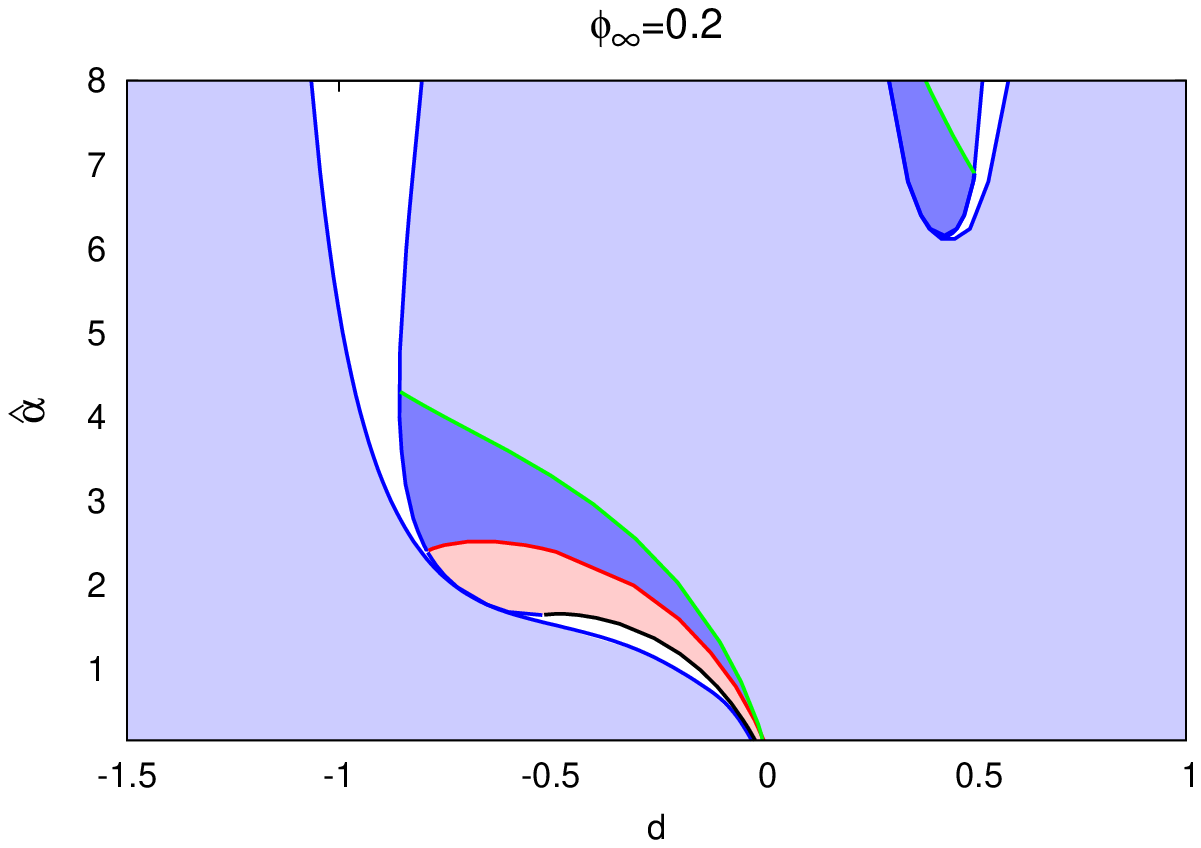}\\ 
(g)\includegraphics[width=.45\textwidth, angle =0]{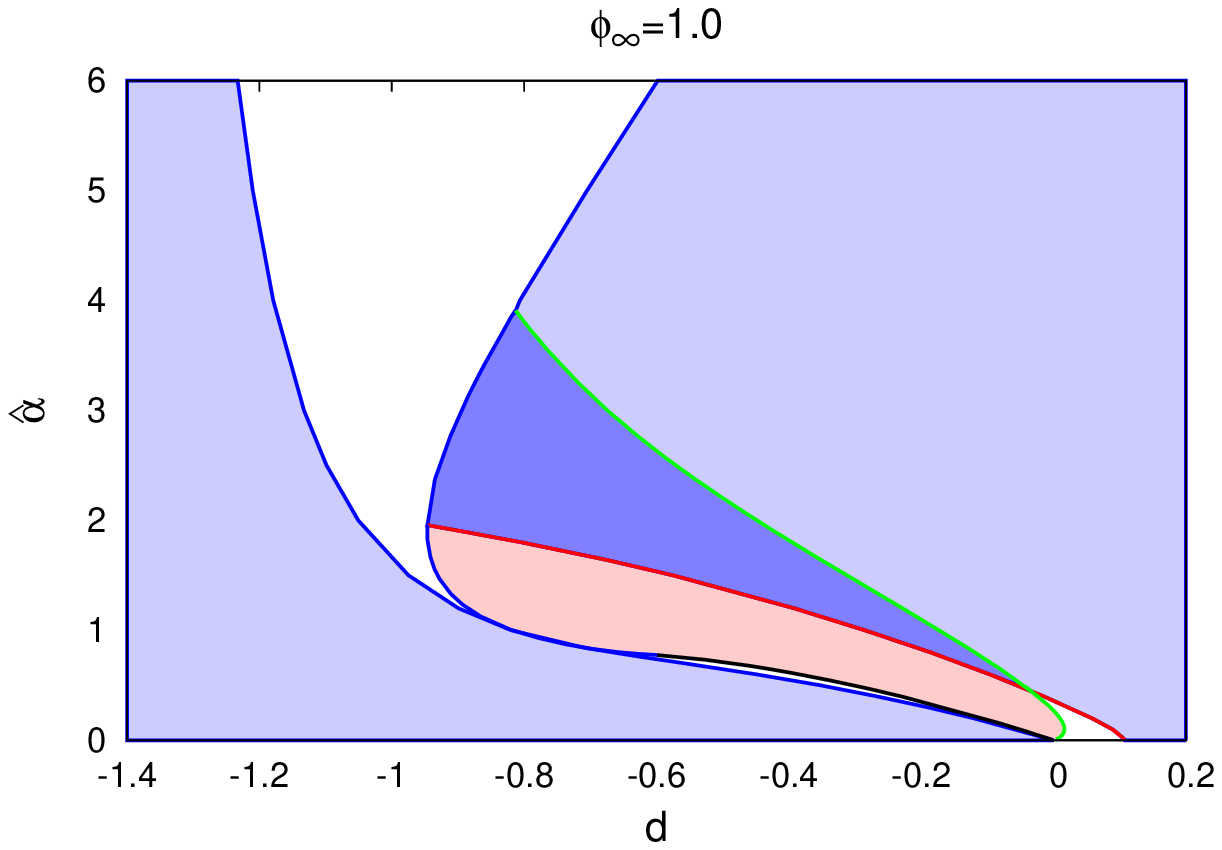}
(h)\includegraphics[width=.45\textwidth, angle =0]{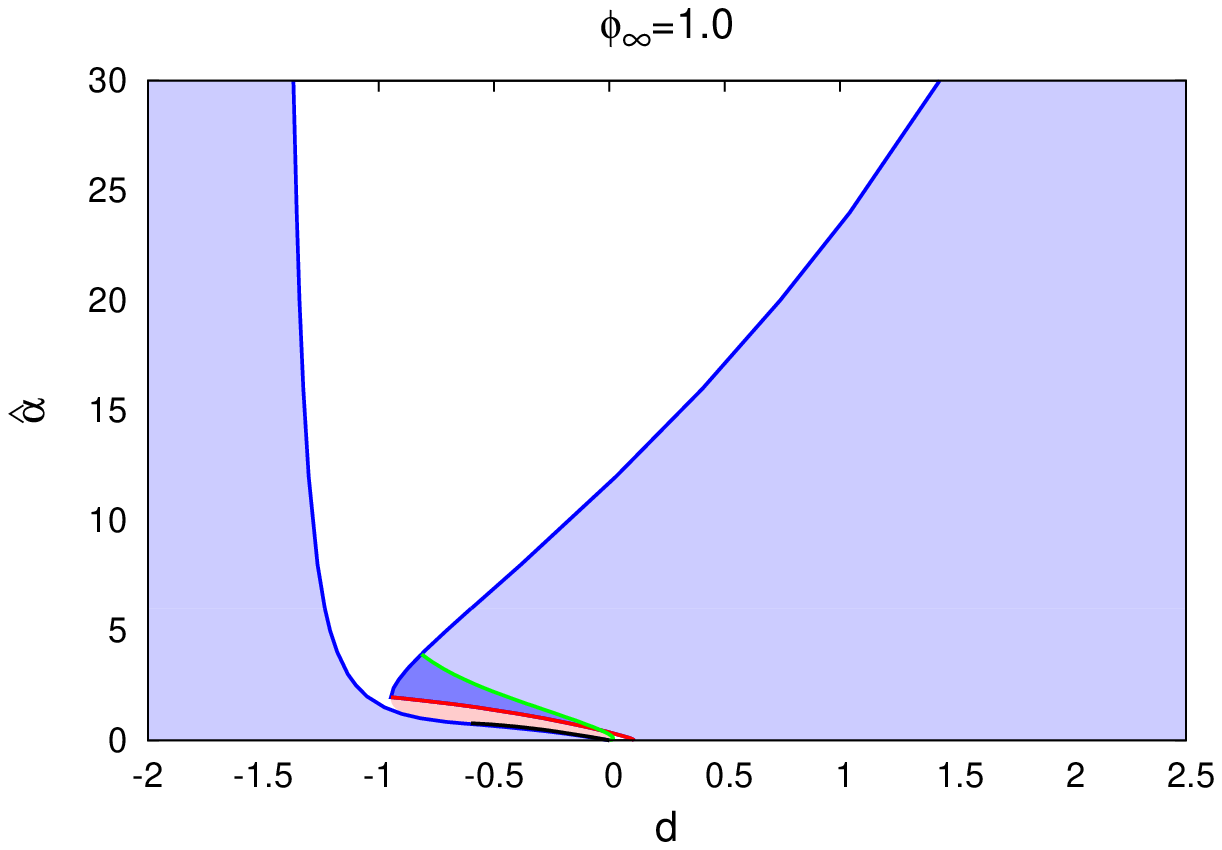}
\end{center}
\caption{
Domain of existence (blue) in the $d-\hat{\alpha}$ plane for 
the quadratic coupling function 
$F=\alpha\phi^2$ with $\phi_\infty=0$ (a), (b) and (c), 
and for $\phi_\infty=0.01$ (d), $\phi_\infty=0.1$ (e), $\phi_\infty=0.2$ (f), 
$\phi_\infty=1$ (g) and (h). 
Also included are the domains of existence of wormholes (dark-blue and rose)
and black holes (black curves).
The boundaries of the domain of existence are marked differently for
cusp singularities (blue) and curvature singularities (red).
The green curves show solutions with degenerate throat.
The domain with both particle-like objects and wormholes 
is indicated in dark blue. The domain with wormholes only (rose)
is in part delimited by scalarized black holes.
The dashed curve indicates the boundary of solutions for which
lightrings exist. Solutions without lightrings are located above and 
to the right of the dashed curve.
}
\label{fig_doe2}
\end{figure}

\subsection{Quadratic EsGB solutions}

We begin by investigating the domain of existence of the quadratic EsGB solutions.
The coupling function $F=\alpha\phi^2$ is symmetric in $\phi$,
and thus the solutions come in degenerate pairs with respect
to the transformation $\phi \to - \phi$,
carrying positive and negative scalar charge,
provided the symmetry is respected
by the boundary conditions as dictated by the choice of $\phi_\infty=0$.
To inspect the domain of existence it is therefore sufficient
to restrict to non-negative scalar charge $d$.

We exhibit the domain of existence in Figs.~\ref{fig_doe2}(a) and (b)
for $\phi_\infty=0$,
employing the scalar charge $d$ 
and the scaled GB coupling constant $\hat{\alpha}$
to delimit its size.
We restrict to $\hat{\alpha}\geq 0$.
As we see in Fig.~\ref{fig_doe2}(a), for a quadratic coupling function,
the domain of existence consists of many disconnected regions
due to the emergence of solutions
with a radially excited scalar field for large values of $\hat{\alpha}$.
In these figures, 
all regions with particle-like solutions are shown in blue color.

Figure \ref{fig_doe2}(b) provides a magnification of the
domain of existence up to the value $\hat \alpha = 6$, and allows one to observe
more carefully the different regions.
The lowest region, corresponding to small values of $\hat \alpha$,
contains particle-like solutions
without nodes of the scalar field. Here, $\phi$ is negative as it approaches the origin.
In the second region, located above and to the left of the first region,
there are particle-like solutions with one node. 
Here, $\phi$ is positive as it approaches the origin.
Parts of these regions, both indicated by blue colour, are seen
in Figs.~\ref{fig_doe2}(a)-(b). In Fig.~\ref{fig_doe2}(a),
also part of the third region is seen, where the scalar field has two nodes
and is again negative as it approaches the origin. 
The regions with more nodes reside at still higher values of $\hat \alpha$.
%in  branch of the scalarized EsGB black holes(b) 

Let us now inspect the boundaries of the domain of existence.
There are various mechanisms that delimit the domain.
The nodeless particle-like solutions are simply delimited by
the occurrence of a cusp singularity, marked by a dark blue line.
The excited particle-like solutions are also delimited
by singular solutions, where $g_{tt}$ vanishes at some point.
These are marked by a red line.
The excited particle-like solutions possess regions of overlap
with the wormhole solutions, indicated by a darker shade of blue.
In these regions, the particle-like solutions possess a throat
and an equator.
In this overlapping region, wormholes can be constructed
from the particle-like solutions by either cutting at the throat or 
cutting at the equator, and then continuing (after a suitable coordinate transformation) with the symmetrically
reflected solution into the second asymptotic region.
Let us note that wormholes can also be constructed from singular solutions 
with throat/equator in the case that the singularity is located 
at a smaller radius than the one of the throat/equator.
The limiting line of the overlap region
of particle-like and wormhole solutions
to the region with particle-like solutions only
%where the circumferential radius $R_c$ develops a saddle point, 
is marked in green.
The region with wormholes only is shown in rose.
The vertical line (at the value $d=0$) represents the Schwarzschild black holes.
The location of the fundamental branch of the scalarized EsGB black holes 
is indicated in black, and likewise the location of the first excited EsGB
black hole branch in Fig.~\ref{fig_doe2}(a).

When we allow for a non-vanishing value of the scalar field at infinity,
$\phi_\infty \ne 0$, the symmetry is broken,
and we have to study both positive and negative values 
of the scalar charge.
In Figs.~\ref{fig_doe2}(c)-(h), we illustrate
how the domain of existence changes as the asymptotic
value $\phi_\infty$ increases.
%displaying a symmetric interval in $d$
In particular, we choose $\phi_\infty=0$ [Fig.~\ref{fig_doe2}(c)],
$\phi_\infty=0.01$ [Fig.~\ref{fig_doe2}(d)], $\phi_\infty=0.1$ [Fig.~\ref{fig_doe2}(e)],
$\phi_\infty=0.2$ [Fig.~\ref{fig_doe2}(f)] and $\phi_\infty=1.0$ [Figs.~\ref{fig_doe2}(g)-(h)].
%The coloring here is intended to guide the eye
%when observing the changes.
%For better clarity also only the particle-like  and black hole solutions are shown.
We note that, when increasing $\phi_\infty$,
dramatic changes arise.
The regions not only change their sizes considerably,
but they change the ways they are connected.
The upper region for negative $d$ becomes connected to the lower region with positive
$d$ when  $\phi_\infty$ is increased from zero. 
Remarkably, the single branch of scalarized black holes splits into two branches
with positive, resp. negative $d$. 
We see that the scalarized black holes for positive values of $d$ have no limit of 
vanishing $\hat{\alpha}$ and, in fact, cease to exist if $\phi_\infty$ is as large as $0.2$.
Consequently, the wormhole solutions have no black-hole limit in this case.
In this region, the domain of existence is limited by cusp-singularities only.
Increasing $\phi_\infty$ further
to the value $\phi_\infty=1$,
%and we have resumed including
%also information on the wormholes and black holes, employing 
%the coloring and labeling of Figs.~\ref{fig_doe2}(a) and (b).
we note that the particle-like solutions with throat and equator (and wormholes)
of the negative $d$ region have moved to smaller values of $\hat \alpha$,
%whereas the ones of the positive $d$ region have
%moved up to higher $\hat \alpha$ and are no longer visible in
%the figure.
whereas the ones of the positive $d$ region are no longer visible in the figure.
By increasing again the interval in $\hat{\alpha}$, as seen in Fig.~\ref{fig_doe2}(h),
we see that this region has disappeared.
Finally, we note that for quadratic coupling function and $\phi_\infty=1$, the boundary of existence
does not include the interval $0< d <0.1$ for $\hat{\alpha} \to 0$.
Note that the domain of existence for negative values of $\phi_\infty$ can be obtained from 
the one with positive values of  $\phi_\infty$ by reflection with respect to the $d=0$ axis.

\subsection{Cubic EsGB solutions}

\begin{figure}
\begin{center}
%
%(a)\includegraphics[width=.45\textwidth, angle =0]{doeBH_3_0_aM2vDM.eps}
(a)\includegraphics[width=.45\textwidth, angle =0]{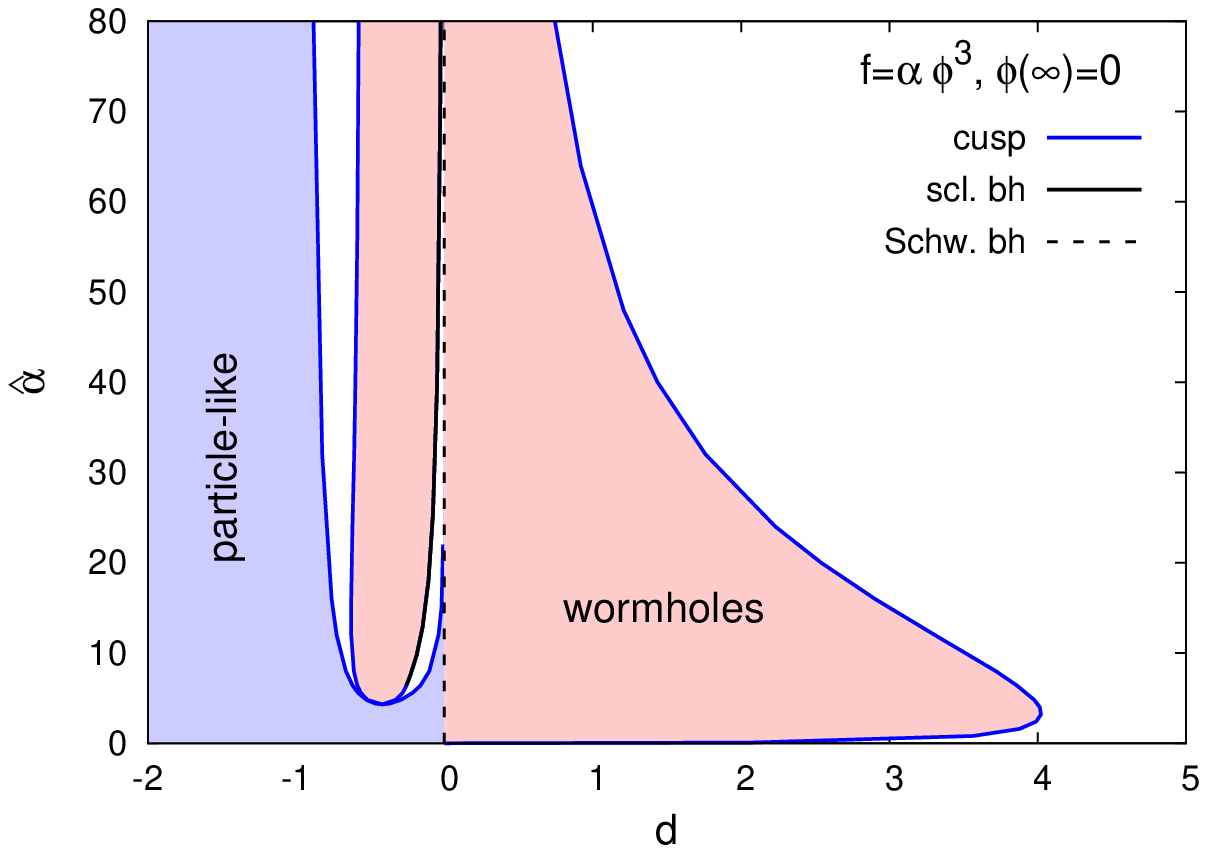}
(b)\includegraphics[width=.45\textwidth, angle =0]{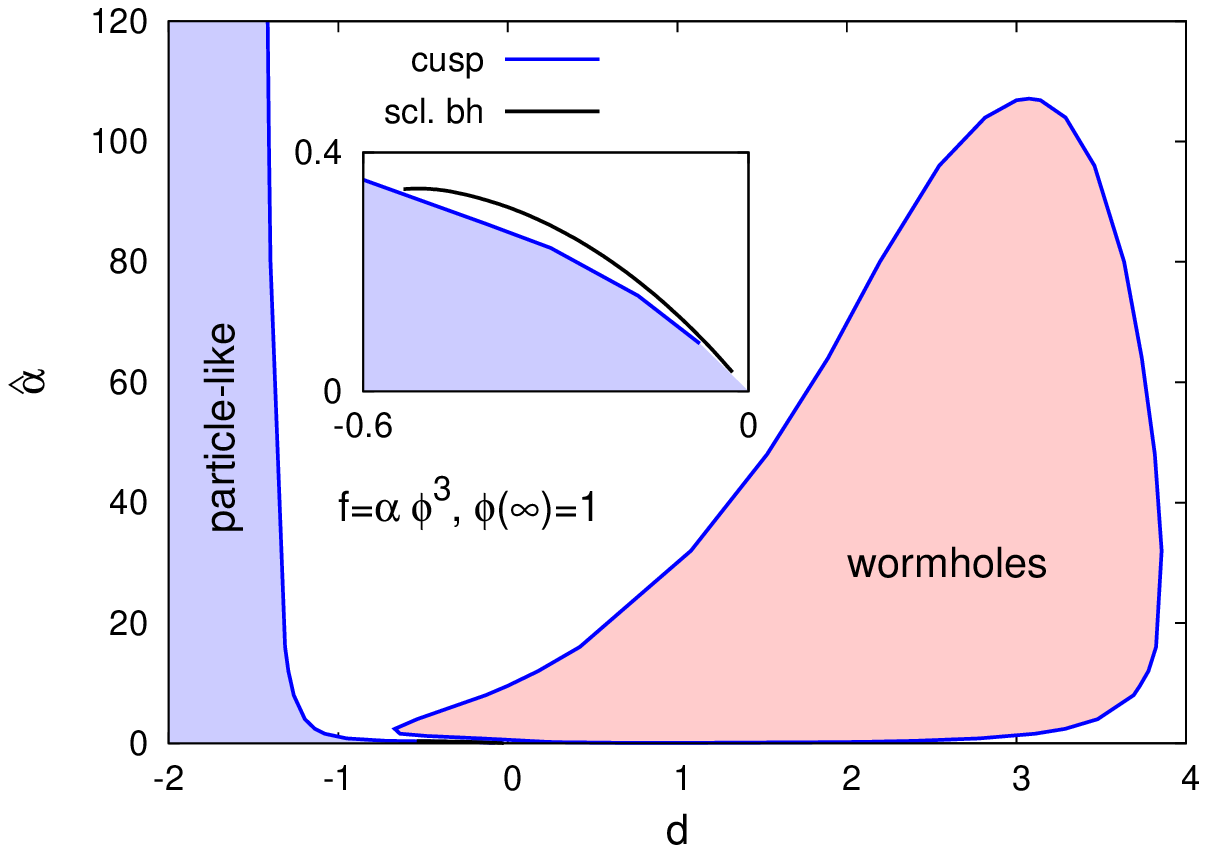}\\
(c)\includegraphics[width=.45\textwidth, angle =0]{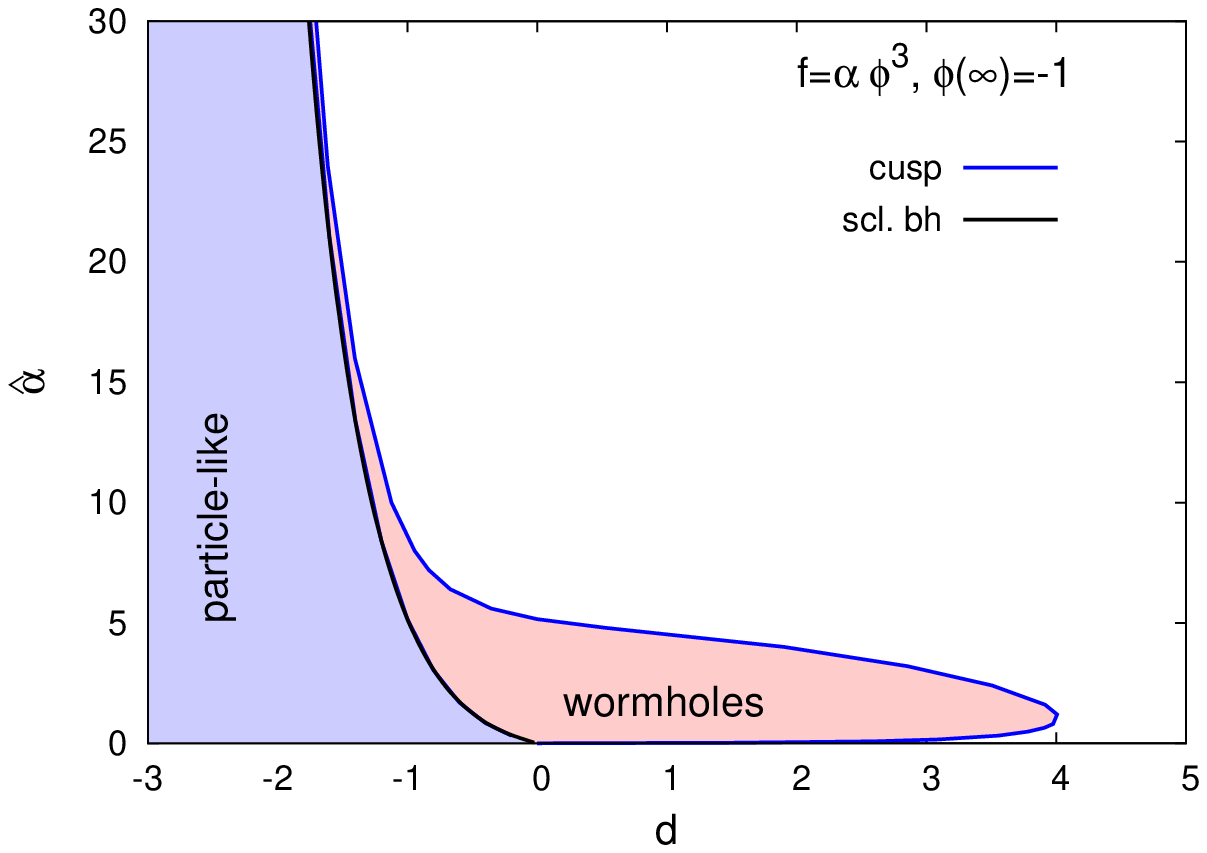}
(d)\includegraphics[width=.45\textwidth, angle =0]{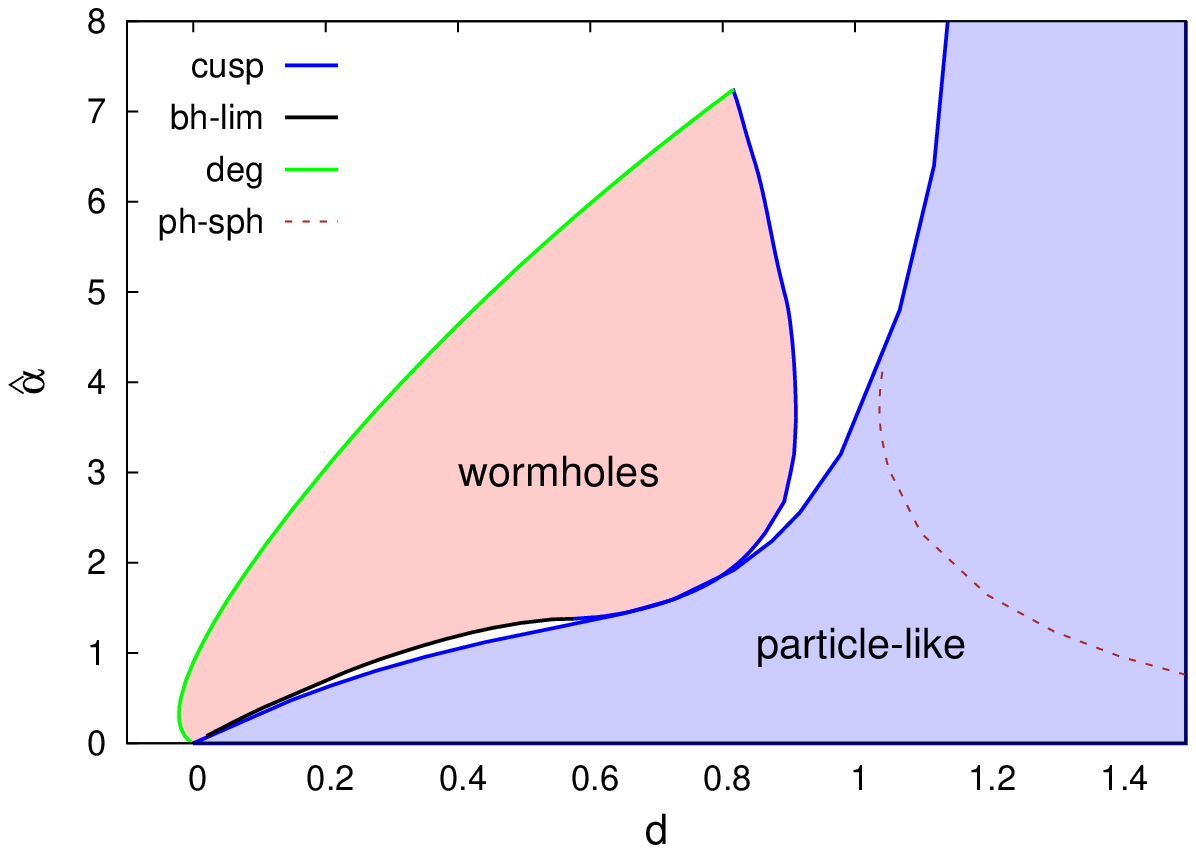}
%(b)\includegraphics[width=.45\textwidth, angle =0]{doe_3_1_aM2vDMph.eps}\\
%(c)\includegraphics[width=.45\textwidth, angle =0]{doe_3_-1_aM2vDMph.eps}
%(d)\includegraphics[width=.45\textwidth, angle =0]{doer_dil_aM2vDMph.eps}
\end{center}
\caption{
(a)-(c) 
Domain of existence of particle-like solutions (in blue colour) in the
$d-\hat{\alpha}$ plane for the cubic coupling function 
$F=\alpha\phi^3$ with $\phi_\infty=0$ (a), 
$\phi_\infty=1$ (b), and $\phi_\infty=-1$ (c);
and for the dilatonic coupling function $F=\alpha e^{-\phi}$ with $\phi_\infty=0$ (d).
The boundaries of their domain of existence are given by
cusp singularities (dark blue line) and black holes 
(solid grey line).
The domain of wormholes is also indicated (in rose colour).
The dashed red curve indicates the boundary of solutions for which
lightrings exist. Solutions without lightrings are located above the dashed 
red curve.
}
\label{fig_doe3}
\end{figure}

We now turn to the cubic coupling function $F=\alpha\phi^3$.
Here, the domain of existence of particle-like solutions  resides
strictly in the negative $d$ region (for positive coupling constant $\hat \alpha$).
We exhibit their domain of existence (in blue colour) in Figs.~\ref{fig_doe3}(a)-(c),
employing again the scaled scalar charge $d$ 
and the scaled GB coupling constant $\hat{\alpha}$
to delimit its size.

In Fig.~\ref{fig_doe3}(a), we have chosen a vanishing boundary
value for the scalar field at infinity, $\phi_\infty=0$. We have also included the region of
wormholes (in rose colour) which resides in the
positive $d$ region. As a result, there is no overlap between the particle-like
solutions and the wormholes for the cubic coupling function.
The boundaries of the domain of existence of particle-like solutions
are again mainly determined by the occurrence of cusp singularities 
(denoted again by the dark blue line). In this case, 
Schwarzschild black holes (denoted by the dashed black line) are encountered
for vanishing scalar charge $d$. The domain of existence of wormholes is also
delimited by cusp singularities (blue line) and Schwarzschild black holes (black line),
but also by scalarized black holes (denoted by the solid grey line).
The latter solutions also reside strictly in the negative $d$ region (for positive $\alpha$).

As was illustrated in the case of the quadratic coupling, the domain of
existence of particle-like solutions changes when the boundary
value of the scalar field at infinity $\phi_\infty$ also changes.
We demonstrate that this holds also for the cubic coupling 
function by choosing $\phi_\infty=1$ in Fig.~\ref{fig_doe3}(b) 
and $\phi_\infty=-1$ in Fig.~\ref{fig_doe3}(c).
From Fig. \ref{fig_doe3}(b), we see that, 
when adopting a positive value for $\phi_\infty$, 
the domain of existence gets restricted towards larger
(absolute) values of $d$.
%For sufficiently negative $\phi_\infty$, on the other hand,
%the domain is no
%longer bounded by cusp singularities but only by scalarized
%black holes, as seen in Fig.~\ref{fig_doe3}(c). 
For sufficiently negative $\phi_\infty$, on the other hand,
it seems that the domain is no
longer bounded by cusp singularities but only by scalarized
black holes. 
However, there is a tiny gap between the boundary of the particle-like solutions
and the scalarized black hole solutions, not visible in  Fig.~\ref{fig_doe3}(c).
%We note that in these two figures, we depict only 
%the negative $d$ region, and have not therefore included the
%respective wormhole solutions.

\subsection{Dilatonic EsGB solutions}

We finally consider the domain of existence of particle-like solutions
in the case of the dilatonic coupling function $F=\alpha e^{-\phi}$, in the plane again
of the scale-invariant quantities $d=D/M$ and $\hat{\alpha}=8\alpha/M^2$.
We exhibit their domain of existence (in blue colour) in Fig.~\ref{fig_doe3}(d)
for $\phi_\infty=0$, restricting again to $\alpha\geq 0$. 
Now, particle-like solutions exist only in the positive $d$ region.
At the boundary of the domain, a cusp singularity (indicated
by the blue line) is reached.

Also shown in Fig.~\ref{fig_doe3}(d) are the domain of existence of
wormholes (again in rose colour)
and the scalarized black holes (solid grey line),
which form part of the boundary of the domain of existence
of wormholes. 
Again, there is no overlap between the domains of existence of
particle-like solutions and wormholes, although their
boundaries lie quite close.
For the dilatonic coupling function, all wormholes can be constructed
from singular solutions.

\section{Observational effects}

We now discuss possible observational effects for these particle-like solutions.
We first address their redshift. Then, we consider the effective
stress-energy tensor to elucidate the high compactness
of many of these solutions. Finally, we study the motion of
both massive and massless particles in their gravitational background, and
show that many particle-like solutions possess lightrings
and can give rise to echoes in gravitational wave signals.

\subsection{Gravitational redshift}

\begin{figure}
\begin{center}
(a)\includegraphics[width=.45\textwidth, angle =0]{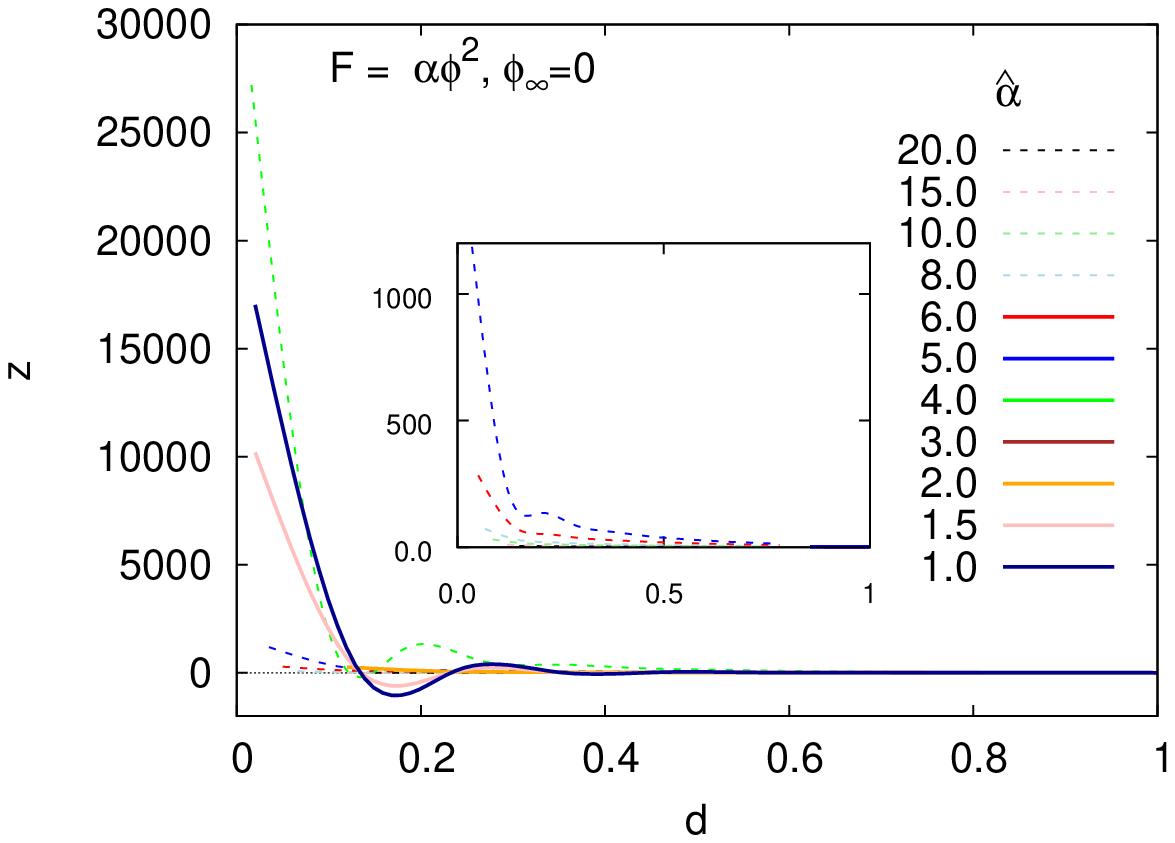}
(b)\includegraphics[width=.45\textwidth, angle =0]{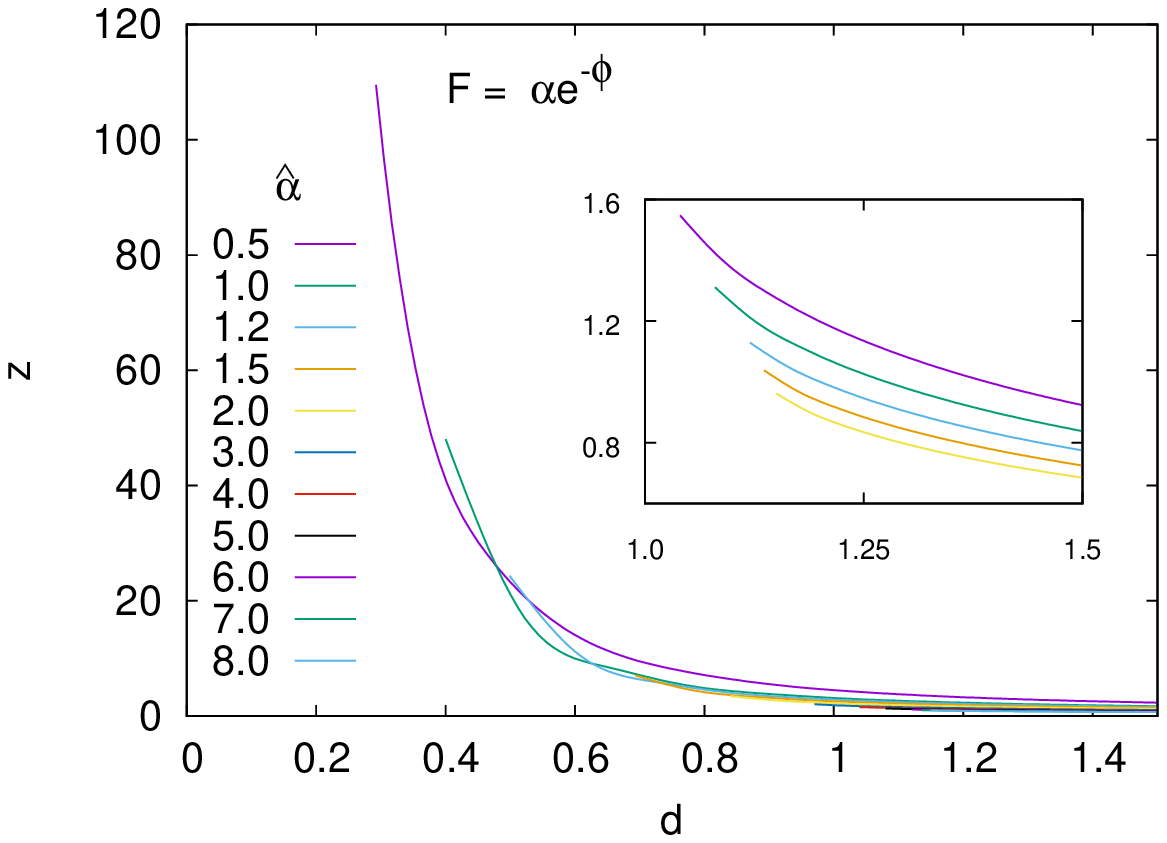}
\end{center}
\caption{
The redshift $z$ vs the dimensionless scalar charge $d$ for the coupling functions 
$F=\alpha\phi^2$ (a) and $F=\alpha e^{-\phi}$ (b) 
with $\phi_\infty=0$, and for several values of $\hat{\alpha}$.}
\label{fig_redshift}
\end{figure}

The redshift of a photon emitted by a source at the origin 
and seen by an observer in the asymptotic region
is given by the redshift factor $z$, Eq.~(\ref{redshift}).
It is therefore purely determined by
the metric component $g_{tt}(0)$ at the origin. 
We exhibit a set of examples for the redshift factor $z$ in Fig.~\ref{fig_redshift},
choosing the coupling function to be $F=\alpha\phi^2$ (a) and $F=\alpha e^{-\phi}$ (b) 
with $\phi_\infty=0$,
varying the scalar charge $d$ and selecting several values of 
the GB coupling constant $\hat{\alpha}$.

Interestingly, as seen in Fig.~\ref{fig_redshift}(a) for particle-like solutions with 
quadratic coupling function and $\phi_\infty=0$ the redshift factor can take both
positive and negative
values. The latter then correspond to a gravitational blueshift.
For a quadratic coupling function with
$\phi_\infty=1$, we observe that the redshift factor increases with
decreasing $\hat{\alpha}$, and takes very large values as the GR limit (Fisher solution)
is approached.
The analogous behavior is observed for a dilatonic coupling function.

\subsection{Stress-energy tensor and compactness}

\begin{figure}
\begin{center}
(a)\includegraphics[width=.45\textwidth, angle =0]{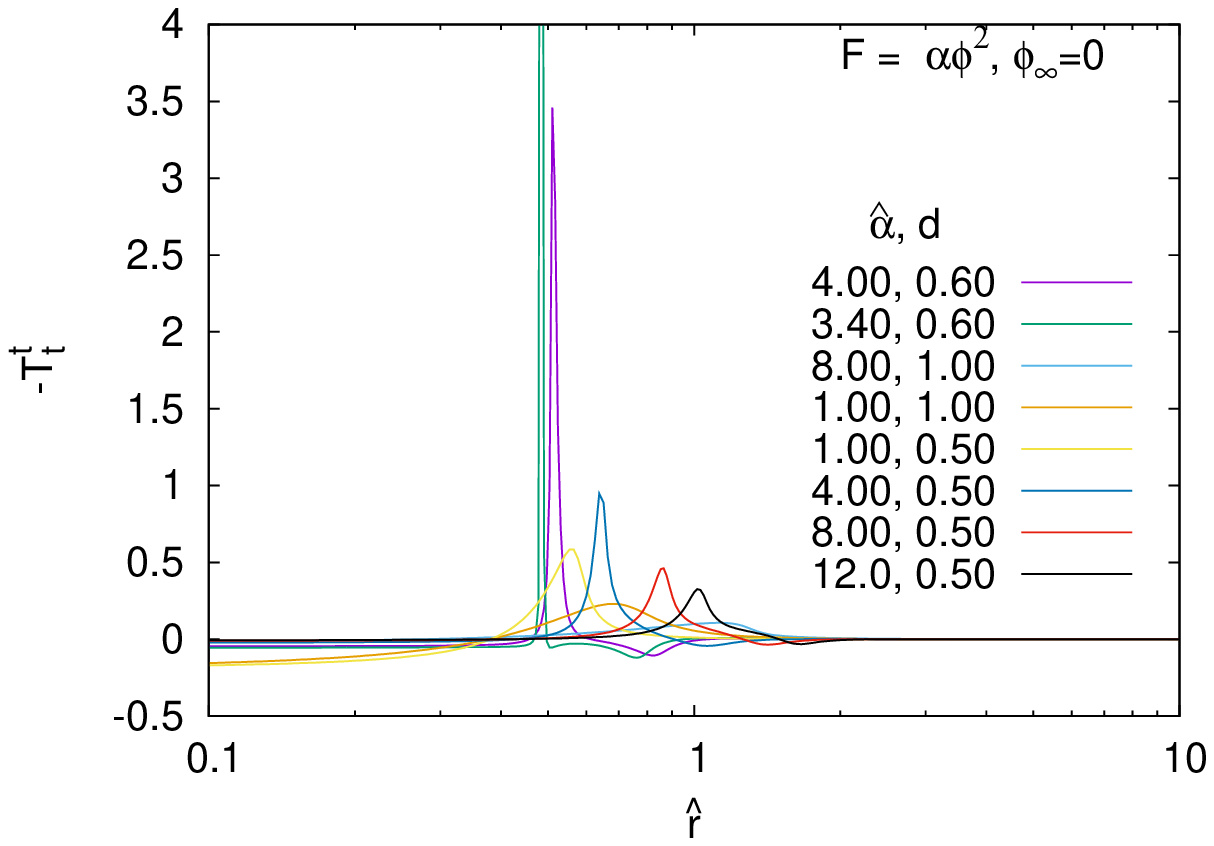}
(b)\includegraphics[width=.45\textwidth, angle =0]{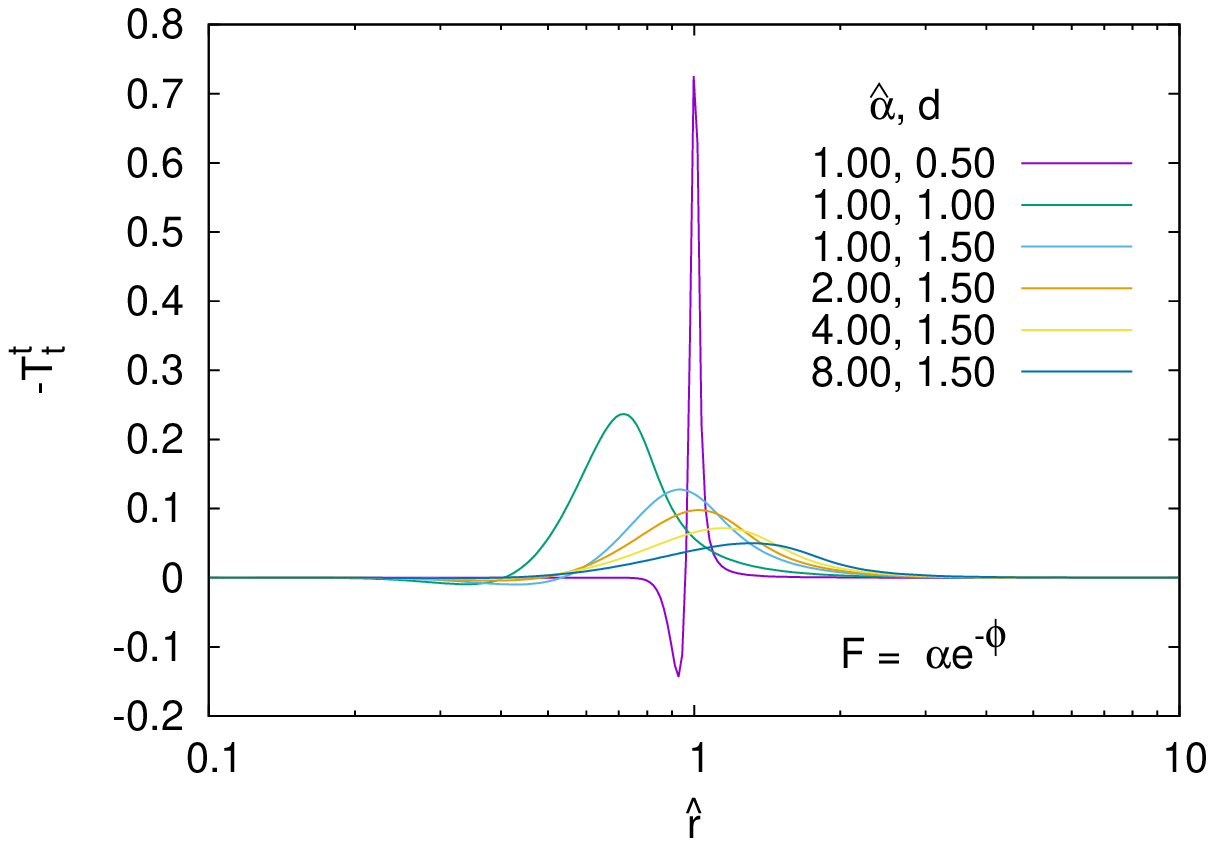}
\\
(c)\includegraphics[width=.45\textwidth, angle =0]{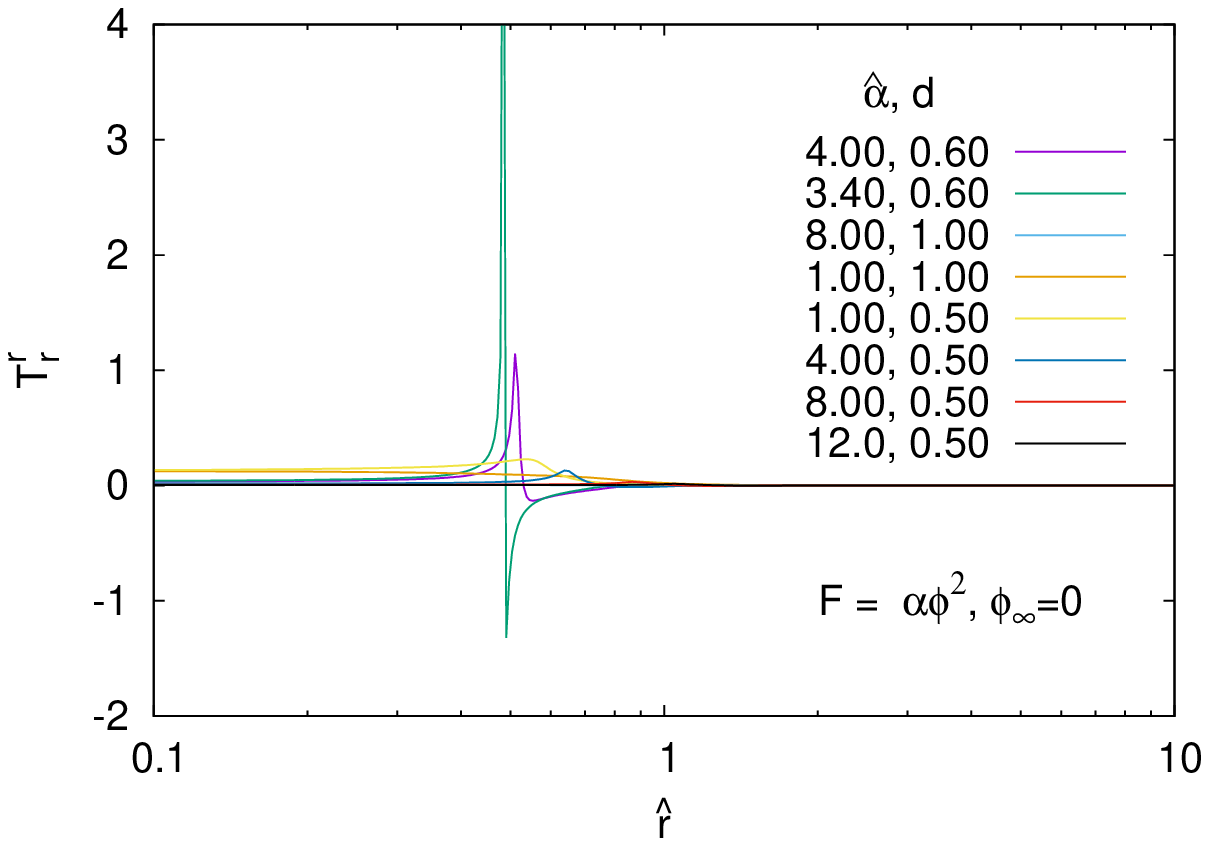}
(d)\includegraphics[width=.45\textwidth, angle =0]{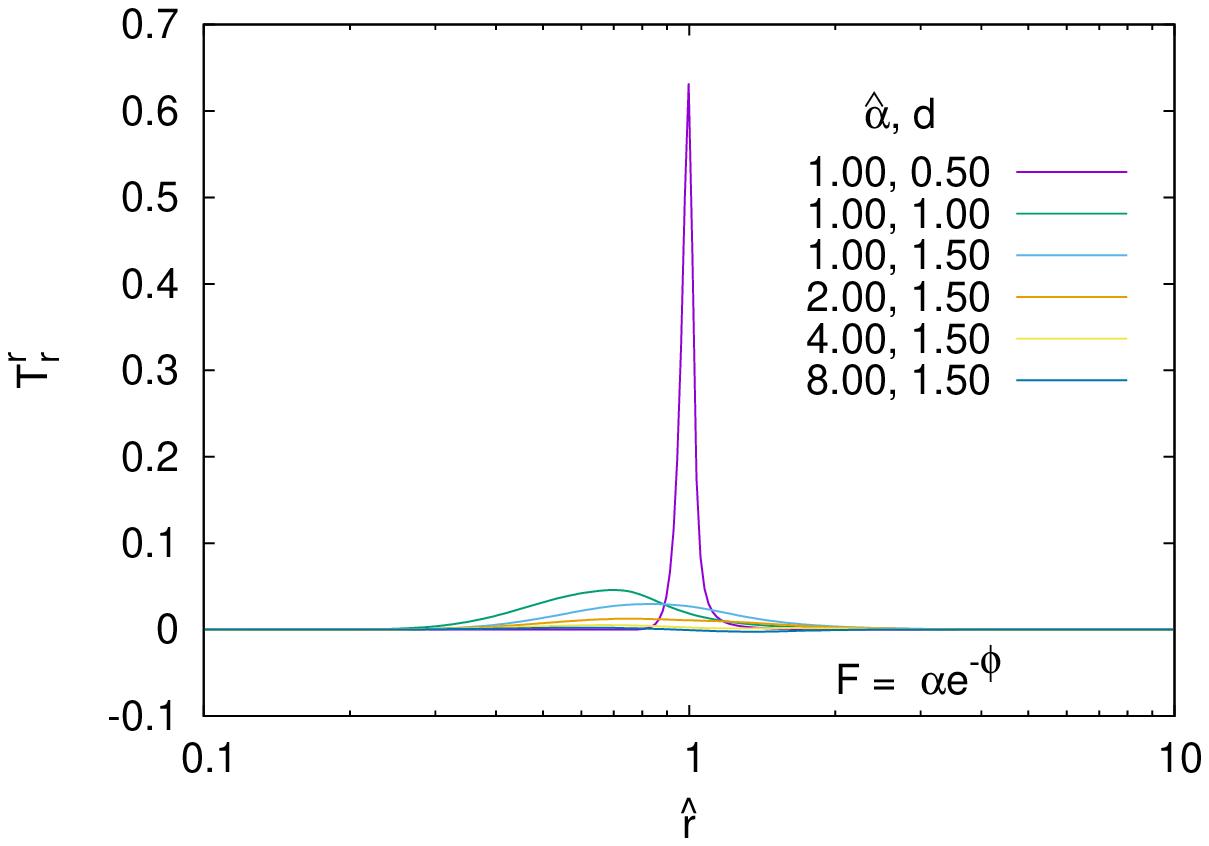}
\\
(e)\includegraphics[width=.45\textwidth, angle =0]{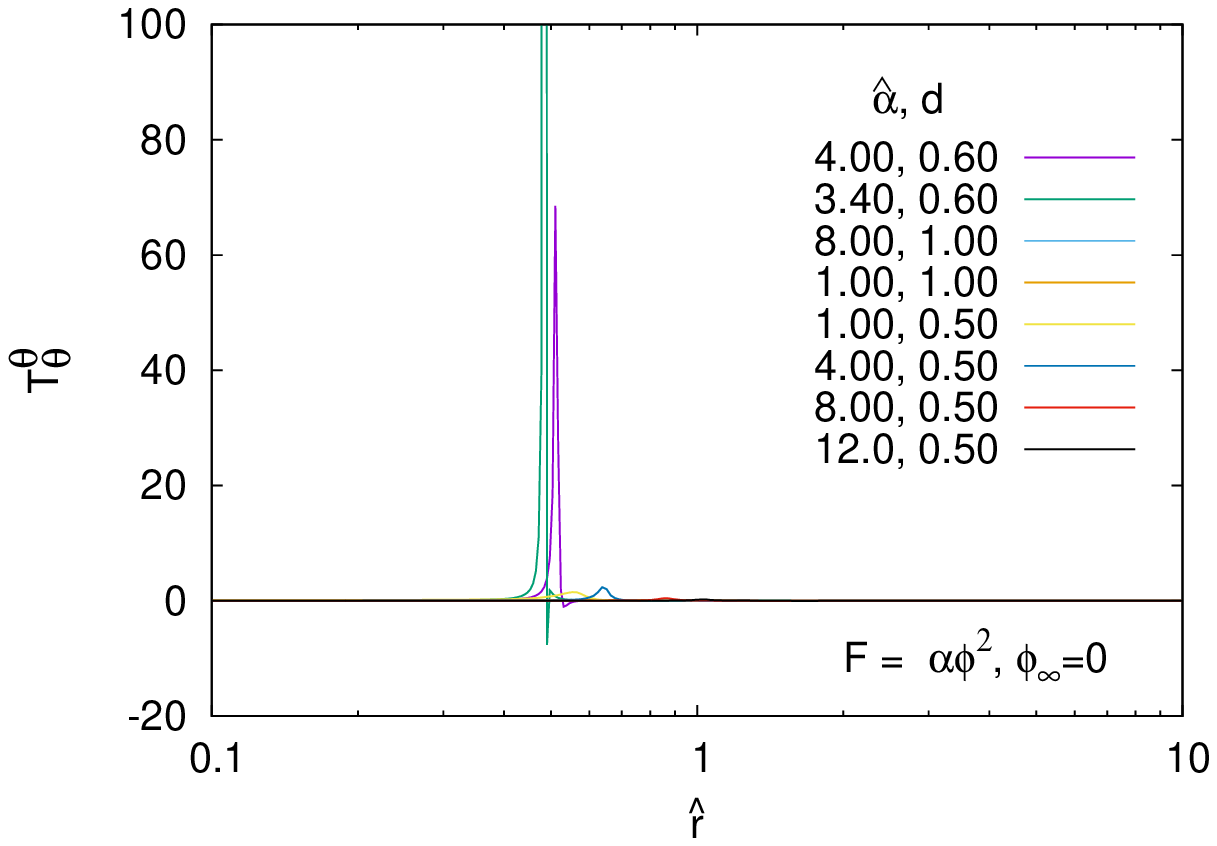}
(f)\includegraphics[width=.45\textwidth, angle =0]{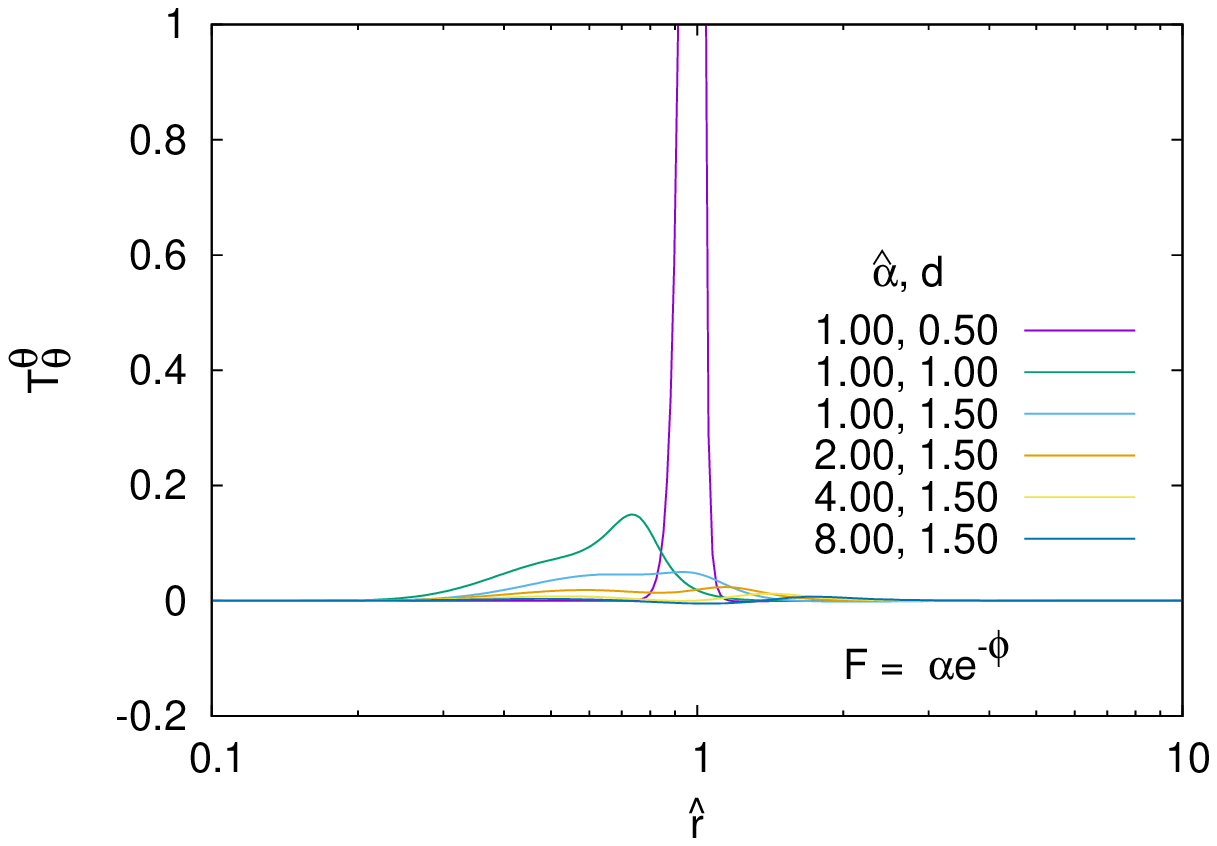}
\\
(g)\includegraphics[width=.45\textwidth, angle =0]{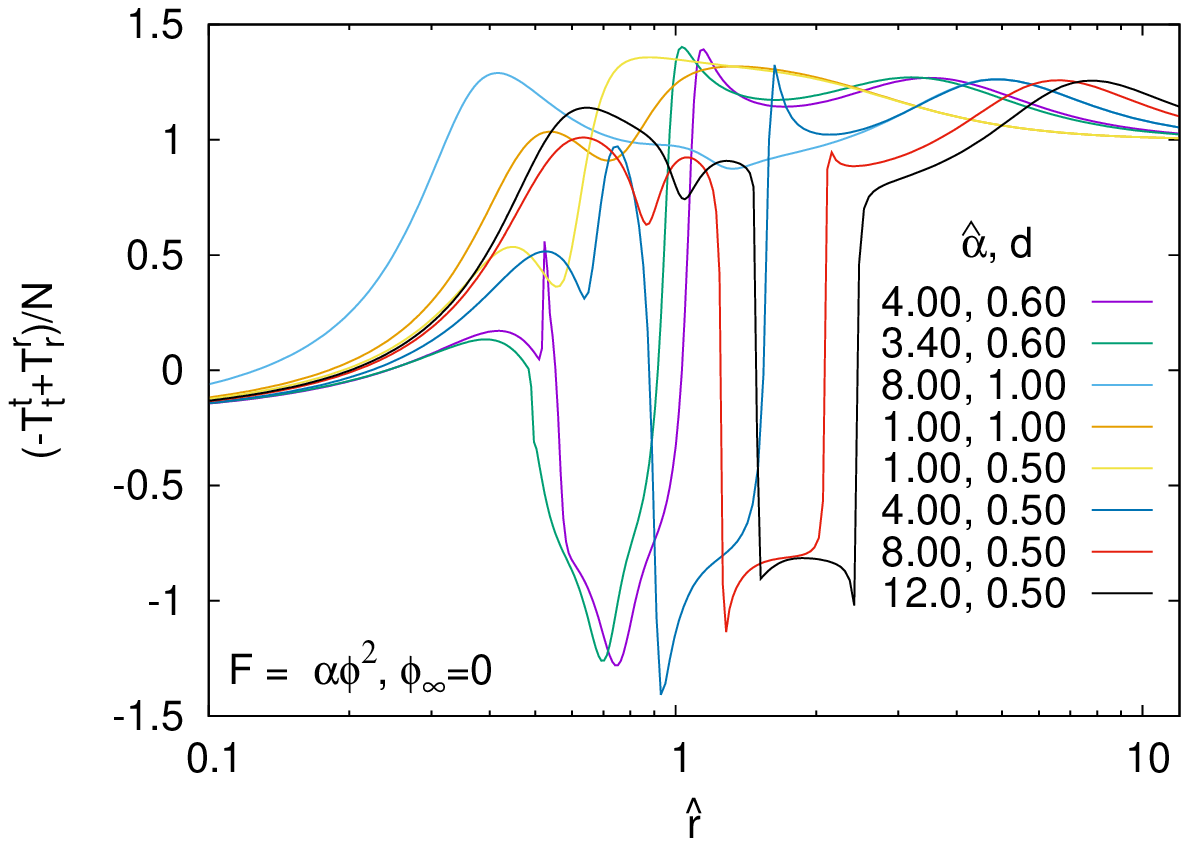}
(h)\includegraphics[width=.45\textwidth, angle =0]{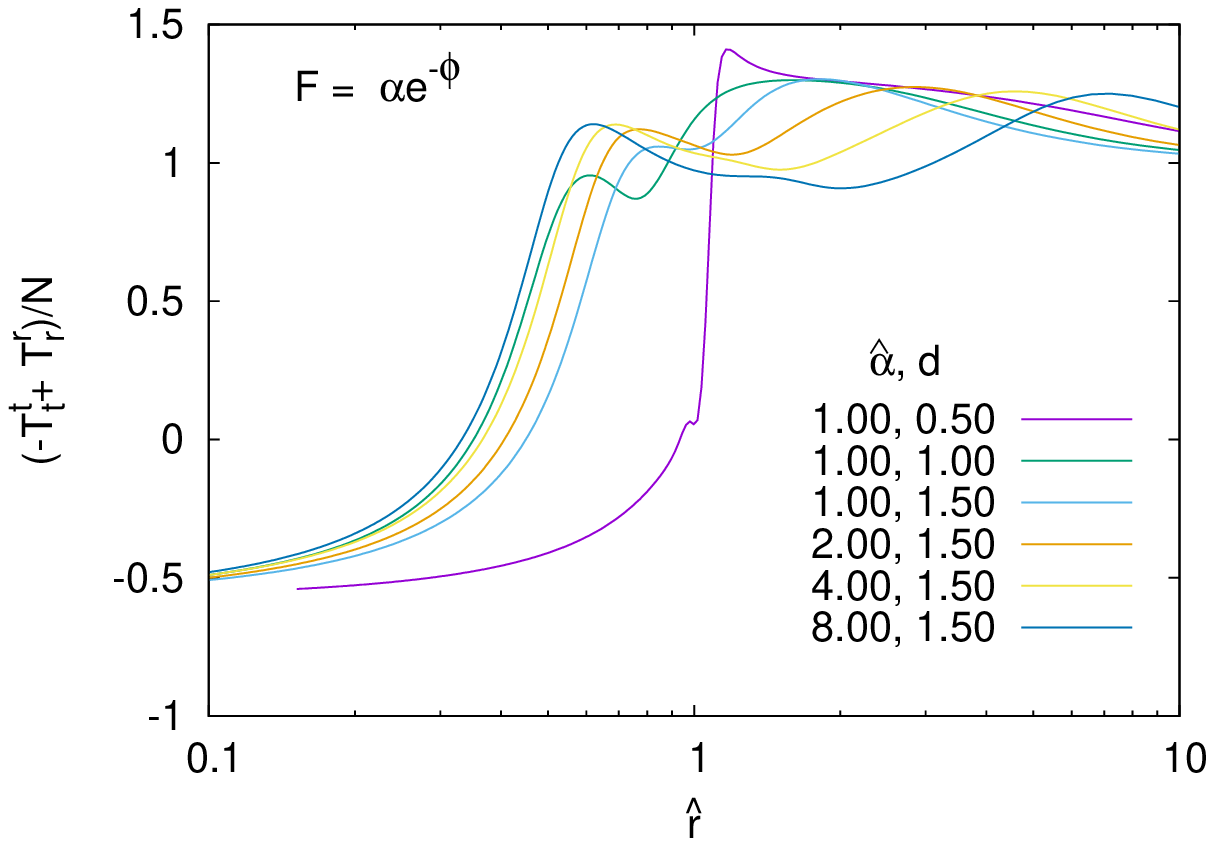}
\end{center}
\caption{
(a)-(h) The components of the effective stress-energy tensor
$- T^t_t$, $T^r_r$, and $T^\theta_\theta$
and the quantity $(- T^t_t+T^r_r)/N$ with 
$N=\sqrt{( T^t_t)^2+(T^r_r)^2+2(T^\theta_\theta)^2}$
vs the scaled radial coordinate $\hat{r}$ for the coupling functions 
$F=\alpha\phi^2$ with $\phi_\infty=0$ [left column: (a), (c), (e), (g)] and 
$F=\alpha e^{-\phi}$ [right column: (b), (d), (f), (h)], and several values of 
$\hat{\alpha}$ and  $d$. }
\label{fig_stress}
\end{figure}

\begin{figure}
\begin{center}
(a)\includegraphics[width=.45\textwidth, angle =0]{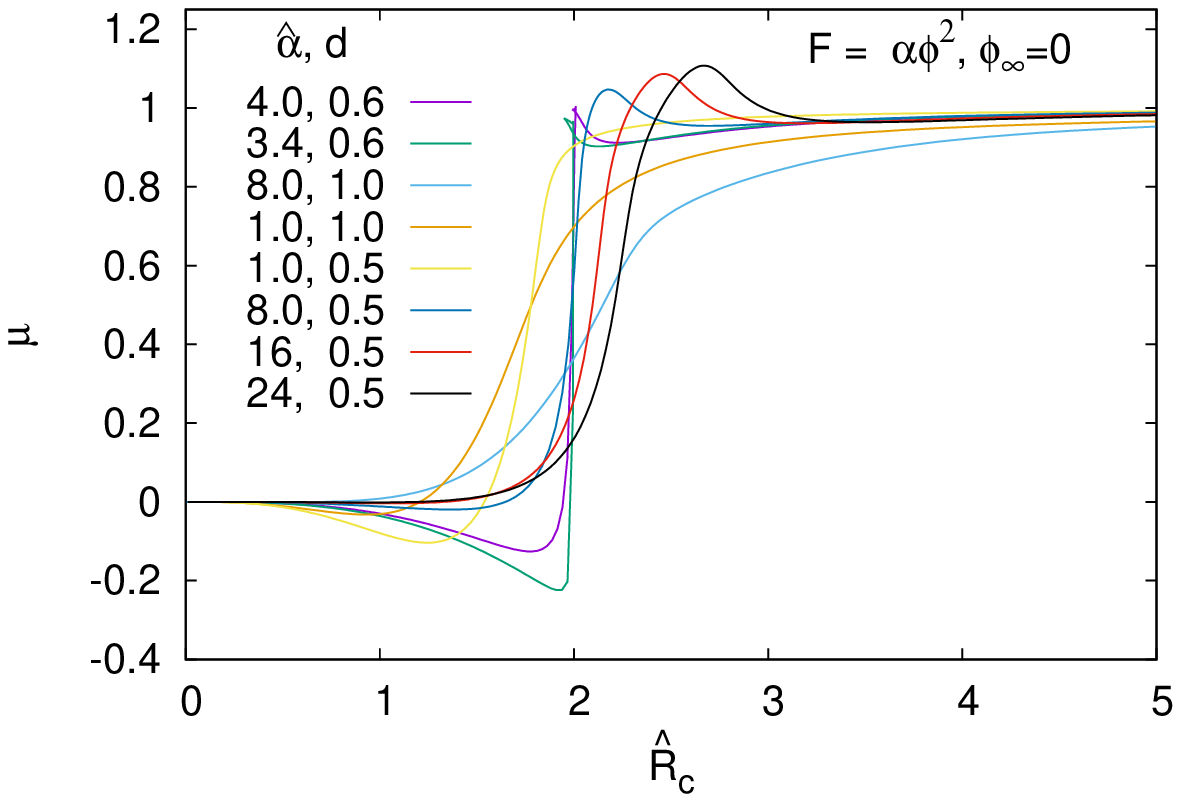}
(b)\includegraphics[width=.45\textwidth, angle =0]{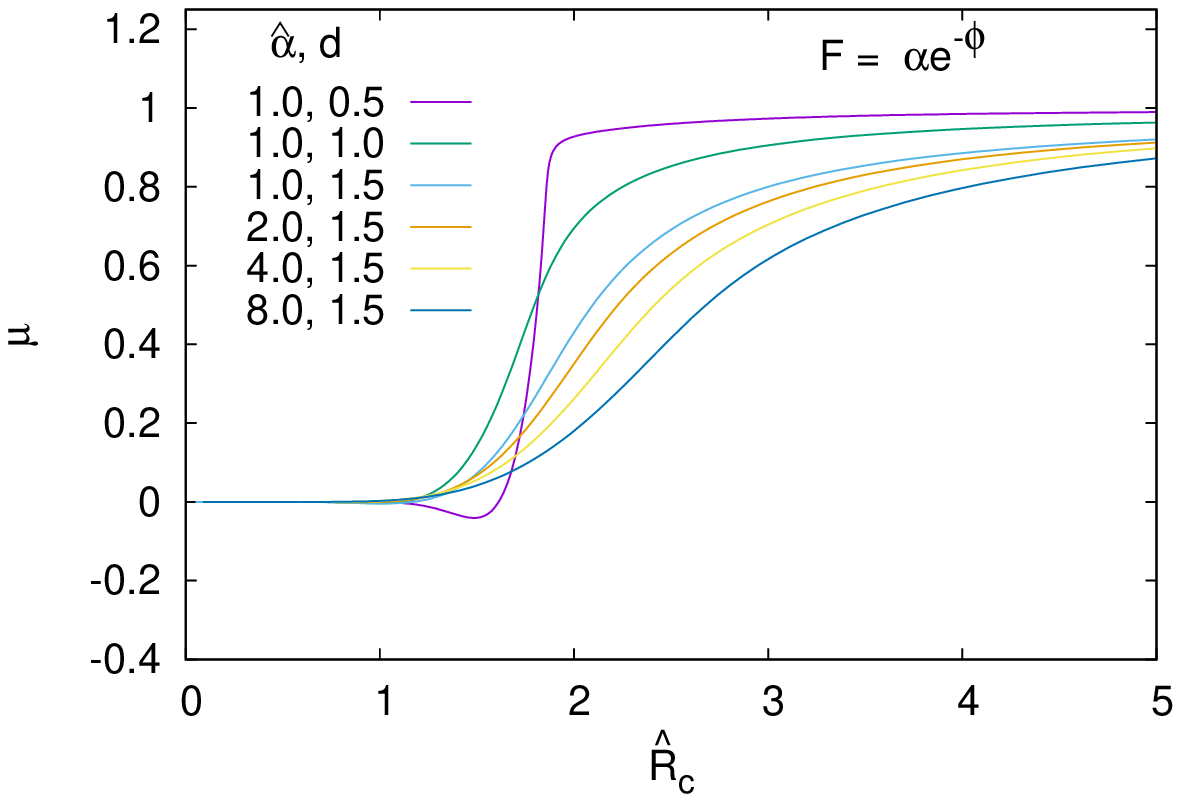}
\end{center}
\caption{
The mass function $\mu(\hat R_c)$ vs the scaled circumferential 
radius $\hat R_c$ for the coupling functions 
$F=\alpha\phi^2$ with $\phi_\infty=0$ (a) and
$F=\alpha e^{-\phi}$ (b), and several values of 
$\hat{\alpha}$ and $d$. }
\label{fig_stress2}
\end{figure}

Let us now consider the effective stress-energy tensor of the particle-like solutions.
We exhibit the components $T^t_t$, $T^r_r$, and $T^\theta_\theta$
in Figs.~\ref{fig_stress}(a)-(d) again for the coupling functions 
$F=\alpha\phi^2$ with $\phi_\infty=0$  and 
$F=\alpha e^{-\phi}$ for several values of $\hat{\alpha}$ and $d$.
The figures show an interesting behavior of these components.
Many particle-like solutions feature huge peaks in their
energy density $\rho=- T^t_t$ and pressures $T^r_r$, and $T^\theta_\theta$.
The energy density and the pressures are then strongly localized
in a shell with a scaled circumferential radius $R_c$ 
on the order of one. These particle-like solutions
seem like (almost) empty bags, with the contributions
from the scalar field and the GB term (almost) cancelling each other.

Figure \ref{fig_stress} also shows the quantity
${\cal Q} =(- T^t_t+T^r_r)/N$ with 
$N=\sqrt{( T^t_t)^2+(T^r_r)^2+2(T^\theta_\theta)^2}$.
If ${\cal Q} <0$ the null energy condition (NEC) is violated.
We recall that the NEC is given by $ T_{\mu\nu} n^\mu n^\nu \geq 0$, where
$n^\mu$ is any null vector satisfying the condition $n^\mu n_\mu=0$. 
Employing the null vector $n^\mu=\left(1,\sqrt{-g_{tt}/g_{rr}},0,0\right)$
yields the condition $- T^t_t+T^r_r <0$ for violation of the NEC to occur,
while choosing $n^\mu=\left(1,0,\sqrt{-g_{tt}/g_{\theta \theta}},0\right)$
leads to the condition $- T^t_t + T_\theta^\theta \geq 0 $.
Inspection of Figs.~\ref{fig_stress}(g) and (h) shows that the presence of
the GB term leads to violations of the NEC for the particle-like solutions,
just as it does for wormholes \cite{Antoniou:2019awm}.
 The NEC is violated at small values of the scaled
radial coordinate $\hat r$ while it is restored as we move toward the asymptotic
infinity.

We will now relate this very characteristic shell-like behavior 
of the particle-like solutions to their compactness. 
To that end, we define the mass function $\mu(\hat R_{\rm c})$,
which is obtained by integrating the energy density $\rho=- T^t_t$
over a sphere with radius equal to
the circumferential radius coordinate $\hat R_{\rm c}$,
\begin{equation}
\mu(\hat R_{\rm c}) = - 1/2 \int_0^{\hat R_{\rm c}}  T^t_t
\hat R_{\rm c}'^2 d \hat R_{\rm c}' \ .
\label{massfun}
\end{equation}
For these spherically symmetric solutions, the limit 
$\hat R_{\rm c} \to \infty$ then yields precisely the mass of the solutions.
The mass function $\mu(\hat R_{\rm c})$ is illustrated in Fig.~\ref{fig_stress2}
for the coupling functions 
$F=\alpha\phi^2$ with $\phi_\infty=0$ (a) and
$F=\alpha e^{-\phi}$ (b), and for several values of 
$\hat{\alpha}$ and $d$.

Figure \ref{fig_stress2} shows that the mass function $\mu(\hat R_{\rm c})$
exhibits a characteristic steep rise towards
its asymptotic value in the vicinity of $\hat R_{\rm c}=2$
for many of the particle-like solutions.
Here such a rise is, in fact, expected from the shell-like behaviour of the
energy density, as seen in Fig.~\ref{fig_stress}.
The mass of the solutions then resides to a large extent
in a very small region, 
qualifying these particle-like solutions as
highly compact objects. 
Let us recall that for Schwarzschild black holes
$\hat R_{\rm c}=2$ corresponds precisely to the scaled horizon radius.
This characteristic behavior holds for both the quadratic 
and the dilatonic coupling functions.
We will next show that these highly compact solutions will then also
feature lightrings.

\subsection{Lightrings}

\begin{figure}
\begin{center}
(a)\includegraphics[width=.45\textwidth, angle =0]{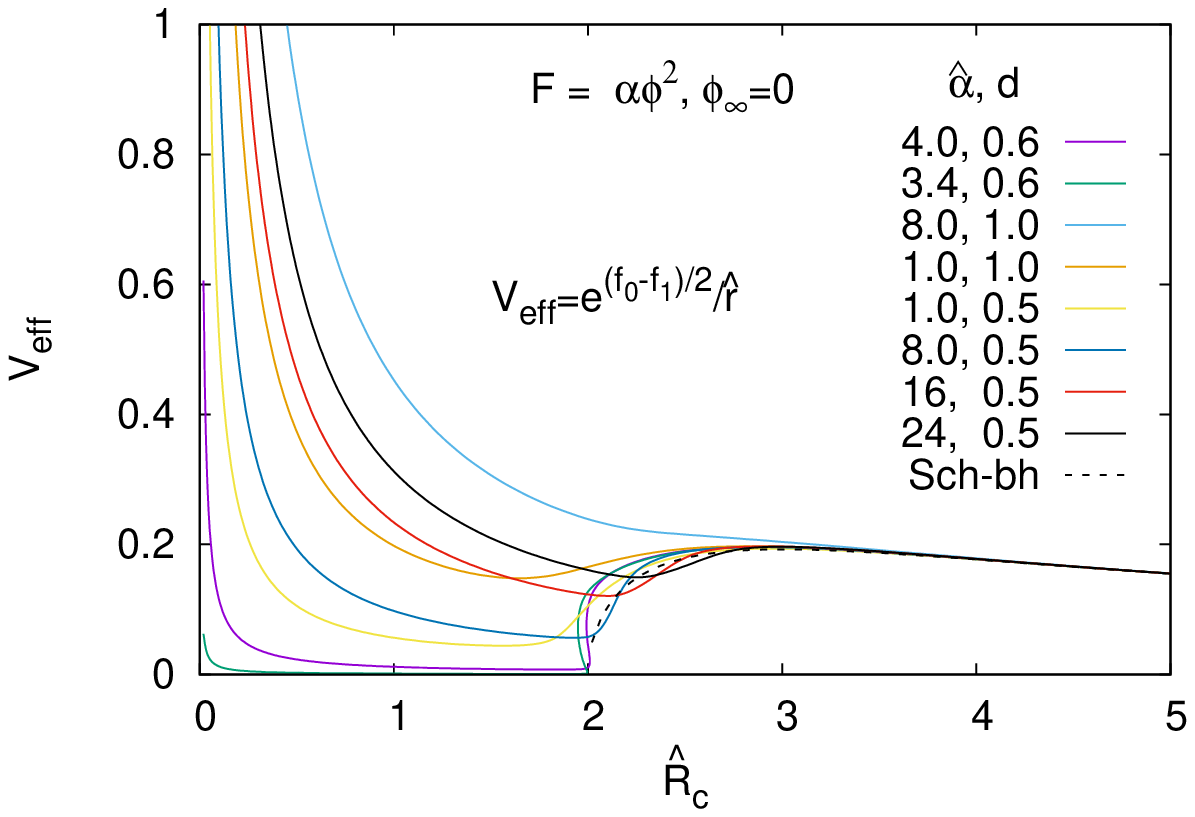}
(b)\includegraphics[width=.45\textwidth, angle =0]{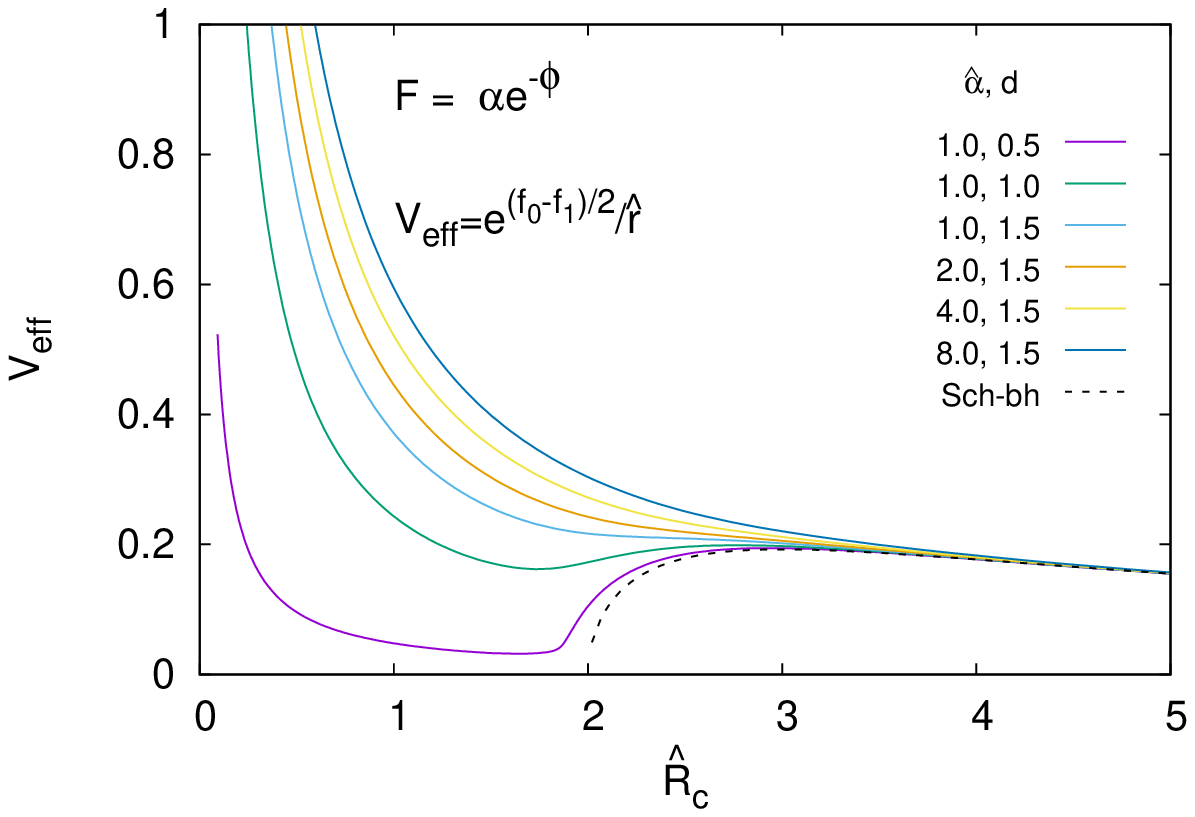}
\\
(c)\includegraphics[width=.45\textwidth, angle =0]{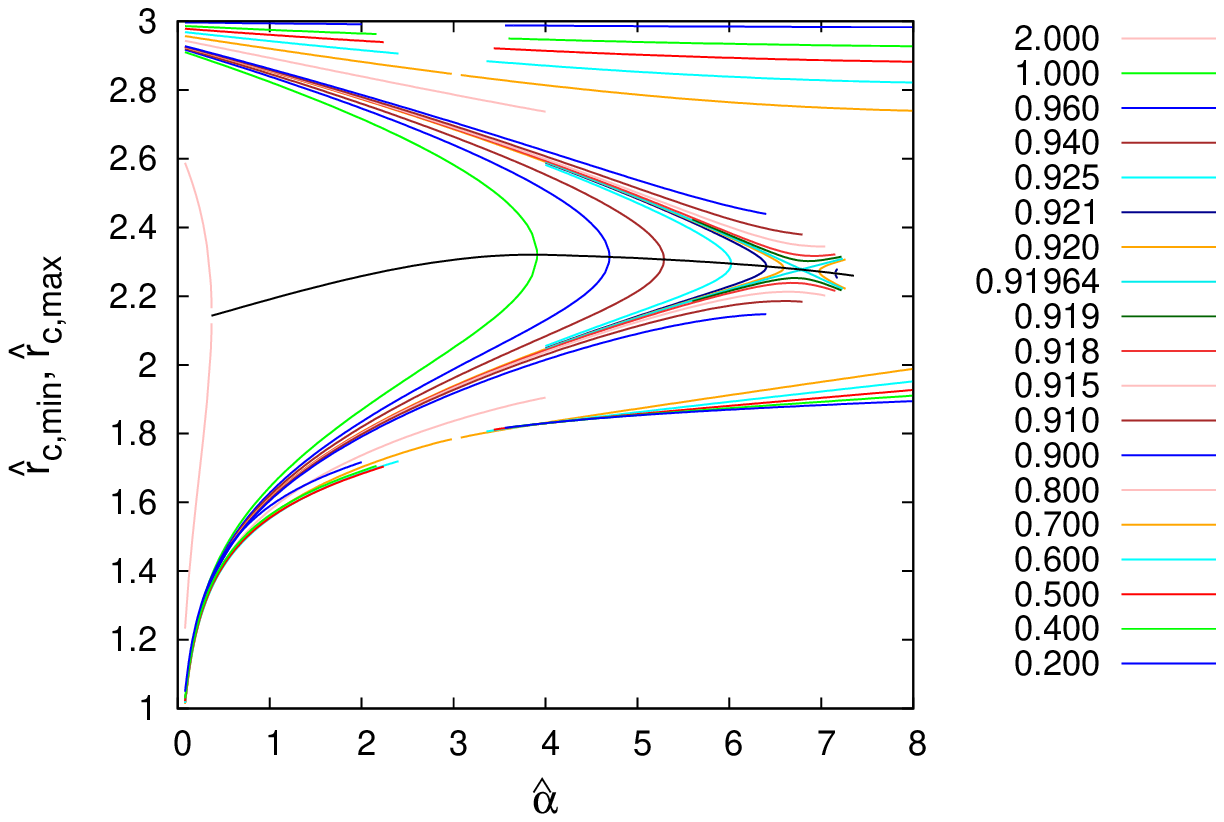}
(d)\includegraphics[width=.45\textwidth, angle =0]{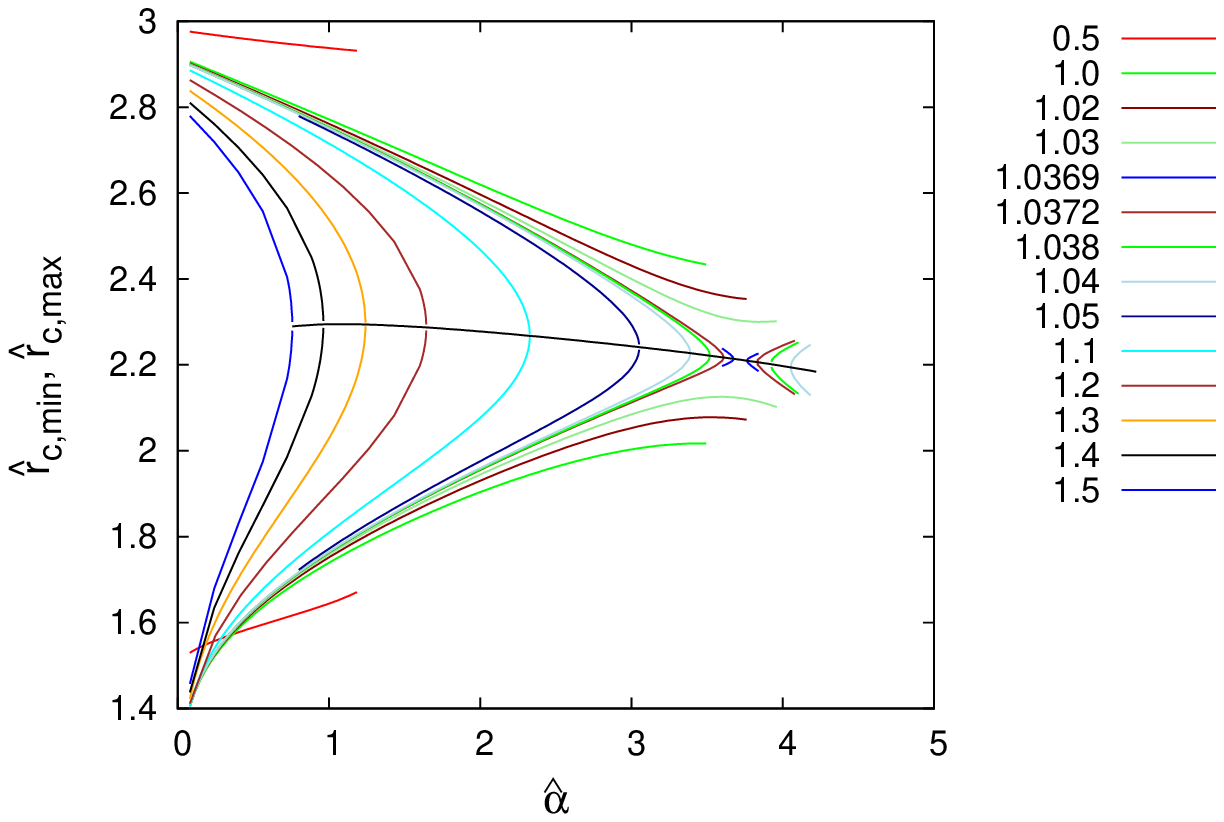}
\\
(e)\includegraphics[width=.45\textwidth, angle =0]{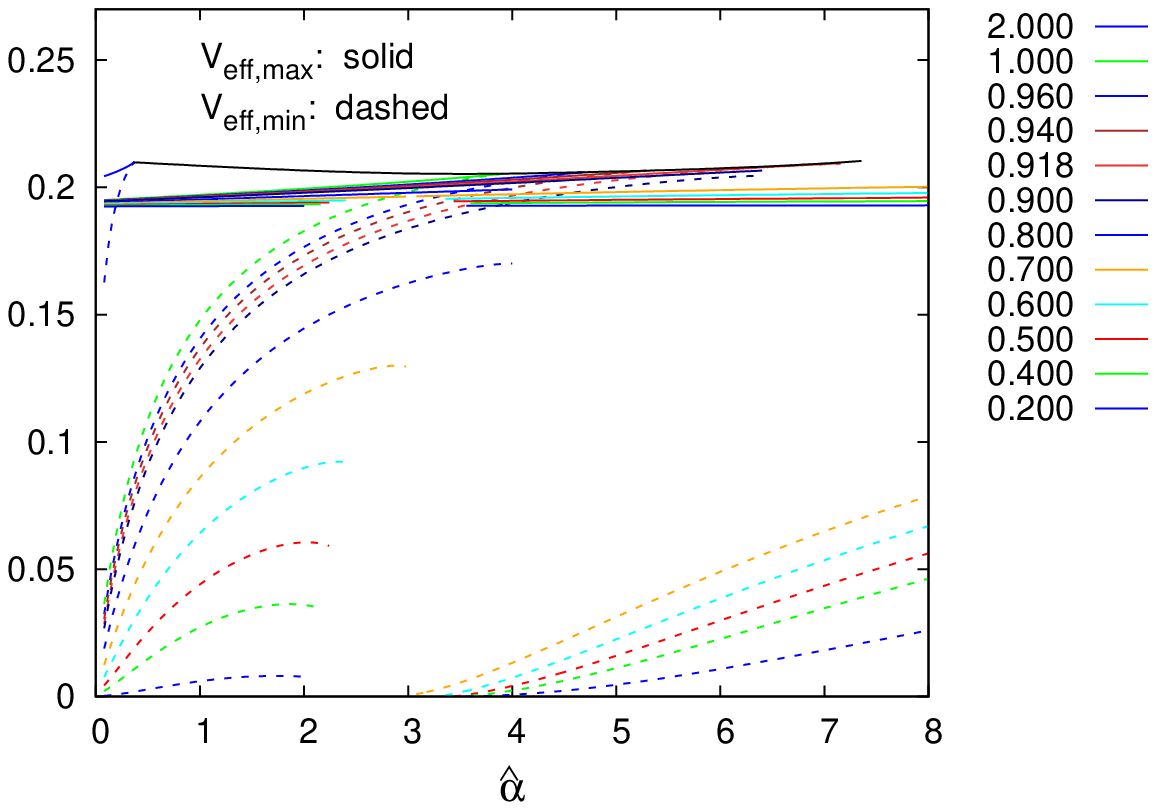}
(f)\includegraphics[width=.45\textwidth, angle =0]{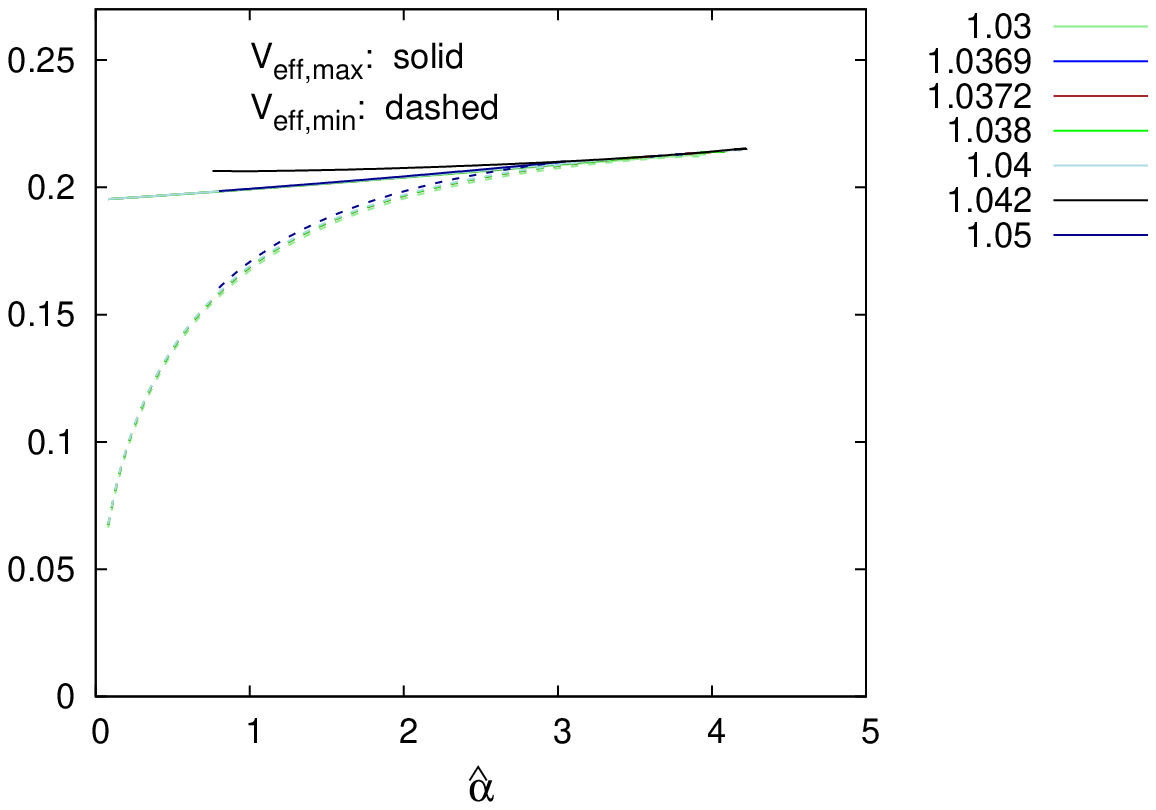}
\\
(g)\includegraphics[width=.45\textwidth, angle =0]{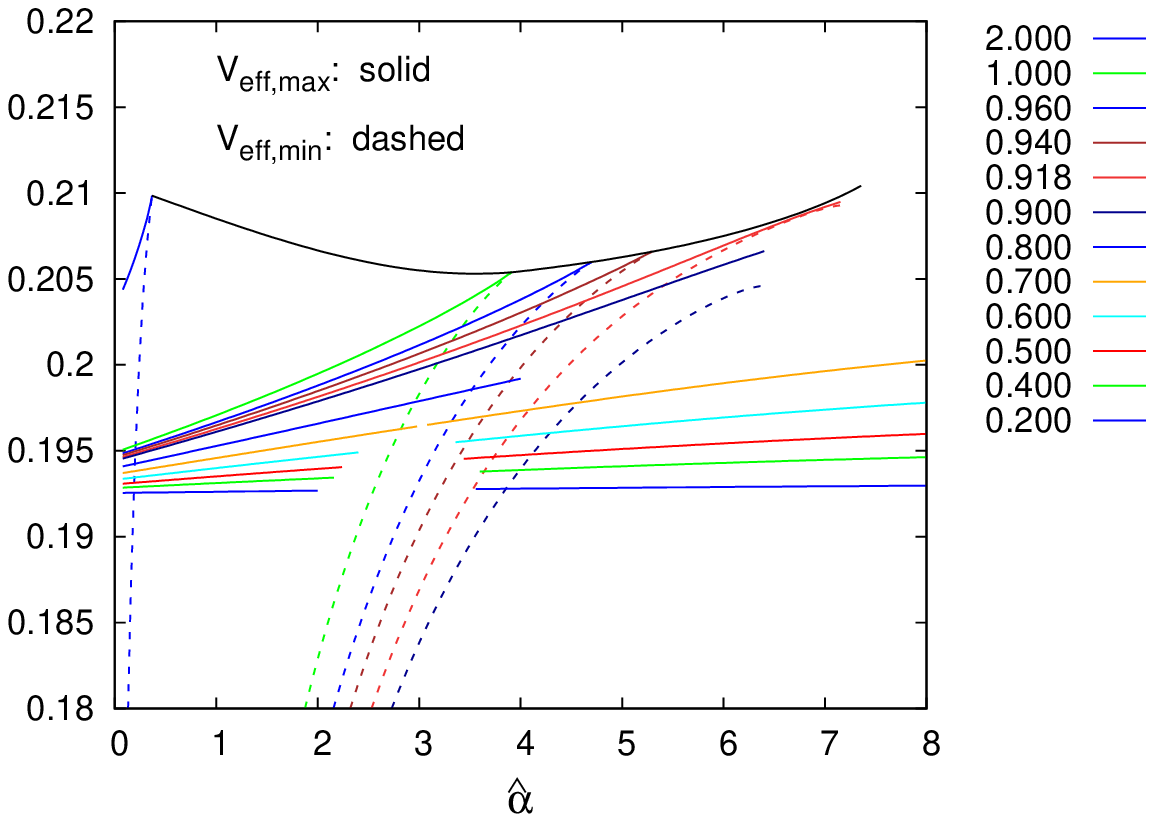}
(h)\includegraphics[width=.45\textwidth, angle =0]{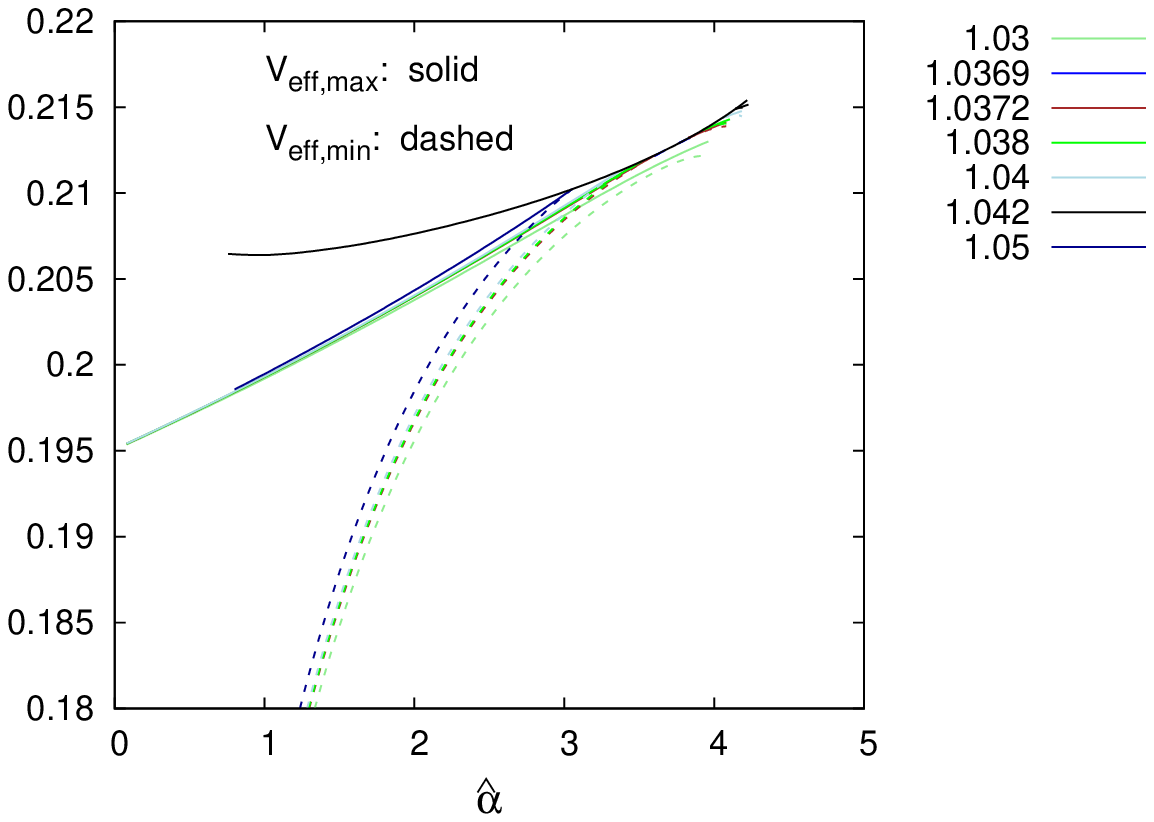}
\end{center}
\caption{
(a)-(b) The photon effective potential $V_{\rm eff}$ vs
the scaled circumferential coordinate $\hat{R}_c$,
for several values of $\hat{\alpha}$ and  $d$;
(c)-(d) the locations of the local maxima $\hat{R}_{\rm c,max}$ and 
minima $\hat{R}_{\rm c,min}$ vs $\hat{\alpha}$ ,for 
several values of $d$;
(e)-(f) the local maxima (solid lines) and minima (dashed
lines) of the effective potential
vs $\hat{\alpha}$, for several values of $d$;
(g)-(h) a zoom of (e)-(f). 
The coupling functions are
$F=\alpha\phi^2$ with $\phi_\infty=0$ [left column: (a), (c), (e), (g)] and 
$F=\alpha e^{-\phi}$ [right column: (b), (d), (f), (h)].}
\label{fig_light}
\end{figure}

In order to obtain the lightrings formed around our
particle-like solutions, we need to consider
the geodesics of light in these static, spherically-symmetric spacetimes.
We may obtain the geodesics for both null and timelike
particles from their Lagrangian ${\cal L}$ 
given by the expression
\begin{equation}
2 {\cal L}=g_{\mu\nu} \dot x^\mu \dot x^\nu 
=  -e^{f_0} \dot t^2 +e^{f_1}\left[\dot r^2 
+r^2\left( \dot \theta^2+\sin^2\theta \dot \varphi^2\right) \right]  = - \epsilon \ ,
\label{lag}
\end{equation}
where $\epsilon=0$ and 1 for massless and massive particles, respectively.
Here, we assume that these test particles possess no direct coupling to
the scalar field.
The independence of ${\cal L}$ of the coordinates $t$ and $\varphi$
leads to two conserved quantities, namely the energy $E$ and
the angular-momentum $L$ of the particle 
\begin{eqnarray}
E &=&-e^{f_0} \dot t \ , \\
L &=&e^{f_1} r^2 \dot \varphi \ .
\label{EL}
\end{eqnarray}
Inserting these expressions into the Lagrangian
and considering motion in the equatorial plane ($\theta=\pi/2$),
we obtain for the radial coordinate the equation
\begin{equation}
e^{f_0+f_1} \dot r^2=E^2 -\epsilon\,e^{f_0} - \frac {e^{f_0-f_1} L^2}{r^2} \ .
\label{geod}
\end{equation}

Let us first consider the radial equation for photons ($\epsilon=0$). In this case, we find
\begin{equation}
e^{f_0+f_1} \dot r^2=(E+L V_{\rm eff})(E-L V_{\rm eff}),
\label{Veff_photons}
\end{equation}
where $V_{\rm eff}$ is the effective potential for photons
\begin{equation}
V_{\rm eff}=e^{(f_0-f_1)/2}/r \ .
\label{Veff_photons2}
\end{equation}
The effective potential $V_{\rm eff}$ vs
the scaled circumferential coordinate $\hat{R}_c$ is shown in Fig.~\ref{fig_light},
for several values of $\hat{\alpha}$ and  $d$, and employing the coupling functions
$F=\alpha\phi^2$ with $\phi_\infty=0$ (a) and $F=\alpha e^{-\phi}$ (b).
For the less compact particle-like solutions, the effective potential is monotonic.
As the energy density gets more localized, the effective potential $V_{\rm eff}$ 
develops a saddle point, which splits into a pair of extrema
upon further change of the parameters, leading to a still stronger
localization of the energy density and an increasing compactness of the solutions.

Lightrings correspond to extrema of the effective potential $V_{\rm eff}$,
since they correspond to photon geodesics with fixed radial coordinate.
Our analysis shows, that the particle-like solutions either possess no lightring at all,
or they possess a pair of lightrings, consisting of a local maximum $\hat{R}_{\rm c,max}$ 
at larger $r$ and a local minimum $\hat{R}_{\rm c,min}$ at smaller $r$.
In Figs.~\ref{fig_light}(c) and (d), the locations of the local maxima $\hat{R}_{\rm c,max}$ 
and minima $\hat{R}_{\rm c,min}$ are shown versus $\hat{\alpha}$, for several values of $d$
and the coupling functions
$F=\alpha\phi^2$ with $\phi_\infty=0$ (c) and $F=\alpha e^{-\phi}$ (d).
Also indicated are the saddle points, where $\hat{R}_{\rm c,max}$
and $\hat{R}_{\rm c,min}$ emerge.
In the figures showing the domains of existence, Figs.~\ref{fig_doe2} and Figs.~\ref{fig_doe3}, 
we have marked, by using dashed curves, where particle-like solutions with
lightrings emerge.

The local maxima and minima of the effective potential $V_{\rm eff}$
are exhibited in Figs.~\ref{fig_light}(e) and (f) (with Figs.~\ref{fig_light}(g) and (h)
representing amplifications).
Clearly, photons with energies smaller than the value of $L\,V_{\rm eff}$ at its
local maximum, that reside inside the potential well,
remain in bound orbits in the spacetimes of the particle-like solutions. 
Therefore, the lightrings at the minima of the potentials represent stable photon orbits.
In contrast, the maxima represent unstable bound photon orbits.
We emphasize that
the presence of lightrings qualifies the respective particle-like solutions
as UCOs \cite{Cardoso:2017cqb}.

The fact that lightrings can only be present in pairs in the spacetime of our
particle-like solutions is in accordance with a theorem on lightrings of UCOs
\cite{Cunha:2017qtt}. There, it is also shown
that, in a smooth spherically-symmetric metric,
one of the lightrings should be stable,
which also agrees with our findings for the particle-like solutions. % \cite{Cunha:2017qtt}.
Let us finally mention that it has recently been pointed out 
that the presence of a stable lightring 
might lead to nonlinear spacetime instabilities
\cite{Keir:2014oka,Cardoso:2014sna}.
So far the arguments involved are in a preliminary stage,
and will need further investigation.
%Clearly, if such instabilities would indeed be present,
%astrophysical observations of such UCOs would require 
%their lifetime to be large on relevant astrophysical timescales.
%{\color{blue} Further investigation along these lines is,
%however, beyond the scope of the present paper.}

\subsection{Particle motion}

\begin{figure}
\begin{center}
(a)\includegraphics[width=.45\textwidth, angle =0]{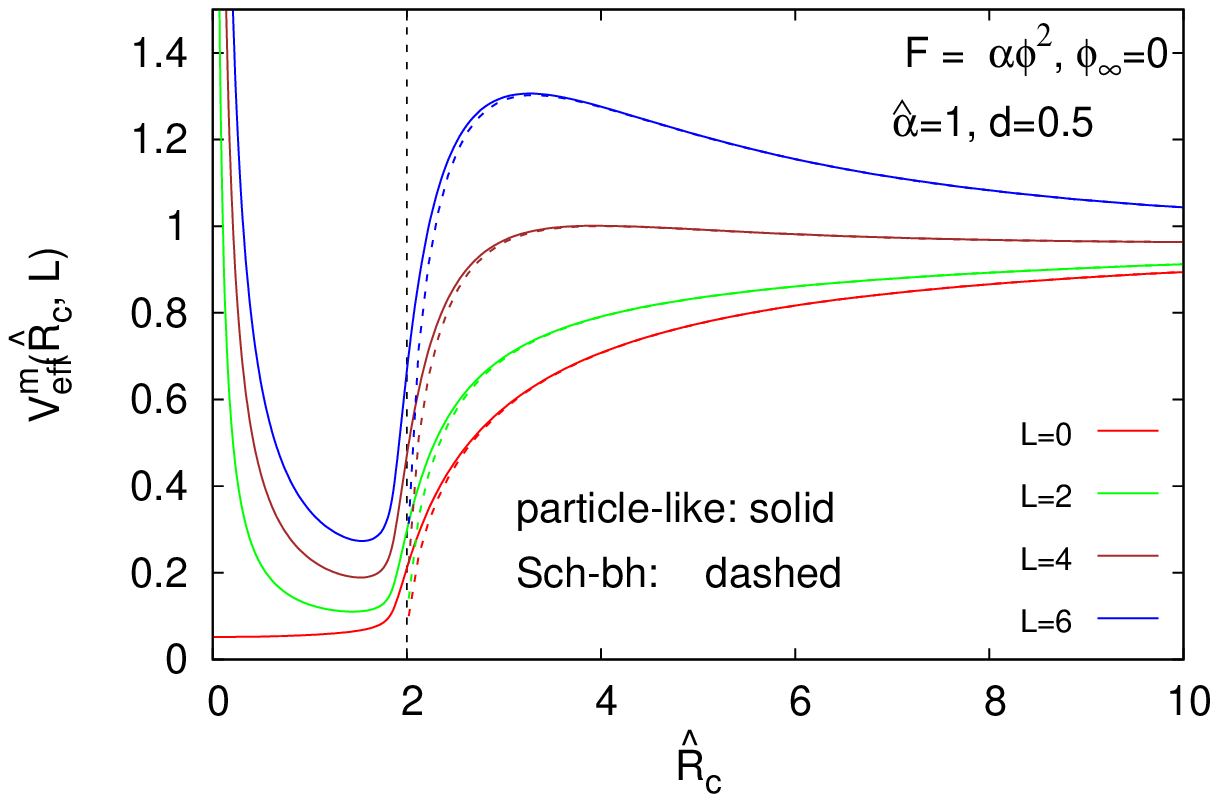}
(b)\includegraphics[width=.45\textwidth, angle =0]{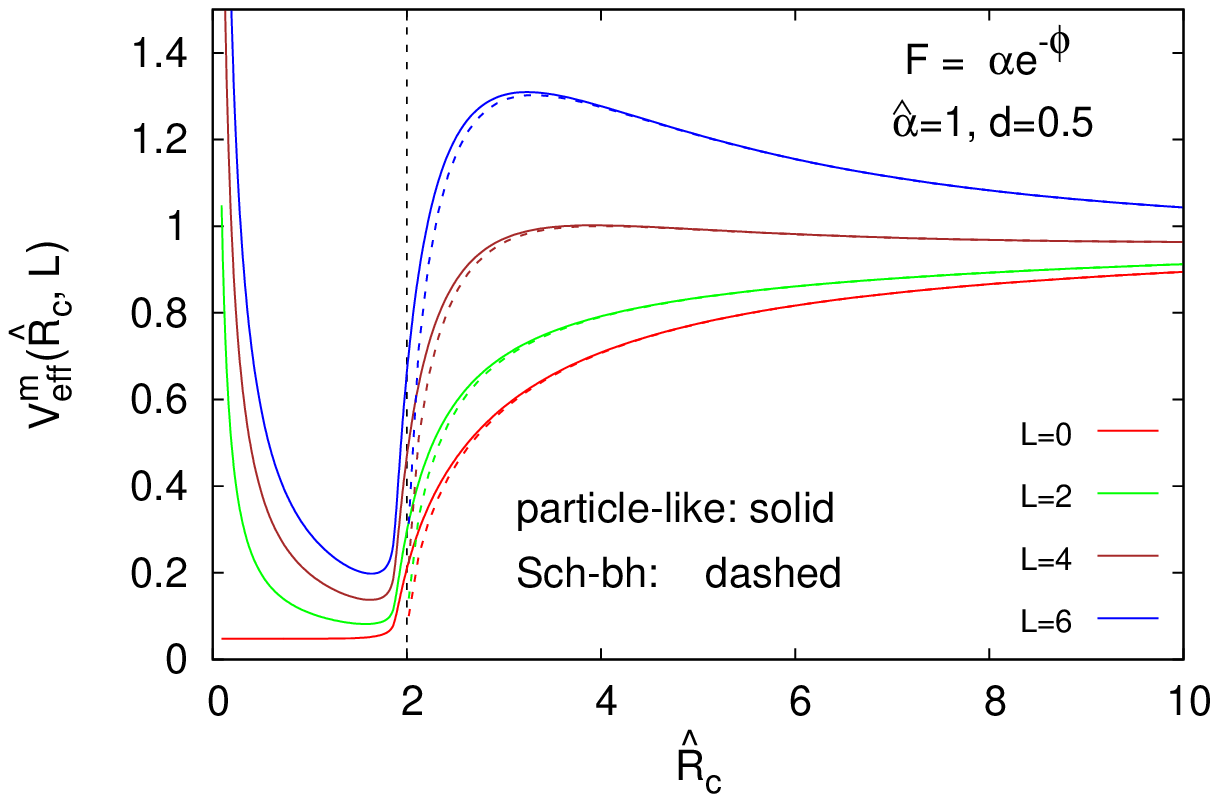}
\\
(c)\includegraphics[width=.45\textwidth, angle =0]{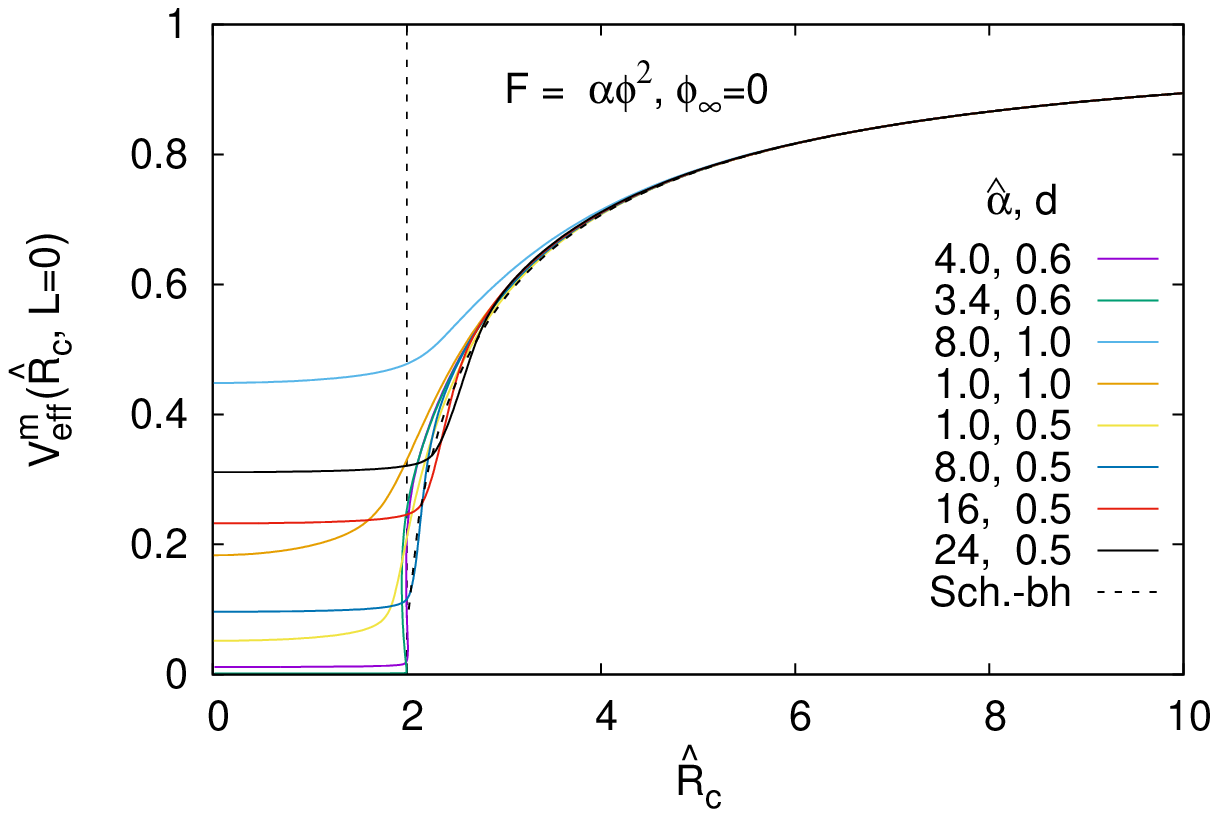}
(d)\includegraphics[width=.45\textwidth, angle =0]{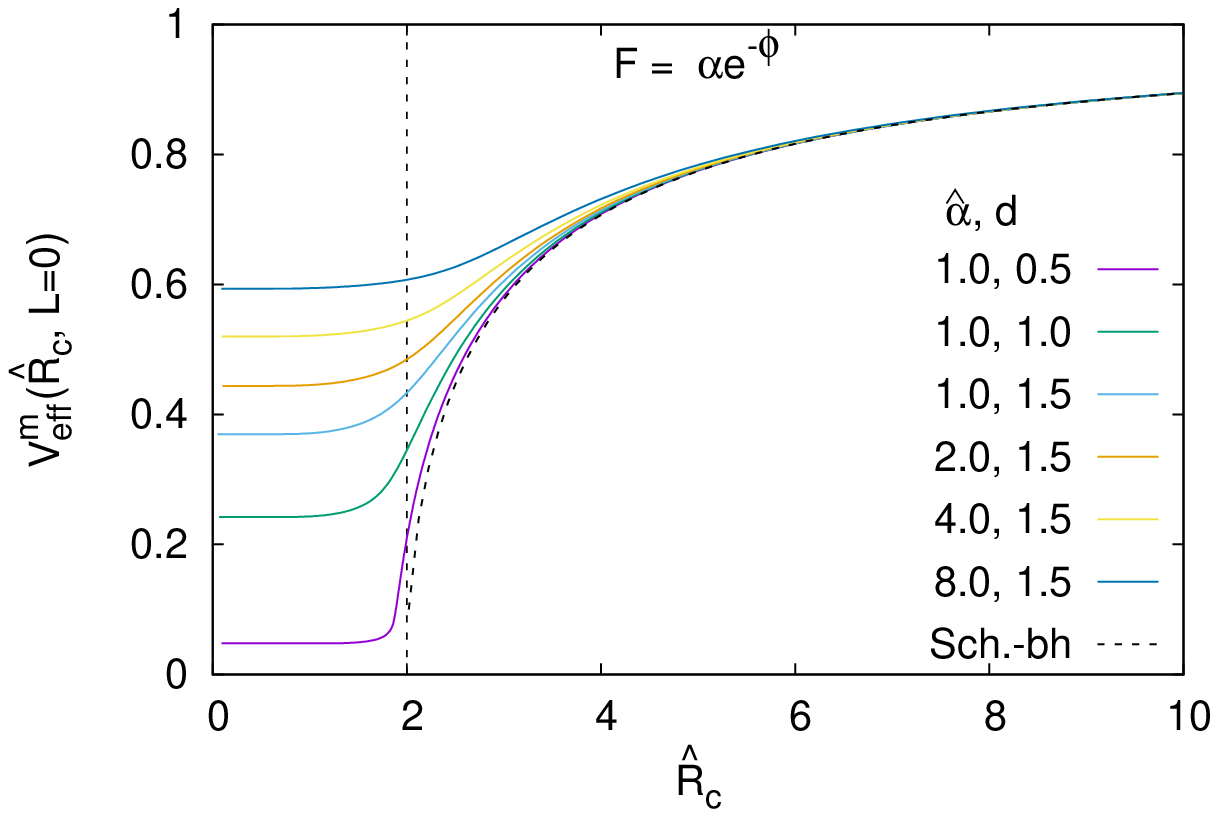}
\\
(e)\includegraphics[width=.45\textwidth, angle =0]{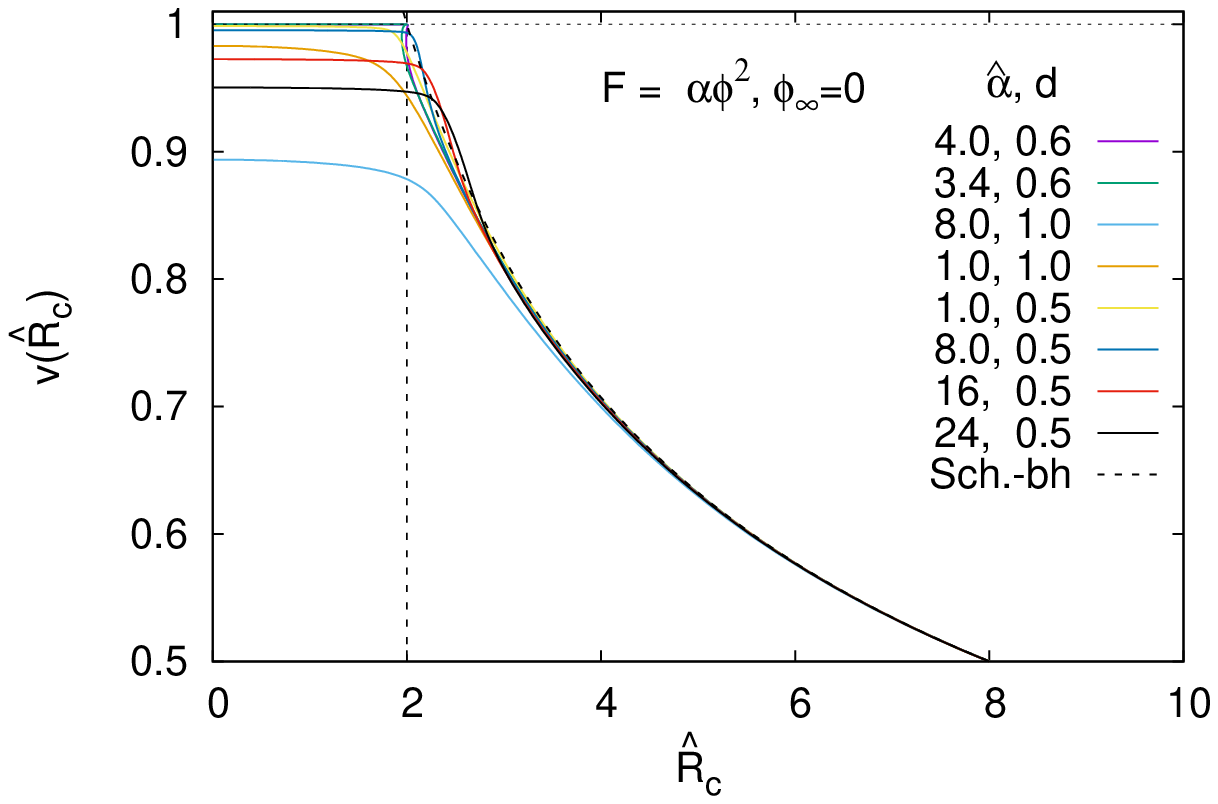}
(f)\includegraphics[width=.45\textwidth, angle =0]{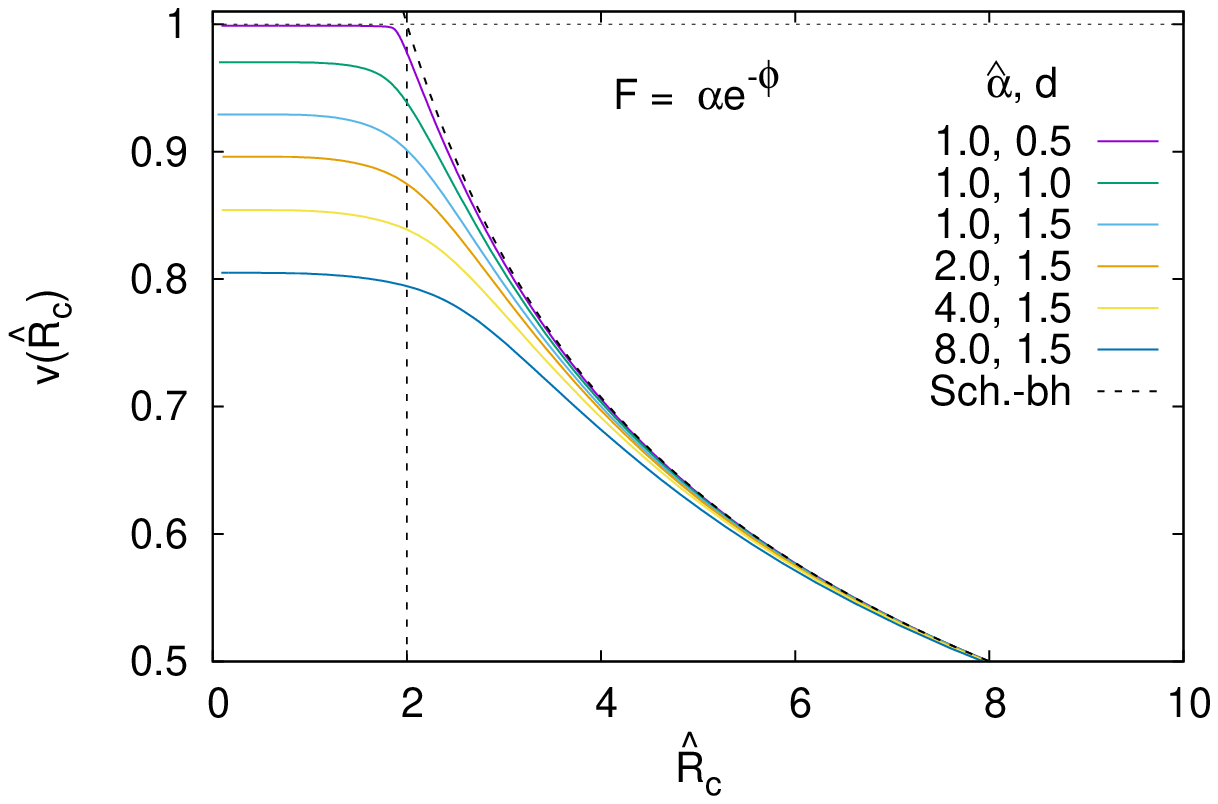}
\\
(g)\includegraphics[width=.45\textwidth, angle =0]{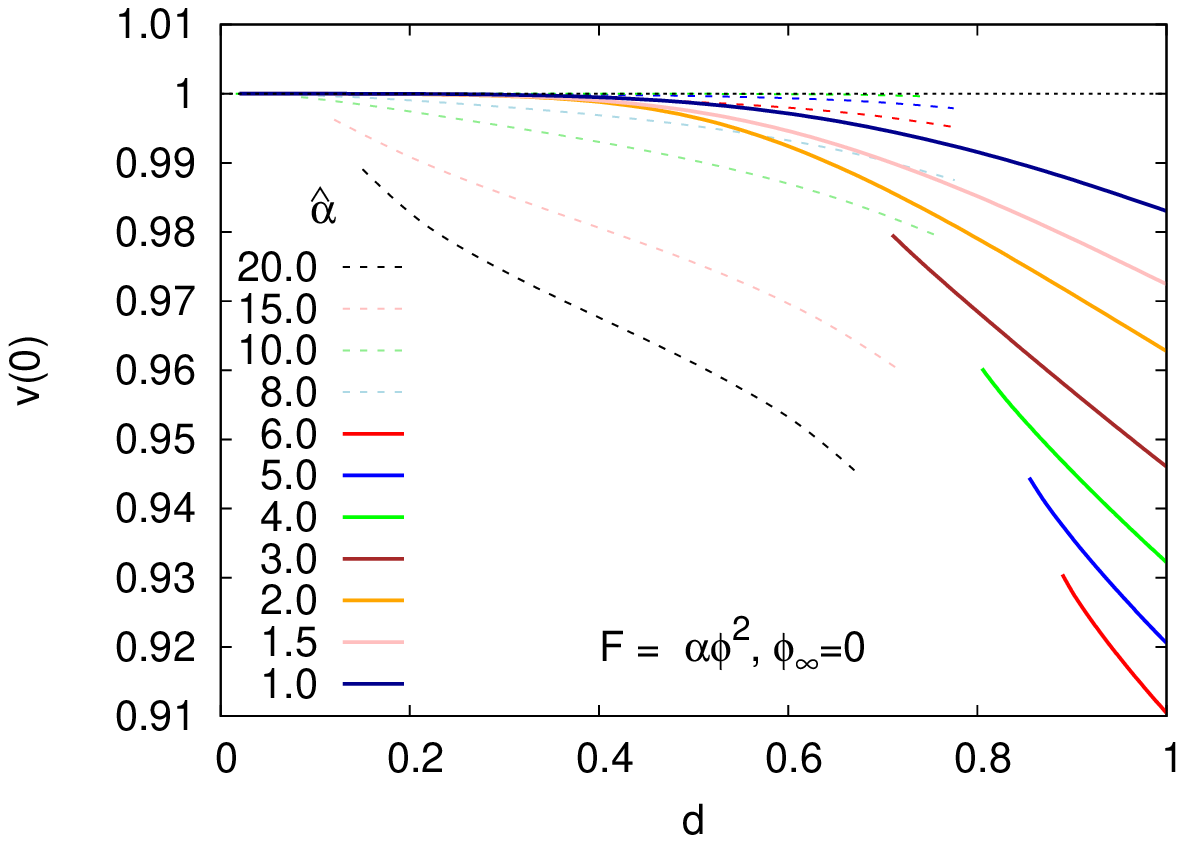}
(h)\includegraphics[width=.45\textwidth, angle =0]{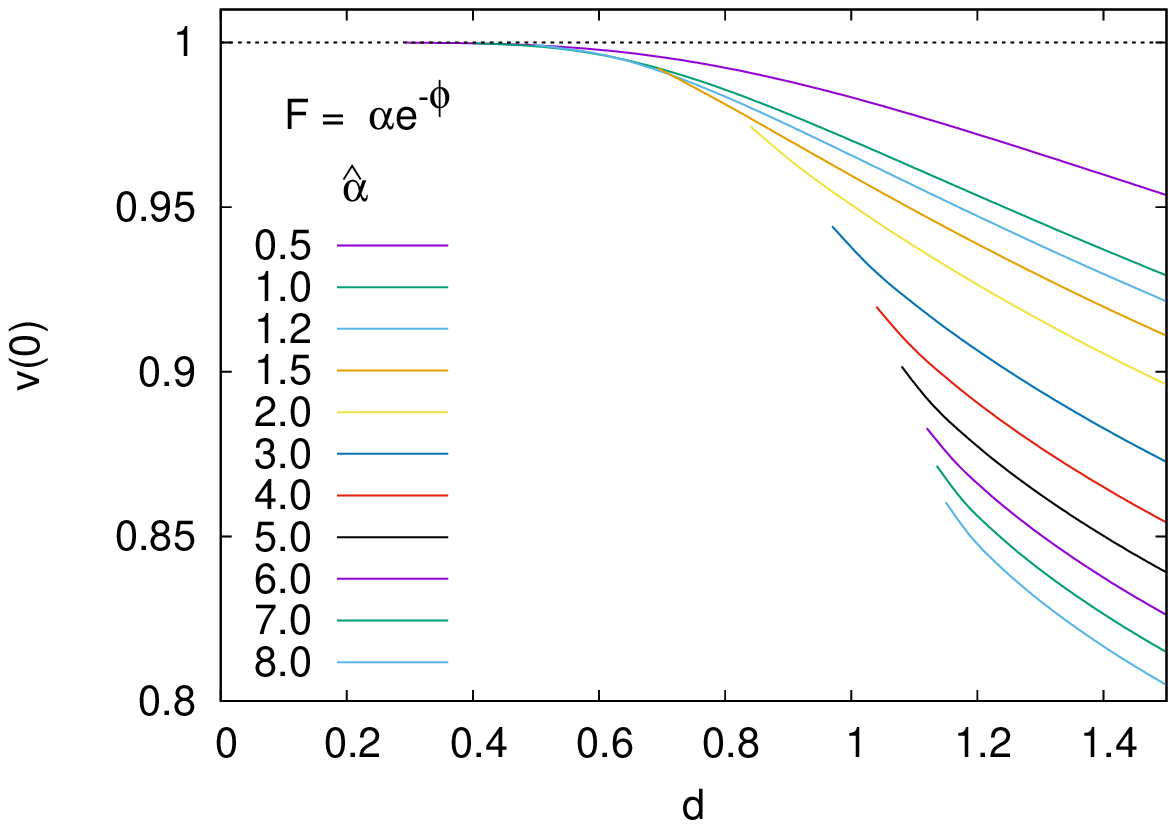}
\end{center}
\caption{
%(a)-(b) The effective potential for massive test particles is shown
%as function of the circumferential radius for several values of the
%angular momentum $L$ in the spacetime of a particle like solution 
%with $\hat{\alpha}=1$ and scaled scalar charge $d=1$. 
%The dashed curves correspond to the Schwarzschild spacetime with the
%same mass ($M=1$). The vertical line corresponds to the event horizon.
%(b) The same as (a) for radial motion ($L=0$) in several spacetimes.
%(c) The speed of radially infalling test particles ($L=0$) is shown
%as function of the circumferential radius for several spacetimes.
%(d) The maximal speed $v_{\rm max}=v(0)$ of radially infalling test particles ($L=0$) 
%is shown as function of the scaled scalar charge for several values of $\hat{\alpha}$.
%
(a)-(b) The effective potential $V_{\rm eff}^m$ for massive particles vs
the scaled circumferential coordinate $\hat{R}_c$,
%for several values of $\hat{\alpha}$ and  $d$
for several values of the angular momentum $L$
and for $\hat{\alpha}=1$, $d=1$
(Schwarzschild: dashed lines, event horizon: vertical line);
(c)-(d) $V_{\rm eff}^m$ for radial motion ($L=0$) in several spacetimes;
(e)-(f) the speed $v$ of radially infalling test particles ($L=0$);
(g)-(h) the maximal speed $v_{\rm max}=v(0)$ of radially infalling test particles ($L=0$).
The coupling functions are
$F=\alpha\phi^2$ with $\phi_\infty=0$ [left column: (a), (c), (e), (g)] and 
$F=\alpha e^{-\phi}$ [right column: (b), (d), (f), (h)].
}
\label{fig_massive}
\end{figure}

For massive test particles, we introduce the effective potential $V_{\rm eff}^m$
by starting again from the radial equation (\ref{geod}) and setting $\epsilon=1$.
In that case, we obtain
\begin{equation}
e^{f_0+f_1} \dot r^2=(E+ V_{\rm eff}^m)(E- V_{\rm eff}^m) \ .
\label{Veff_massive0}
\end{equation}
Then, $V_{\rm eff}^m$ is given by
\begin{equation}
V_{\rm eff}^m= \sqrt{ e^{f_0} + \frac{e^{f_0-f_1} L^2}{r^2} } \ .
\label{Veff_massive}
\end{equation}
The effective potential $V_{\rm eff}^m$ is shown in Figs.~\ref{fig_massive}(a) and (b)
as a function of the scaled circumferential coordinate $\hat{R}_c$,
choosing the coupling functions
$F=\alpha\phi^2$ with $\phi_\infty=0$ (a) and  $F=\alpha e^{-\phi}$ (b)
and the parameters $\hat{\alpha}=1$ and $d=1$,
corresponding to highly compact particle-like solutions.

In these figures, the particle angular momentum $L$ is varied,
to illustrate the interesting features of the angular momentum barrier for these solutions.
The dashed curves correspond to particle motion in the spacetime of a Schwarzschild black hole with the
same mass ($M=1$). The vertical line corresponds to the event horizon.
We note that the effective potential $V_{\rm eff}^m$ follows very closely the 
effective potential in the Schwarzschild spacetime in most of the outer region.
Only close to the black hole horizon, which resides at the scaled circumferential radius
$\hat R_c=2$, the respective effective potentials for the massive test
particle in the black-hole and particle-like spacetimes start to deviate.
Inside $\hat R_c=2$, the effective potential $V_{\rm eff}^m$ of the particle-like solutions
then exhibits a completely different behavior. Instead of turning negative,
the effective potential $V_{\rm eff}^m$ remains always positive. 
In fact, it reaches a minimum,
and then diverges at the origin when $L$ is non-zero.
Thus, while the motion of particles beyond $\hat R_c=2$ will be very similar
to the motion in a black-hole spacetime, the absence of an event horizon
and the regularity of the spacetime at the center
allow for a very different type of motion in the inner region.
In particular, there are always bound orbits of massive particles,
although these may be confined to reside close to the center of the spacetime.

To see that the divergence of the scalar field at the origin does not affect the motion
in the spacetime, and that a particle may reach and pass over the origin without
encountering any influence of this divergence, we now set the angular momentum
of the particle to zero.
The effective potential $V_{\rm eff}^m$ for radial motion ($L=0$) 
in a variety of spacetimes is illustrated in Figs.~\ref{fig_massive}(c) and (d).
We see that the effective potential is always regular at the center, where it also assumes
its minimal value.
Thus, a particle could sit at rest right at the center.
Also bound oscillating motion across the center is possible,
as observed also in other regular spacetimes \cite{Teodoro:2020gps}.
In Figs.~\ref{fig_massive}(e) and (f), we illustrate,
for the same set of spacetimes,
the speed $v$ of radially infalling test particles ($L=0$),
which all possess a speed of half the velocity of light
($v=0.5$) at a circumferential radius $\hat R_c=8$.
Again, comparison with the Schwarzschild black hole is made in the figures.
Figures \ref{fig_massive}(g) and (h) finally show
the maximal speed $v_{\rm max}=v(0)$ of radially infalling test particles ($L=0$) 
as a function of the scaled scalar charge $d$, for several values of $\hat{\alpha}$.

\subsection{Echoes of UCOs}

\begin{figure}
\begin{center}
(a)\includegraphics[width=.45\textwidth, angle =0]{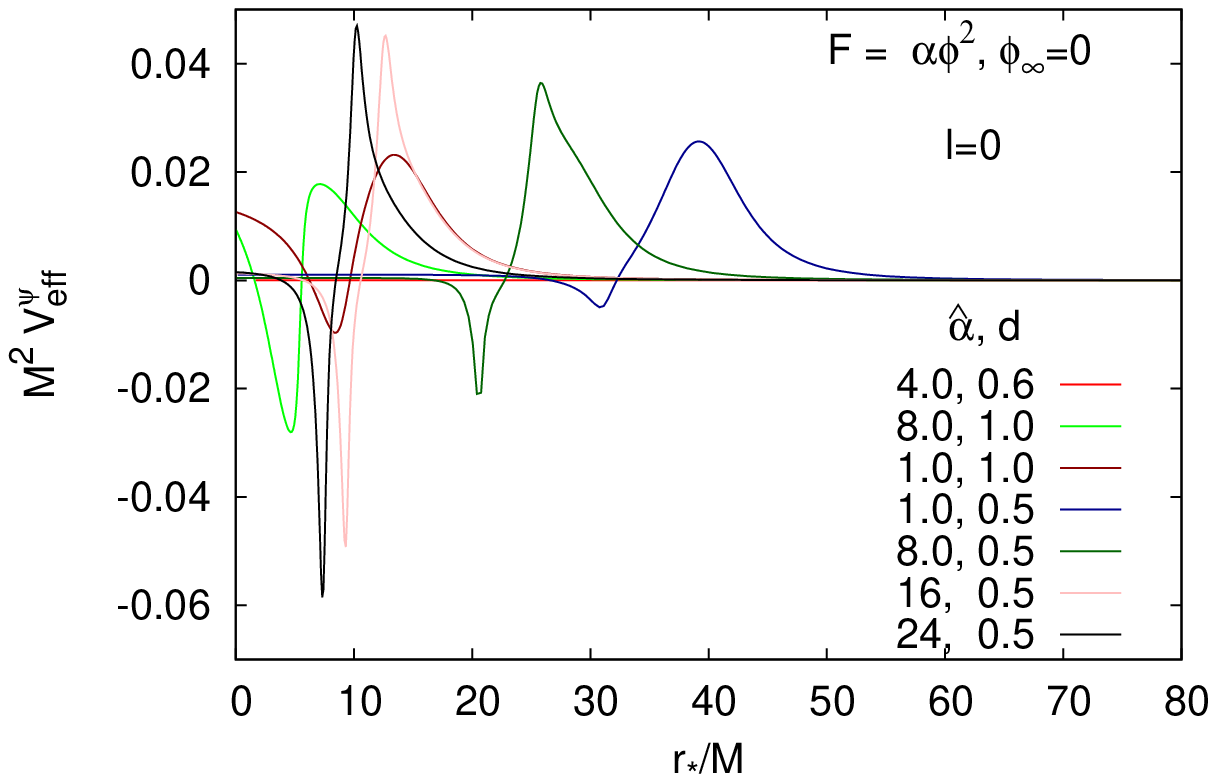}
(b)\includegraphics[width=.45\textwidth, angle =0]{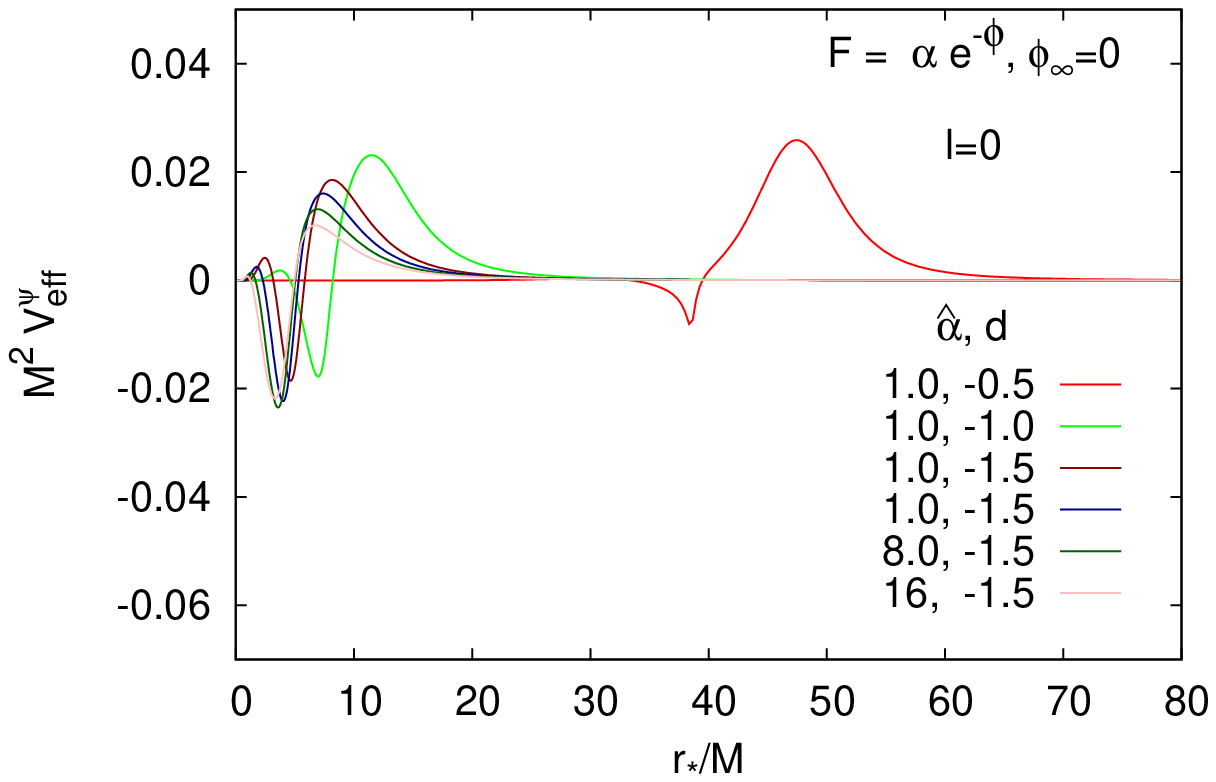}
\\
(c)\includegraphics[width=.45\textwidth, angle =0]{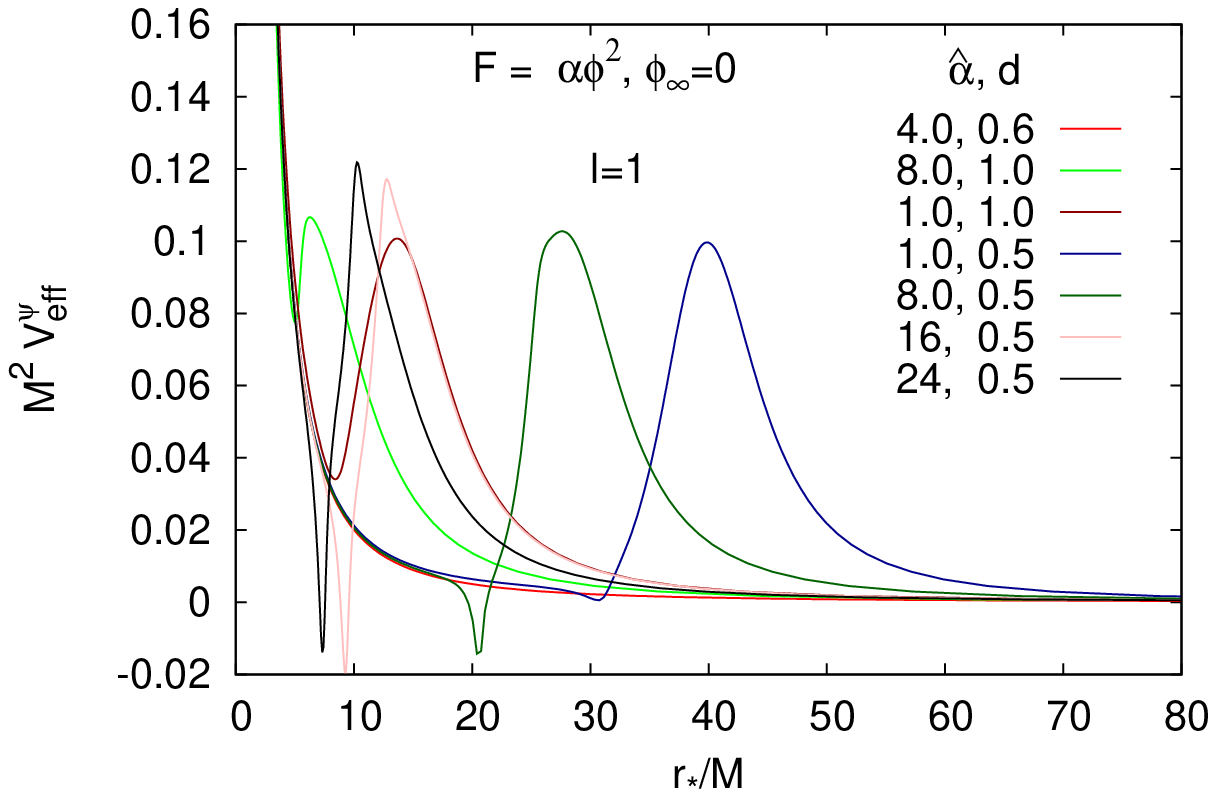}
(d)\includegraphics[width=.45\textwidth, angle =0]{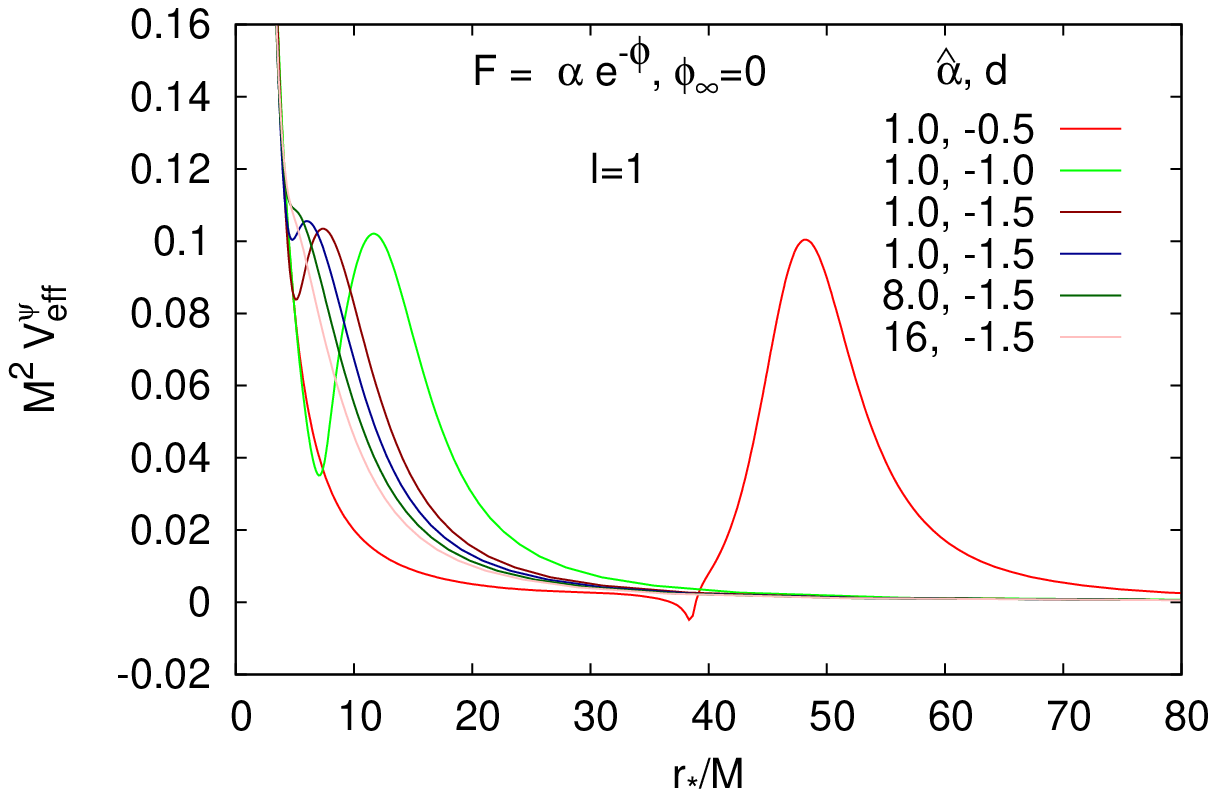}
\\
(e)\includegraphics[width=.45\textwidth, angle =0]{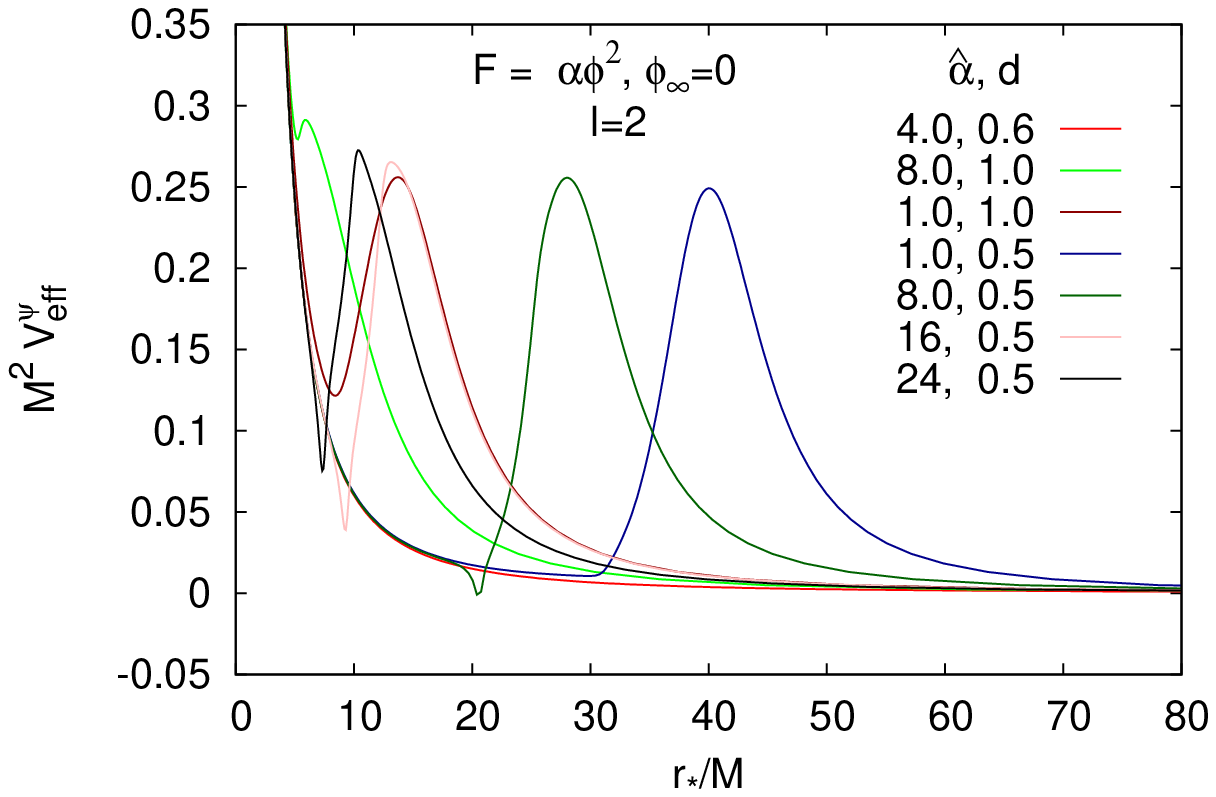}
(f)\includegraphics[width=.45\textwidth, angle =0]{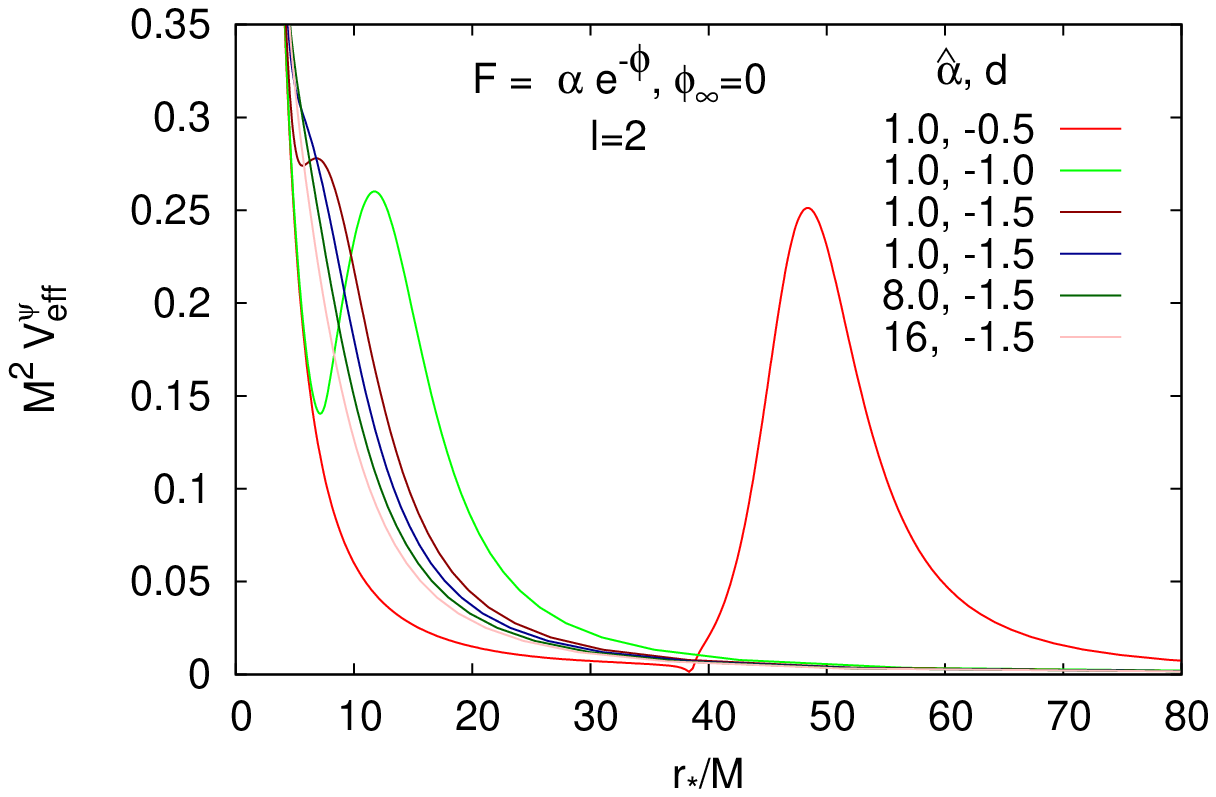}
\\
(g)\includegraphics[width=.45\textwidth, angle =0]{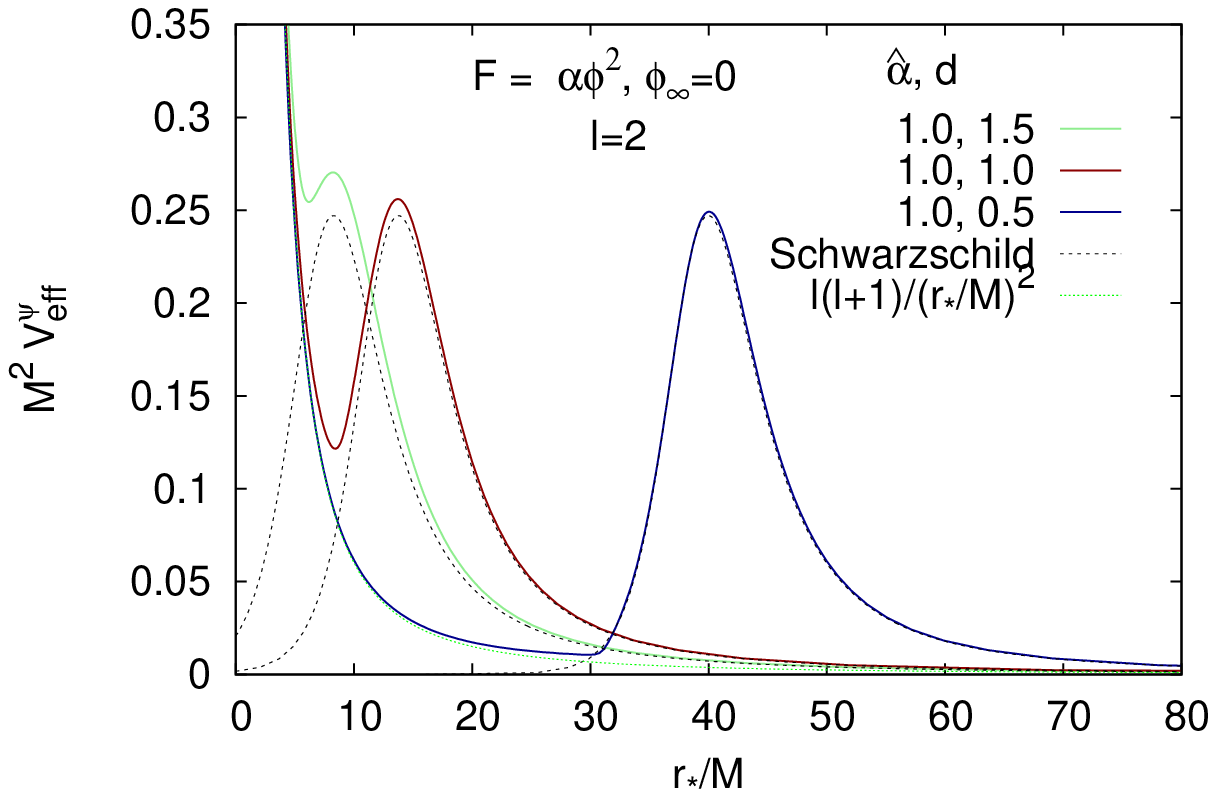}
(h)\includegraphics[width=.45\textwidth, angle =0]{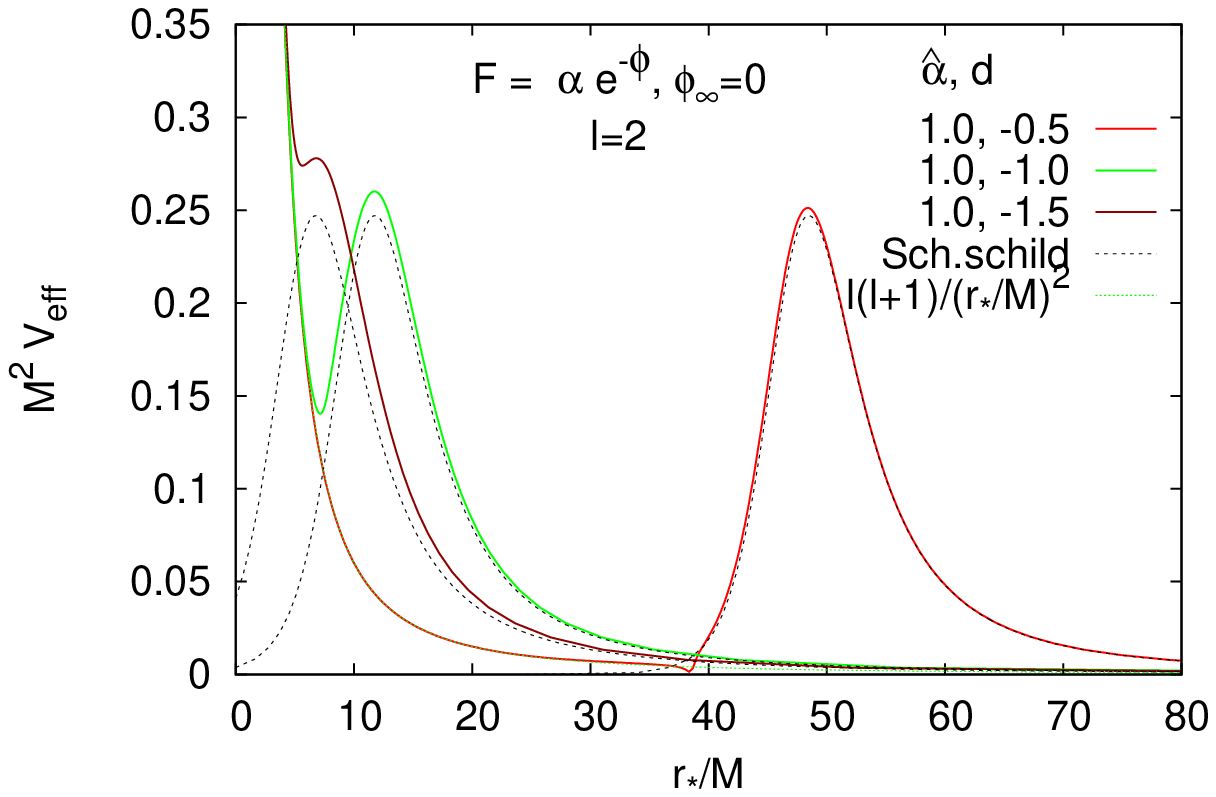}
\end{center}
\caption{
(a)-(h) The scaled effective potential $M^2 V^\psi_{\rm eff}$ 
vs the scaled tortoise coordinate $r_*/M$ for 
a test scalar particle with angular momentum numbers $l=0$ (a)-(b),
$l=1$ (c)-(d), and $l=2$ (e)-(h)
for a set of particle-like solutions
with coupling functions 
$F=\alpha\phi^2$ with $\phi_\infty=0$ [left column: (a), (c), (e), (g)] and 
$F=\alpha e^{-\phi}$ [right column: (b), (d), (f), (h)].
In (g) and (h), also the effective potential in the Schwarzschild black-hole
spacetime is shown (black dashed line).
}
\label{fig_echo}
\end{figure}

Lastly, we will address possible signatures of the ultra-compact
particle-like solutions in the framework of
gravitational-wave spectroscopy.
As pointed out in \cite{Cardoso:2016rao,Cardoso:2016oxy,Cardoso:2017cqb},
because of the absence of an event horizon,
UCOs might reveal themselves in the ringdown phase of a merger event,
since they will give rise to secondary pulses in the ringdown waveform,
i.e., to a sequence of echoes.

To demonstrate the occurrence of echoes for the ultra-compact particle-like solutions,
we consider a simple scattering process of a test scalar wave $\Psi$ by the 
gravitational potential of the UCO. In particular, we expand the scalar wave
in spherical harmonics $Y^m_l(\theta, \varphi)$, 
\begin{equation}
\Psi(t,r,\theta,\varphi)=\sum_{l,m} \psi_{l,m}(t,r)\,e^{-f_1/2} Y^m_l(\theta, \varphi)/r \ ,
\label{Psi}
\end{equation}
and insert this expansion into the free Klein-Gordon equation,
\begin{equation}
\partial_\mu (\sqrt{-g}\,\partial^\mu \Psi)=0 \ .
\end{equation}
%Separation of the angular coordinates t
This then results in the following equation for $\psi_{l,m}(t,r)$
\begin{equation}
(\partial_t^2- \partial^2_{r_*} +V^{\psi}_{\rm eff})\,\psi_{l,m}(t,r)=0 \ .
\end{equation}
Here, we have introduced the tortoise coordinate $r_*$ defined as
\begin{equation}
r_*= \int_0^r e^{(f_1-f_0)/2} dr \ ,
\end{equation}
where the integration constant is chosen such that $r_*=0$ corresponds to 
the origin. 
We have also defined the $l$-dependent effective potential $V^{\psi}_{\rm eff}$,
\begin{equation}
V_{\rm eff}^{\psi}=e^{f_0-f_1} \left[\frac{l(l+1)}{r^2}+ \frac{2 (f_1'+f_0') +
r f_1'f_0'+2r f_1 ''}{4r}\right] \ .
\end{equation}

Let us now inspect the profile of $V_{\rm eff}^{\psi}$.
We exhibit the scaled effective potential $M^2 V^\psi_{\rm eff}$ 
versus the scaled tortoise coordinate $r_*/M$ in Figs.~\ref{fig_echo}
for a set of particle-like solutions with coupling functions 
$F =\alpha\phi^2$ with $\phi_\infty=0$ [left column: (a), (c), (e), (g)] and 
$F=\alpha e^{-\phi}$ [right column: (b), (d), (f), (h)].
For the test scalar wave, we have chosen the angular momentum numbers 
$l=0$ [Figs. ~\ref{fig_echo}(a)-(b)], $l=1$ 
[Figs. ~\ref{fig_echo}(c)-(d)], and $l=2$ 
[Figs. ~\ref{fig_echo}(e)-(h)].
From Figs. ~\ref{fig_echo}(a)-(b), we observe that, for a scalar
wave with $l=0$, the effective potential $V^\psi_{\rm eff}$  acquires a finite value
near the origin and for both coupling functions. However, $V^\psi_{\rm eff}$ 
diverges at the origin $r=0$,
when $l>0$, as seen in Figs.~\ref{fig_echo}(c)-(h).
Thus, also for a test scalar wave with $l > 0$
an infinite angular momentum barrier resides at the origin.
In addition, there is the usual finite local barrier, located at a larger
value of $r_*$, that is also present for black holes.

We now consider an incoming test scalar-wave with modes with $l>0$.
These modes will then be partially transmitted through
the finite local barrier and partially reflected back to infinity. 
The transmitted modes will then be  fully reflected by 
the infinite angular momentum barrier. Upon reaching the
local barrier, the reflected modes will be partially transmitted through
the finite local barrier and partially reflected again.
Thus, we observe a perpetual process of full and partial reflection 
between the angular-momentum and the local barrier, respectively. 
As a result, we will see an infinite number of echoes, which possess an amplitude
that is decreasing every time that the
wave gets partially transmitted through the local barrier,
as it moves outwards.

In Figs.~\ref{fig_echo}(g) and (h) we compare 
the effective potential $V^\psi_{\rm eff}$ in several
UCO spacetimes with the effective potential in the Schwarzschild spacetime
(dashed black line).
Here the integration constant for the tortoise coordinate for the 
Schwarzschild solution is chosen such that the local maxima 
for $V^\psi_{\rm eff}$ in the particle-like and Schwarzschild spacetimes coincide.
Again, we note that in the outer region
the effective potentials agree very well, whereas in the inner region
the behavior is very different, since the black hole effective potential
features only the finite local barrier,
and any modes passing this barrier will disappear behind the horizon.
So only a single reflection will be seen for a wave impinging on a black hole.
The observation of the presence of echoes in a wave signal
will thus reveal the lack of a horizon of the compact object.

\section{Conclusions}

We have investigated a new type of solutions of EsGB theories, 
employing a variety of coupling functions for the scalar field
\cite{Kleihaus:2019rbg}.
These particle-like solutions are asymptotically flat,
and possess a globally regular metric.
While the scalar field diverges as $1/r$ at the origin,
this divergence is cancelled in the effective stress-energy tensor
by contributions from the GB term, yielding a
regular effective stress-energy tensor and thus a regular
source term in the Einstein equations.
When taking the GB coupling constant to zero, on the other hand,
the singular Fisher (JNWW) solution of GR is recovered.
Let us add, that regular particle-like solutions have also
been found in Einstein-scalar Maxwell theories
with various coupling functions
\cite{Herdeiro:2019iwl}.

We have presented the domain of existence of these
particle-like solutions in detail for quadratic, cubic and
dilaton coupling functions, allowing for either a vanishing or
finite (cosmological) values of the scalar field at asymptotic infinity, $\phi_\infty$.
For quadratic coupling and vanishing $\phi_\infty$,
the solutions are symmetric with respect to $\phi \to -\phi$,
and thus the domain of existence is symmetric with respect to
positive and negative scalar charge.
The symmetry is broken when $\phi_\infty \ne 0$, or when
a non-symmetric coupling is employed.

In the case of a quadratic coupling and $\phi_\infty=0$,
the domain of existence of particle-like solutions 
has an overlap with the domain of wormhole solutions. The reason is the occurrence
of particle-like solutions which feature a throat and an equator.
In this case, wormhole solutions can be constructed
by continuing symmetrically at the throat or equator (after a suitable coordinate transformation)
with a second asymptotically flat spacetime \cite{Antoniou:2019awm}.
The domain of existence of particle-like solutions consists of a tower of
distinct regions, that differ in the number of nodes of the
scalar-field function. 
We note that both the tower of distinct regions and the overlap
between particle-like and wormhole solutions are characteristic
features of the domain of existence for the quadratic coupling function - the
respective domains for the cubic or dilatonic coupling functions feature no
distinct regions, as the coupling constant increases, and no overlap is found
between the domains of the particle-like and wormhole solutions (which are
again present in the solution space of the theory as shown in \cite{Antoniou:2019awm}.)

The boundary of the domain of existence of the particle-like solutions is formed by
solutions with singularities. Mostly a cusp singularity is encountered
at the boundary, but also an ordinary curvature singularity 
can be encountered. Only in the case of the cubic coupling
with negative $\phi_\infty$, we have encountered a boundary formed by
scalarized black holes.

However, the focus of the paper has been the study of the properties of 
these particle-like solutions. Here we have first investigated the
redshift function $z$ of these solutions, which can assume
very large values. Interestingly, for the quadratic coupling function
with $\phi_\infty=0$ also negative values of $z$ are found
corresponding to blueshift.

Next, we have addressed the compactness of these particle-like solutions.
By investigating the effective stress-energy tensor, we have seen that
the energy density and pressure can be highly localized.
This yields a shell-like structure of the solutions corresponding
to almost empty bags in terms of the energy density.
For these solutions, the mass function rises very steeply
in the vicinity of the circumferential radius $R_c = 2M$,
which would correspond to the horizon of a Schwarzschild black hole.
In this case, then, the solutions are highly compact.

Subsequently, we have considered the geodesics in these spacetimes.
To show that many of the solutions qualify as %ultra-compact objects, 
UCOs, we have investigated the presence of lightrings.
As predicted on general grounds \cite{Cunha:2017qtt},
the lightrings of these particle-like solutions always come in pairs. The outer unstable lightring
corresponds to the lightring of the Schwarzschild spacetime.
But an inner stable lightring occurs as well, due to the
diverging (at the origin) angular momentum barrier of these
regular spacetimes. Thus, there are stable bound photon orbits
in these spacetimes.

Ordinary massive particles moving in these spacetimes do not feel the
divergence of the scalar field at the origin. They simply see
the gravitational field as given by the regular spacetime metric.
Thus these particles can pass over the center of the spacetime unhindered, 
they can oscillate across the center or even remain stationary at the center. 
If they carry angular momentum,
the presence of the infinite angular momentum barrier at the center 
will always allow for bound particle motion.

Lastly, we have addressed possible gravitational-wave signals for these
ultra-compact spacetimes, showing the presence of echoes
in the asymptotic wave signal. This was to be expected on general grounds
because of the absence of horizons in these particle-like solutions
 \cite{Cardoso:2016rao,Cardoso:2016oxy,Cardoso:2017cqb}.
In particular, we have studied the effective potential for a scalar wave,
which again features two barriers, a finite local barrier
and an infinite angular momentum barrier at the center.
An incoming wave will thus be fully reflected every time it encounters
the infinite central barrier, but only partly reflected, every time it encounters
the finite outer barrier. This sequence of reflections
will then give rise to a sequence of echoes with decreasing amplitude in the
gravitational wave signal, not present in signals from black holes.

Open questions to be answered in future work concern
the stability of these solutions. In particular, a quasi-normal mode analysis
could be performed, as has already been (partly) performed for 
EsGB black holes and wormholes.
It would also be interesting to address the existence of rotating
generalizations of these particle-like solutions
and of the associated wormholes.
So far only rotating EsGB black holes and their properties,
including their shadow, have been studied.

\section{Acknowledgments}

BK and JK gratefully acknowledge support by the
DFG Research Training Group 1620 {\sl Models of Gravity}
and the COST Actions CA15117 and CA16104. 
BK and PK acknowledge helpful
discussions with Eugen Radu and Athanasios Bakopoulos, respectively.

\end{document}